\documentclass[11pt]{article}
\usepackage[dvipdfmx]{graphicx}
\usepackage{amsmath,amssymb,color}
\usepackage{tikz}
\usetikzlibrary{intersections, calc, arrows}

\begin{document}

\begin{titlepage}
\begin{flushright}
    July 2025
\end{flushright}
\begin{center}
  \vspace{3cm}
  {\bf \Large Effective Action in M-Theory: $(DF)^4$ Superinvariants}
  \\  \vspace{2cm}
  Yoshifumi Hyakutake
  \\ \vspace{1cm}
  {\it College of Science, Ibaraki University, \\
   Bunkyo 2-1-1, Mito, Ibaraki 310-8512, Japan}
\end{center}

\vspace{2cm}
\begin{abstract}
 We construct superinvariants in M-theory via local supersymmetry, which include $(DF)^4$ terms and fermionic bilinear terms with mass dimension eight. The bases of the $(DF)^4$ and the fermionic bilinear terms are classified and the variations under the local supersymmetry are evaluated by using the Mathematica codes. By imposing the local supersymmetry, we find that 24 independent terms of $(DF)^4$ are governed by 10 parameters. The result is consistent with the effective action obtained by analyzing scattering amplitudes of superparticles or supermembranes.
\end{abstract}
\end{titlepage}

\section{Introduction} \label{sec:Introduction}

Requirement of local supersymmetry is a powerful tool to construct supergravities, and effective actions of superstring theories and M-theory\cite{Green:1987sp,Polchinski:1998rr}. By imposing the local supersymmetry, low energy effective action of the superstring theories are described by corresponding ten dimensional supergravities, and that of the M-theory is uniquely expressed by eleven dimensional supergravity\cite{Salam:1989fm}. In this paper, we will construct the effective action of the M-theory beyond the supergravity approximation, which is often called higher derivative terms, via the local supersymmetry.

There is a long history on the effective actions of the superstring theories and the M-theory, and several methods are  complementarily used to investigate their structures. Famous one is to evaluate scattering amplitudes in the superstring theories, which include exchange of massive string states as well as massless ones of a graviton, a Kalb-Ramond field (NS-NS $B$-field) and a dilaton field $\phi$\,\cite{Schwarz:1982jn,Gross:1986iv}. In the case of the type II superstring theories, nontrivial contributions to one particle irreducible action arise from 4 points amplitudes of gravitons, and the effective actions consist of quartic order in the Riemann tensor, which is symbolically written as $e^{-2\phi} t_8 t_8 R^4$ ($R^4$ terms)\cite{Gross:1986iv}\footnote{Calculations of tree level closed string amplitudes are simplified by employing the KLT relation, which states that a closed string amplitude can be obtained by a tensor product of two open strings ones\cite{Kawai:1985xq}.}. Generalization of the Riemann tensor with Kalb-Ramond field was done in ref.~\cite{Gross:1986mw}, and it is also shown that 3 points amplitudes give quadratic terms of the Riemann tensor ($R^2$ terms) as well as $R^4$ terms in the type I superstring theories. The effective action for the type II superstring theories at 1-loop 4 points amplitudes was studied in ref.~\cite{Sakai:1986bi}, and the result is  expressed by the same structure as $t_8 t_8 R^4$.

Around the same time, loop calculations of non-linear sigma models also played crucial roles to determine the structure of the effective actions of the superstring theories\cite{Grisaru:1986px}-\cite{Freeman:1986zh}. The non-linear sigma models describe closed strings which propagate in a background of massless fields, and the loop calculations on two dimensional world sheet give corrections to $\beta$-functions. It is necessary to set the $\beta$-functions to be zero because of the conformal invariance of the strings, and these give equations of motion for the massless background fields. Nontrivial corrections to the equations of motion for the supergravities arise from 4-loop calculations of the $\beta$-functions in the non-linear sigma models\cite{Grisaru:1986px}-\cite{Grisaru:1986kw}. The existence of $e^{-2\phi} t_8 t_8 R^4$ term is confirmed, and furthermore, it is clarified that the structure of the corrections should be $e^{-2\phi} (t_8 t_8 R^4 + \tfrac{1}{8}\epsilon_{10}\epsilon_{10} R^4)$\cite{Grisaru:1986vi,Freeman:1986zh}. It  is also expected that the effective actions become $t_8 t_8 R^4 \mp \tfrac{1}{8}\epsilon_{10}\epsilon_{10} R^4$ in the type II superstring theories ($-$ for IIA, $+$ for IIB), which are confirmed by evaluating the 1-loop 5 points amplitudes\cite{Peeters:2000qj,Richards:2008jg}.  With respect to $H^2R^3$ and $H^2(D H)^2 R$ terms, where $H$ represents a 3-form field strength of the $B$-field, similar structures of the effective actions also appear from 5 points amplitudes at 1-loop level\cite{Peeters:2001ub,Richards:2008sa,Liu:2013dna}, and additional kinematical structures show up at the tree level\cite{Grimm:2017okk,Liu:2019ses}. The effective action of the type II superstring theories including R-R massless fields are discussed in refs.~\cite{Peeters:2003pv,Policastro:2006vt}.

The effective actions of the type I and the heterotic superstring theories contain $\epsilon_{10}t_8 BR^4$ term to cancel an anomaly of the local Lorentz transformation\cite{Green:1984sg}. And from a duality between the type I superstring theory and the type IIA superstring theory on K3 surface, it also arises in the effective action of the type IIA superstring theory\cite{Vafa:1995fj}. Since the strong coupling limit of the type IIA superstring theory is described by the M-theory in eleven dimensions, the effective action of the M-theory is expressed as $t_8 t_8 R^4 - \tfrac{1}{4!}\epsilon_{11}\epsilon_{11} R^4 - \tfrac{1}{6} \epsilon_{11}t_8 AR^4$, where $A$ is a 3-from massless field\cite{Duff:1995wd}-\cite{Tseytlin:2000sf}. The structures of the effective actions of the type II superstring theories and M-theory are confirmed by imposing the local supersymmetry in eleven dimensions\cite{Peeters:2000qj},\cite{deRoo:1992sm}-\cite{Hyakutake:2007sm}. Requirement of the duality for the type II superstring theories also give strong restrictions on the structure of the effective action\cite{Codina:2020kvj}-\cite{Garousi:2022ghs}. See ref.~\cite{Ozkan:2024euj} for recent summary of the effective actions in the superstring theories and the M-theory.

The purpose of this paper is to reveal the structure of the effective action in the M-theory via the local supersymmetry. As mentioned above, there are several works to construct superinvariants which contain $t_8 t_8 R^4 - \tfrac{1}{4!}\epsilon_{11}\epsilon_{11} R^4 - \tfrac{1}{6} \epsilon_{11}t_8 AR^4$ terms\cite{Peeters:2000qj,Hyakutake:2005rb,Hyakutake:2006aq}, and there is some extensions to $F^2R^3$ terms, where $F$ is a field strength of the 3-form field $A$\,\cite{Hyakutake:2007sm}. The approach of this paper is opposite to these works. Namely we start from $(DF)^4$ and super partners of gravitino bilinear terms to impose the local supersymmetry. Since the structure of the $R^4$ terms is well understood, it is natural to investigate whether these are supersymmetric or not. If we wish to construct bosonic parts of the structure completely, however, it is efficient to start from $(DF)^4$ terms. There are several technical reasons, but one thing we should notice is that in order to check the local supersymmetry, we need to deal with variations with covariant derivatives. These terms will become commutators of the covariant derivatives after partial integrals, and contain the Riemann tensor.  Thus it is natural to start from $(DF)^4$ terms at $\mathcal{O}(R^0)$, and after completing the cancellation of the variations at this order, we consider higher powers of the Riemann tensor step by step. Although the method of imposing the local supersymmetry is straightforward, at least we should classify all possible terms in the effective action with making some ansatz, and also classify their variations under the local supersymmetry. To execute these calculations, it is inevitable to build original Mathematica codes.  

The bosonic terms at $\mathcal{O}(R^0)$ consist of $[e(DF)^4]$, $[e\epsilon_{11}F^2(DF)^3]$, $[eF^4(DF)^2]$, $[e\epsilon_{11}F^6DF]$ and $[eF^8]$. Here $e$ is a volume factor and, throughout this paper, the notation $[X]$ is used to ignore indices of the tensor $X$. There are so many terms (more than million) at this level, so we only examine $[e(DF)^4]$ terms in the above list. There are 24 (or 56, if we include terms which partially contain the equations of motion for the eleven dimensional supergravity) independent terms for $[e(DF)^4]$ and we will restrict the structure of these terms by imposing the local supersymmetry. As a final result, 24 coefficients of these bosonic terms are governed  by 10 free parameters. This result is consistent with the effective action obtained by the calculations of scattering amplitudes of superstrings, superparticles or super membranes\cite{Gross:1986mw,Deser:2000xz,Peeters:2005tb}.

The paper is organized as follows. In section \ref{sec:Overview}, we fix notations by reviewing the eleven dimensional supergravity, and overview a cancellation mechanism of the variations under the local supersymmetry. The ansatz for the effective action at the level of $\mathcal{O}(R^0)$ is explicitly given, and the variations of these terms under the local supersymmetry are also listed. In section \ref{sec:ActionVariation}, we concentrate on a part of the effective action at $\mathcal{O}(R^0)$, and explain a scheme to classify terms of $B_1=[e(DF)^4]$, $B_2=[e\epsilon_{11}R(DF)^3]$, $F_1=[e(DF)^2 \overline{\psi_2} \gamma \mathcal{D} \psi_2]$, $F_2=[e(DF)^3 \bar{\psi} \gamma \psi_2]$ and $F_3=[eF(DF)^3 \bar{\psi} \gamma \psi]$. Here $[\psi]$ represents a Majorana gravitino, $[\psi_2]$ does its field strength and $[\gamma]$ does a gamma matrix. We also classify their variations under the local supersymmetry, which are represented by $V_1=[e(DF)^2 DDF \bar{\epsilon} \gamma \psi_2]$, $V_2=[e(DF)^3 \bar{\epsilon} \gamma \mathcal{D} \psi_2]$, $V_4=[e(DF)^4 \bar{\epsilon} \gamma \psi]$ and $V_5=[eF(DF)^2 DDF \bar{\epsilon} \gamma \psi]$. In section \ref{sec:SusyVariation}, the variations of $B_1$, $B_2$, $F_1$, $F_2$ and $F_3$ under the local supersymmetry are performed in detail. In section \ref{sec:Cancel}, we consider the cancellation mechanism of $V_1$, $V_2$, $V_4$ and $V_5$. At this stage, we solve Bianchi identities and dimension dependent ones to reduce the number of independent terms in $V_1$, $V_2$, $V_4$ and $V_5$. It should be carefully executed because some terms in $V_2$ and $V_4$ are related with each other by the Bianchi identities. We also explain field redefinition ambiguities in the effective action and corrections to the local supersymmetry transformations. The latter also gives nontrivial relations among terms in $V_2$ and $V_4$, so it should be carefully treated. Finally the structure of the effective action for $B_1$ and $B_2$ are explicitly shown and these are expressed by 10 parameters. In section \ref{sec:Scattering}, we show that the effective action of the M-theory obtained by imposing the local supersymmetry is consistent with that obtained by scattering amplitudes of supermembranes. Section \ref{sec:Conclusion} is devoted to conclusions and discussions. In appendix \ref{app:appHDT}, we show technical calculations of higher derivative terms in detail. In appendix \ref{app:anothersol}, we give another solution for the effective action which is obtained by assuming that there are no corrections to the local supersymmetry transformations. The result is not consistent with that of the scattering amplitudes, so this means that there should be corrections to the local supersymmetry transformations. Since we employ the Mathematica codes throughout this paper, we write tips for the codes in appendix \ref{app:Code}.

\section{Overview of the Effective Action in the M-Theory} \label{sec:Overview}

\subsection{Notation of the eleven dimensional supergravity} \label{subsec:Notation}

The M-theory preserves $\mathcal{N}=1$ local supersymmetry which is expected to be strong enough to determine the structure of the theory. In fact, the low energy limit of the M-theory is uniquely described by the eleven dimensional supergravity. In this subsection, we briefly review these well-known results while fixing notations. The prescription in this subsection is also useful to investigate a strategy for finding higher derivative terms in the M-theory.

The field content of the eleven dimensional supergravity is simple and consists of a vielbein $e^\mu{}_a$, a Majorana gravitino $\psi_\mu$ and a three-form potential $A_{\mu\nu\rho}$. Here Greek indices $\mu, \nu, \rho, \cdots$ and Latin ones $a,b,c,\cdots$, which run from 0 to 10, label space-time coordinates and local Lorentz ones, respectively. Spinor index of the Majorana gravitino is neglected as usual. Since the Lagrangian density of the eleven dimensional supergravity should be invariant under the general coordinate transformation and $U(1)$ gauge one for the 3-form field, the building blocks are the volume factor $e$, the Riemann tensor $R_{abcd}$, the covariant derivative $D_a$, the 4-form field strength $F_{abcd}$, the Majorana gravitino and its field strength. The 4-form field strength is defined as $F_{\mu\nu\rho\sigma}=4\partial_{[\mu}A_{\nu\rho\sigma]}$, and the notation $[\mu_1 \cdots \mu_n]$ is used to antisymmetrize the indices completely with the weighting factor $\frac{1}{n!}$. The field strength of the Majorana gravitino is defined by $\psi_{\mu\nu} = 2 \mathcal{D}_{[\mu} \psi_{\nu]}$ with
\begin{alignat}{3}
  \mathcal{D}_\mu \psi_\nu = D_\mu \psi_\nu + F_\mu \psi_\nu, \qquad 
  F_\mu = - \tfrac{1}{36} F_{\mu ijk}\gamma^{ijk} 
  + \tfrac{1}{288} F_{ijkl}\gamma_\mu{}^{ijkl}. \label{eq:curlD}
\end{alignat}
Here $D_\mu$ represents an ordinary covariant derivative which acts only on the local Lorentz indices as follows.
\begin{alignat}{3}
  D_\mu &= \partial_\mu + \tfrac{1}{2} \omega_{\mu ab} T^{ab}, &\qquad& 
  T^{ab} = 2 \delta^{[a}_i \delta^{b]}_j \; \text{(vector)}, \quad
  T^{ab} = \tfrac{1}{2} \gamma^{ab} \; \text{(spinor).} \label{eq:LorentzD}
\end{alignat}
Gamma matrices are denoted by $\gamma^i (i=0,1,\cdots,10)$ and $\gamma^{i_1\cdots i_n} = \gamma^{[i_1} \cdots \gamma^{i_n]}$. 

We require the following parity invariance for the action.
\begin{alignat}{3}
  x^{10} \;\to\; -x^{10}, \qquad A \;\to\; -A. \label{eq:11parity}
\end{alignat}
Because of this requirement, the 3-form field should appear together with the antisymmetric tensor $\epsilon_{11}^{a_1 \cdots a_{11}}$. Up to $\mathcal{O}(\psi^3)$, the action of the eleven dimensional supergravity is given by
\begin{alignat}{3}
  S_0 &= \frac{1}{2 \kappa_{11}^2} \int d^{11}x \, \mathcal{L}_0, \notag 
  \\
  \mathcal{L}_0 &= e \big\{ R - \tfrac{1}{2} \overline{\psi_c} \gamma^{cab} \psi_{ab}
  - \tfrac{1}{2 \cdot 4!} F_{abcd}F^{abcd} 
  - \tfrac{1}{(3!\,4!)^2} \epsilon_{11}^{a_1\cdots a_{11}} A_{a_1 a_2 a_3} F_{a_4\cdots a_7} F_{a_8\cdots a_{11}}
  \big\}, \label{eq:sugra}
\end{alignat}
where $2\kappa_{11}^2 = (2\pi)^8\ell_p^9$ and $\ell_p$ is the eleven dimensional Planck length.
We choose the notations of the antisymmetric tensor and gamma matrices as $\epsilon_{11}^{012\cdots 10}=1$ and $\gamma^{10}=-\gamma^0\gamma^1\cdots\gamma^{9}$, which mean $\epsilon_{11}^{012\cdots 10} {\bf 1}=-\gamma^{012\cdots 10}$. In general, gamma matrices satisfy the relation of
\begin{alignat}{3}
  \gamma^{i_1\cdots i_n} = \frac{(-1)^\frac{n(n-1)}{2}}{(11-n)!}
  \epsilon_{11}^{i_1\cdots i_n i_{n+1} \cdots i_{11}} \gamma_{i_{n+1} \cdots i_{11}}.
  \label{eq:gammadual}
\end{alignat}
A general formula for a product of two gamma matrices are given in the appendix \ref{app:gamma}.

Let us consider a variation of the Lagrangian density $\mathcal{L}_0$ up to total derivative terms. This is written as
\begin{alignat}{3}
  \delta \mathcal{L}_0 &= e E(e)^a{}_\mu \delta e^\mu{}_a + e \overline{\delta \psi_\mu} E(\psi)^\mu
  + e E(A)^{\mu\nu\rho} \delta A_{\mu\nu\rho}, \label{eq:deltaL0}
\end{alignat}
where
\begin{alignat}{3}
  E(e)_{ab} &= 
  2 R_{ab} - \eta_{ab} R - \tfrac{1}{6} F_{aijk} F_{bijk} 
  + \tfrac{1}{48} \eta_{ab} F_{ijkl} F^{ijkl} + \mathcal{O}(\psi^2), \notag
  \\
  E(\psi)^a &= - \gamma^{abc} \psi_{bc} + \mathcal{O}(\psi^3), \label{eq:EoM}
  \\
  E(A)^{abc} &= \tfrac{1}{6} D_d F^{dabc} 
  - \tfrac{1}{48 \cdot 144} \epsilon_{11}^{abcijklmnop} F_{ijkl} F_{mnop} 
  + \mathcal{O}(\psi^2). \notag
\end{alignat}
$E(e)_{ab}=0$ and $E(A)^{abc}=0$ become field equations for the vielbein and the 3-form field, and $E(\psi)^{a}=0$ does that for the Majorana gravitino.

Let us briefly check that the action (\ref{eq:sugra}) is invariant under $\mathcal{N}=1$ local supersymmetry in eleven dimensions.  Up to $\mathcal{O}(\psi^2)$, the local supersymmetry transformations of the vielbein, the Majorana gravitino and the 3-form field are written by
\begin{alignat}{3}
  &\delta_0 e^a{}_{\mu} = \bar{\epsilon} \gamma^a \psi_\mu, \qquad
  \delta_0 \psi_\mu = 2 \mathcal{D}_\mu \epsilon, \qquad
  \delta_0 A_{\mu\nu\rho} = -3 \bar{\epsilon} \gamma_{[\mu\nu} \psi_{\rho]}, \label{eq:susytr}
\end{alignat}
where $\epsilon$ is a space-time dependent parameter which belongs to the Majorana spinor representation. Note that $\delta_0 e^\mu{}_a = - \bar{\epsilon} \gamma^\mu \psi_a$. The variation of the bosonic part in the eq.~(\ref{eq:sugra}) is evaluated as
\begin{alignat}{3}
  &\delta_0 \Big[ e \Big\{ R - \frac{1}{2 \cdot 4!} F_{abcd}F^{abcd} 
  - \frac{1}{(3!\,4!)^2} \epsilon_{11}^{a_1\cdots a_{11}} A_{a_1 a_2 a_3} F_{a_4\cdots a_7} F_{a_8\cdots a_{11}} 
  \Big\}\Big] \notag
  \\
  &= e E(e)^a{}_\mu \delta_0 e^\mu{}_a + e E(A)^{\mu\nu\rho} \delta_0 A_{\mu\nu\rho} \notag
  \\
  &= - e E(e)_{ab} \bar{\epsilon} \gamma^a \psi^b - 3 e E(A)^{abc} \bar{\epsilon} \gamma_{ab} \psi_c, \label{eq:sugraboson}
\end{alignat}
up to $\mathcal{O}(\psi^2)$. In the above we neglected $\delta_0 \omega_{\mu ab}$ part since it should be cancelled by appropriately choosing a torsion part of the spin connection.

In order to calculate a transformation of the fermionic bilinear term in the action, first of all let us examine a variation of the field strength of the Majorana gravitino. Up to $\mathcal{O}(\psi)$, the variation of the gravitino under the local supersymmetry becomes
\begin{alignat}{3}
  \delta_0 \psi_{ab} &= 2 [\mathcal{D}_a, \mathcal{D}_b] \epsilon, \notag
  \\
  &= 2 \big\{ \tfrac{1}{4} R_{ijab} \gamma^{ij} 
  - \tfrac{1}{18} (D_{[a} F_{b] ijk}) \gamma^{ijk} 
  + \tfrac{1}{144} (D_{[a} F^{ijkl} ) \gamma_{b]ijkl} \notag
  \\
  &\quad\, +
  \tfrac{1}{1728} F_{ijkl} F^{ijkl} \gamma _{ab} 
  + \tfrac{1}{216} F^{ijkl} F_{ijk[a}\gamma_{b]l} 
  -\tfrac{1}{48} F_{aikl}F_{bj}{}^{kl} \gamma^{ij} 
  +\tfrac{1}{72} F^{ij}{}_{kl}F_{ijm[a} \gamma_{b]}{}^{klm} \notag
  \\
  &\quad\,
  + \tfrac{1}{108} F_{ab}{}^{ij} F_{iklm} \gamma_j{}^{klm} 
  -\tfrac{1}{576} F^{ijkl} F_{ij}{}^{mn} \gamma_{abklmn} 
  -\tfrac{1}{432}F^{ijkl}F_{imn[a} \gamma_{b]jkl}{}^{mn} \label{eq:comm}
  \\
  &\quad\,
  +\tfrac{1}{864}F_{aijk}F_{blmn} \gamma^{ijklmn} 
  -\tfrac{1}{2592}F_{ijkl}F_{mno[a}\gamma_{b]}{}^{ijklmno} 
  +\tfrac{1}{41472}F^{ijkl}F^{pqrs} \gamma_{abijklpqrs} \big\} \epsilon \notag
  \\
  &\equiv 2 \big( \tfrac{1}{4} R_{ijab} \gamma^{ij} + DF_{ab} + F^2_{ab} \big) \epsilon , \notag
\end{alignat}
and the commutation relation which acts on $\psi_c$ becomes
\begin{alignat}{3}
  [\mathcal{D}_a, \mathcal{D}_b] \psi_c
  &= R_c{}^i{}_{ab} \psi_i + \tfrac{1}{4} R_{ijab} \gamma^{ij} \psi_c + DF_{ab} \psi_c 
  + F^2_{ab} \psi_c. \label{eq:comm2}
\end{alignat}
After some calculations, it is also possible to show following relations of
\begin{alignat}{3}
  \gamma^{b} \big( \tfrac{1}{4} R_{ijab} \gamma^{ij} \!+\! DF_{ab} \!+\! F^2_{ab} \big) 
  &= - \tfrac{1}{4} E(e)_{ai} \gamma^i \!+\! \tfrac{1}{36} E(e)^i{}_i \gamma_a
  \!-\! \tfrac{1}{2} E(A)_{aij} \gamma^{ij} \!+\! \tfrac{1}{12} E(A)^{ijk} \gamma_{aijk}, \notag
  \\
  \gamma^{ab} \big( \tfrac{1}{4} R_{ijab} \gamma^{ij} \!+\! DF_{ab} \!+\! F^2_{ab} \big) 
  &= \tfrac{1}{18} E(e)^i{}_i \!+\! \tfrac{1}{6}E(A)^{ijk} \gamma_{ijk}, \label{eq:RDFF2}
  \\
  \gamma^{cab} \big( \tfrac{1}{4} R_{ijab} \gamma^{ij} \!+\! DF_{ab} \!+\! F^2_{ab} \big) 
  &= \tfrac{1}{2} E(e)^{ci} \gamma_i \!+\! \tfrac{3}{2}E(A)^{cij} \gamma_{ij}. \notag
\end{alignat}
By using the above relations and $\overline{F_c}\gamma^{cab} = - \gamma^{cab} F_c$, the variation of the Majorana gravitino bilinear term in the eq.~(\ref{eq:sugra}) is calculated as
\begin{alignat}{3}
  \delta_0 \Big[ - \frac{1}{2} e \overline{\psi_c} \gamma^{cab} \psi_{ab} \Big] 
  &= - 2 e \overline{\psi_c} \gamma^{cab} [\mathcal{D}_a, \mathcal{D}_b] \epsilon \notag
  \\
  &= - 2 e \overline{\psi_c} \gamma^{cab} 
  \big( \tfrac{1}{4} R_{ijab} \gamma^{ij} + DF_{ab} + F^2_{ab} \big) \epsilon \notag
  \\
  &= e E(e)^{ab} \bar{\epsilon} \gamma_b \psi_a + 3 e E(A)^{abc} \bar{\epsilon} \gamma_{bc} \psi_a, \label{eq:sugrafermion}
\end{alignat}
up to $\mathcal{O}(\psi^2)$. Again, in the above we neglected $\delta_0 \omega_{\mu ab}$ part since it should be cancelled by appropriately choosing the torsion part of the spin connection. Thus the variations of (\ref{eq:sugraboson}) and (\ref{eq:sugrafermion}) under the local supersymmetry can be cancelled up to $\mathcal{O}(\psi^2)$. 

\begin{figure}[htb]
\begin{center}
\includegraphics[width=10cm]{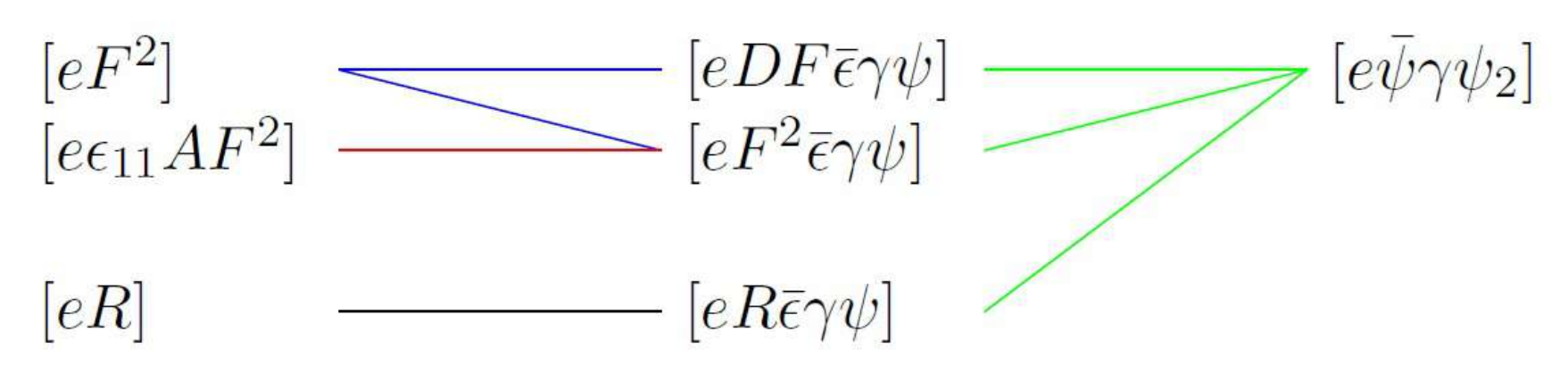}
\caption{Variations of the eleven dimensional supergravity}\label{fig:sugra}
\end{center}
\end{figure}

Throughout this paper, we use the notation $[X]$ to abbreviate local Lorentz indices of the tensor $X$, and $[X]$ represents a collection of terms in general. For example,  $[F\bar{\psi}\gamma \psi]$ consists of terms like
\begin{alignat}{3}
  [eF\bar{\psi} \gamma \psi] = \{e F_{abcd}\overline{\psi_a} \gamma_{bc} \psi_d, \,
  e F_{abcd}\overline{\psi_e} \gamma_{abcdef} \psi_f , \, \cdots \}. \label{eq:FPP}
\end{alignat}
Using this notation, the variations of fields under the local supersymmetry transformations are expressed as
\begin{alignat}{3}
  [\delta_0 e] &= [\bar{\epsilon} \gamma \psi], \qquad
  [\delta_0 \psi] &= [D \epsilon] + [F \gamma \epsilon], \qquad
  [\delta_0 A] &= [\bar{\epsilon} \gamma \psi]. \label{eq:simplesusytr}
\end{alignat}
Then the variation of the Riemann tensor $\delta R_{abcd}$ up to $\mathcal{O}(\psi^2)$ is expressed as
\begin{alignat}{3}
  [\delta_0 R] &= [R \bar{\epsilon} \gamma \psi] + [D(\delta_0 \omega)], \label{eq:trR}
\end{alignat}
and the gravitino field strength, denoted as $\psi_2$, transforms like
\begin{alignat}{3}
  [\delta_0 \psi_2] &= [R \epsilon] + [DF \epsilon] + [F^2\epsilon] + [\delta_0 \omega \gamma \psi]. \label{eq:trpsi2}
\end{alignat}
The variation of the action up to $\mathcal{O}(\psi^2)$ is summarized in fig.~\ref{fig:sugra}. The bosonic terms in the action are located on the left, and the fermionic bilinear term is done on the right. The variations under the local supersymmetry are placed at the center.

As a remark, in the case of the supergravity, the spin connection is dealt as an independent variable, and the variation with respect to the spin connection is set to be zero in order to express the spin connection in terms of the vielbein and bilinears of the Majorana gravitino. On the other hand, in the case of higher derivative corrections, we need to take into account $[D(\delta_0 \omega)]$ in the eq.~(\ref{eq:trR}). Since $[\delta_0 \omega \gamma \psi]$ in the eq.~(\ref{eq:trpsi2}) consists of bilinears of the Majorana gravitino, its contribution is neglected if we examine the variations up to $\mathcal{O}(\psi^2)$. Namely, we use
\begin{alignat}{3}
  [\delta_0 \psi_2] &= [R \epsilon] + [DF \epsilon] + [F^2\epsilon], \label{eq:trpsi2simple}
\end{alignat}
for the higher derivative terms.

\subsection{Ansatz for the effective action up to $\mathcal{O}(\psi^3)$ and variations up to $\mathcal{O}(\psi^2)$} \label{subsec:Ansatz}

In this subsection, we sketch the strategy of finding higher derivative terms in the M-theory via the local supersymmetry. With respect to the effective action in the M-theory, leading corrections to the eleven dimensional supergravity arise from terms with mass dimension eight, which correspond to 1-loop 4 points (or higher points) interactions in the type IIA superstring theory. General ideas for the ansatz of the effective action up to $\mathcal{O}(\psi^3)$, as well as the variations up to $\mathcal{O}(\psi^2)$, are listed as follows.

\noindent
\begin{enumerate}
\item Bosonic terms in the effective action consist of combinations of $[\hat{R}]$, $[\hat{F}]$ and $[D\hat{F}]$. Total number of $[\hat{R}]$ and $[\hat{F}]$ should be equal to or more than four. 

\item Fermionic bilinear terms in the effective action consist of combinations of $[R]$, $[F]$, $[DF]$, $[\bar{\psi}\gamma\psi]$, $[\bar{\psi}\gamma\psi_2]$, $[\overline{\psi_2}\gamma\psi_2]$ and $[\overline{\psi_2}\gamma \mathcal{D} \psi_2]$. The rank of gamma matrix for $[\bar{\psi}\gamma\psi]$ should be chosen so that two gravitinos are antisymmetric under the exchange. Total number of $[R]$ and $[F]$ should be equal to or more than two. Notice that $[\bar{\psi}\gamma \mathcal{D} \psi_2]$, which is antisymmetric with respect to the former gravitino and the covariant derivative, is mapped to $[\overline{\psi_2}\gamma\psi_2]$ and $[\bar{\psi}\gamma\psi_2]$ by partial integrals. Similarly, $[\overline{\mathcal{D} \psi_2}\gamma \mathcal{D} \psi_2]$, which is antisymmetric with respect to two covariant derivatives, is mapped to $[\overline{\psi_2}\gamma \mathcal{D} \psi_2]$ and $[\overline{\psi_2}\gamma\psi_2]$ by partial integrals. 

\item Variations under the local supersymmetry transformations consist of combinations of $[R]$, $[DR]$, $[F]$, $[DF]$, $[DDF]$, $[\bar{\epsilon}\gamma\psi]$, $[\bar{\epsilon}\gamma\psi_2]$ and $[\bar{\epsilon} \gamma\mathcal{D} \psi_2]$. $[DR]$ and $[DDF]$ should not be duplicated.  Total number of $[R]$ and $[F]$ should be equal to or more than two.
\end{enumerate}

There are several remarks on the above ansatz. First, $[\hat{R}]$ and $[\hat{F}]$ will be defined in eqs.~(\ref{eq:spinhat}), (\ref{eq:Rhat}) and (\ref{eq:Fhat}) below, and these do not consist of purely bosonic variables. Note also that we do not use these `hat' variables in the fermionic bilinears in the effective action, since we do not take variations of $[R]$ an $[F]$ in those terms in this paper. Similarly we do not use `hat' variables in the variations under the local supersymmetry, since we examine the cancellation up to $\mathcal{O}(\psi^2)$ in this paper. Second, we made the ansatz so that the terms in the effective action contain minimum number of the covariant derivatives. In other words, we do not consider terms with more covariant derivatives, such as $[D\hat{R}]$ and $[DD\hat{F}]$, in the effective action. The reason is that the inclusion of such terms with more covariant derivatives causes more partial integrals to cancel the variations and makes the cancellation mechanism quite complicate.

Let us justify the above ansatz in detail. In order to check the local supersymmetry for the higher derivative terms, it is useful to define supercovariant quantities for the Riemann tensor and the 4-form field strength. First let us consider the supercovariant expression of the Riemann tensor. The supersymmetry transformation of the spin connection, $\delta_0 \omega_{cab} \equiv e^\mu{}_c \delta_0 \omega_{\mu ab}$, is given by
\begin{alignat}{3}
  \delta_0 \omega^c{}_{ab} = - D^{c} (\bar{\epsilon} \gamma_{[a} \psi_{b]})
  - D_{[a} (\bar{\epsilon} \gamma_{b]} \psi^{c}) - D_{[a} (\bar{\epsilon} \gamma^c \psi_{b]}). \label{eq:deltaspin}
\end{alignat}
This expression is not convenient since the right hand side includes the covariant derivative of the supersymmetry parameter. In order to cancel these terms, the connection should be modified so as to contain a torsion of $T^c{}_{ab} = \tfrac{1}{2} \overline{\psi_{[a}} \gamma^c \psi_{b]}$. Then a supercovariant spin connection is defined as 
\begin{alignat}{3}
  \hat{\omega}_{\mu ab} \equiv 
  \omega_{\mu ab} + \tfrac{1}{2} T_{\mu ab} - \tfrac{1}{2} T_{ab\mu} + \tfrac{1}{2} T_{ba\mu}, \label{eq:spinhat}
\end{alignat}
and it transforms under the local supersymmetry as
\begin{alignat}{3}
  &\delta_0 \hat{\omega}_{cab} = \tfrac{1}{2} \bar{\epsilon} \gamma_a \psi_{bc} 
  - \tfrac{1}{2} \bar{\epsilon} \gamma_b \psi_{ac} - \tfrac{1}{2} \bar{\epsilon} \gamma_c \psi_{ab} 
  - \tfrac{1}{6} F_{abij} \bar{\epsilon} \gamma^{ij} \psi_{c} 
  - \tfrac{1}{144} F^{ijkl} \bar{\epsilon} \gamma_{abijkl} \psi_{c}. \label{eq:deltaspinhat}
\end{alignat}
Here we made use of the relation $\gamma_a F_b + \overline{F_b} \gamma_a = \tfrac{1}{6} F_{abij} \gamma^{ij} + \tfrac{1}{144} F^{ijkl} \gamma_{abijkl}$ and $\overline{\gamma^{\mu_1 \cdots \mu_n}} \equiv (-1)^{n(n+1)/2}\gamma^{\mu_1 \cdots \mu_n}$. As expected, this variation does not include the covariant derivative of the Majorana spinor parameter. 

For the higher derivative terms, we define the Riemann tensor by using the supercovariant spin connection as $R(\hat{\omega})^a{}_{b\mu\nu} \equiv \partial_\mu \hat{\omega}_\nu{}^a{}_b - \partial_\nu \hat{\omega}_\mu{}^a{}_b
+ \hat{\omega}_\mu{}^a{}_c \hat{\omega}_\nu{}^c{}_b - \hat{\omega}_\nu{}^a{}_c \hat{\omega}_\mu{}^c{}_b$. The Riemann tensor $R(\hat{\omega})_{abcd} = R(\hat{\omega})_{ab\mu\nu} e^\mu{}_c e^\nu{}_d$, however, is not symmetric under the exchange of $(a,b)$ pair and $(c,d)$ pair, so we define an alternative Riemann tensor for the higher derivative terms as
\begin{alignat}{3}
  \hat{R}_{abcd} &\equiv \tfrac{1}{2} R(\hat{\omega})_{abcd} 
  + \tfrac{1}{2}R(\hat{\omega})_{cdab}. \label{eq:Rhat}
\end{alignat}
Obviously, $\hat{R}_{abcd}$ is symmetric under the exchange of $(a,b)$ pair and $(c,d)$ pair, and coincides with the ordinary Riemann tensor if the torsion part is ignored. Up to $\mathcal{O}(\psi^2)$, the variation of $\hat{R}_{abcd}$ under the local supersymmetry becomes as
\begin{alignat}{3}
  \delta_0 \hat{R}_{abcd} &= - \tfrac{1}{2} (R_{ebcd} \bar{\epsilon} \gamma^e \psi_a 
  + R_{aecd} \bar{\epsilon} \gamma^e \psi_b + R_{abed} \bar{\epsilon} \gamma^e \psi_c 
  + R_{abce} \bar{\epsilon} \gamma^e \psi_d) \label{eq:deltaRhat}
  \\
  &\quad\,
  - D_{a} (\tfrac{1}{4} \bar{\epsilon} \gamma_b \psi_{cd} 
  + \tfrac{1}{4} \bar{\epsilon} \gamma_c \psi_{bd} 
  + \tfrac{1}{4} \bar{\epsilon} \gamma_d \psi_{cb}
  + \tfrac{1}{12} F_{cdij} \bar{\epsilon} \gamma^{ij} \psi_{b}
  + \tfrac{1}{288} F^{ijkl} \bar{\epsilon} \gamma_{cdijkl} \psi_{b}) \notag
  \\
  &\quad\,
  - D_{b} (\tfrac{1}{4} \bar{\epsilon} \gamma_a \psi_{dc} 
  + \tfrac{1}{4} \bar{\epsilon} \gamma_c \psi_{da} 
  + \tfrac{1}{4} \bar{\epsilon} \gamma_d \psi_{ac}
  - \tfrac{1}{12} F_{cdij} \bar{\epsilon} \gamma^{ij} \psi_{a}
  - \tfrac{1}{288} F^{ijkl} \bar{\epsilon} \gamma_{cdijkl} \psi_{a}) \notag
  \\
  &\quad\,
  - D_{c} (\tfrac{1}{4} \bar{\epsilon} \gamma_d \psi_{ab} 
  + \tfrac{1}{4} \bar{\epsilon} \gamma_a \psi_{db} 
  + \tfrac{1}{4} \bar{\epsilon} \gamma_b \psi_{ad}
  + \tfrac{1}{12} F_{abij} \bar{\epsilon} \gamma^{ij} \psi_{d}
  + \tfrac{1}{288} F^{ijkl} \bar{\epsilon} \gamma_{abijkl} \psi_{d})\notag
  \\
  &\quad\,
  - D_{d} (\tfrac{1}{4} \bar{\epsilon} \gamma_c \psi_{ba} 
  + \tfrac{1}{4} \bar{\epsilon} \gamma_a \psi_{bc} 
  + \tfrac{1}{4} \bar{\epsilon} \gamma_b \psi_{ca}
  - \tfrac{1}{12} F_{abij} \bar{\epsilon} \gamma^{ij} \psi_{c}
  - \tfrac{1}{288} F^{ijkl} \bar{\epsilon} \gamma_{abijkl} \psi_{c}). \notag
\end{alignat}

Next let us consider the supercovariant expression of the 4-form field strength. The variation of the 4-form field is given by
\begin{alignat}{3}
  \delta_0 F_{abcd} &= - 4 F_{e[bcd} \bar{\epsilon} \gamma^e \psi_{a]} 
  - 12 D_{[a} (\bar{\epsilon} \gamma_{bc} \psi_{d]}). \label{eq:deltaF}
\end{alignat}
In order to cancel terms which include $D_a \bar{\epsilon}$ in the above, we define a supercovariant 4-form field strength as
\begin{alignat}{3}
  \hat{F}_{abcd} &= F_{abcd} + 3 \overline{\psi_{[a}} \gamma_{bc} \psi_{d]}. \label{eq:Fhat}
\end{alignat}
Then, up to $\mathcal{O}(\psi^2)$, the vatiation of $\hat{F}_{abcd}$ under the local supersymmetry is expressed as
\begin{alignat}{3}
  \delta_0 \hat{F}_{abcd} &= - 6 \bar{\epsilon} \gamma_{[ab} \psi_{cd]}. \label{eq:deltaFhat}
\end{alignat}
The above variation does not include $D_a \epsilon$ term as expected. Furthermore, it does not include the 4-form flux part explicitly. By using the eq.~(\ref{eq:deltaFhat}), the variation of $D_e \hat{F}_{abcd}$ under the local supersymmetry is evaluated as
\begin{alignat}{3}
  \delta_0 (D_e \hat{F}_{abcd}) &= - D_f F_{abcd} \bar{\epsilon} \gamma^f \psi_e \notag
  \\&\quad\,
  - D_e(\bar{\epsilon} \gamma_{ab} \psi_{cd}) - D_e(\bar{\epsilon} \gamma_{ac} \psi_{db})
  - D_e(\bar{\epsilon} \gamma_{ad} \psi_{bc}) \notag
  \\&\quad\,
  - D_e(\bar{\epsilon} \gamma_{cd} \psi_{ab}) - D_e(\bar{\epsilon} \gamma_{db} \psi_{ac})
  - D_e(\bar{\epsilon} \gamma_{bc} \psi_{ad}) \notag
  \\&\quad\,
  - \tfrac{1}{2} F^f{}_{bcd} \big\{ D_e(\bar{\epsilon}\gamma_a\psi_f) + D_a(\bar{\epsilon}\gamma_f\psi_e) 
  + D_a(\bar{\epsilon}\gamma_e\psi_f) \notag
  \\&\qquad\qquad\quad
  - D_e(\bar{\epsilon}\gamma_f\psi_a) - D_f(\bar{\epsilon}\gamma_a\psi_e)
  - D_f(\bar{\epsilon}\gamma_e\psi_a) \big\} \notag
  \\&\quad\,
  - \tfrac{1}{2} F_a{}^f{}_{cd} \big\{ D_e(\bar{\epsilon}\gamma_b\psi_f) + D_b(\bar{\epsilon}\gamma_f\psi_e)
  + D_b(\bar{\epsilon}\gamma_e\psi_f) \label{eq:deltaDFhat}
  \\&\qquad\qquad\quad
  - D_e(\bar{\epsilon}\gamma_f\psi_b) - D_f(\bar{\epsilon}\gamma_b\psi_e)
  - D_f(\bar{\epsilon}\gamma_e\psi_b) \big\} \notag
  \\&\quad\,
  - \tfrac{1}{2} F_{ab}{}^f{}_d \big\{ D_e(\bar{\epsilon}\gamma_c\psi_f) + D_c(\bar{\epsilon}\gamma_f\psi_e)
  + D_c(\bar{\epsilon}\gamma_e\psi_f) \notag
  \\&\qquad\qquad\quad
  - D_e(\bar{\epsilon}\gamma_f\psi_c) - D_f(\bar{\epsilon}\gamma_c\psi_e)
  - D_f(\bar{\epsilon}\gamma_e\psi_c) \big\} \notag
  \\&\quad\,
  - \tfrac{1}{2} F_{abc}{}^f \big\{ D_e(\bar{\epsilon}\gamma_d\psi_f) + D_d(\bar{\epsilon}\gamma_f\psi_e)
  + D_d(\bar{\epsilon}\gamma_e\psi_f) \notag
  \\&\qquad\qquad\quad
  - D_e(\bar{\epsilon}\gamma_f\psi_d) - D_f(\bar{\epsilon}\gamma_d\psi_e)
  - D_f(\bar{\epsilon}\gamma_e\psi_d) \big\}. \notag
\end{alignat}
In our notation, the variations (\ref{eq:deltaspin}), (\ref{eq:deltaspinhat}), (\ref{eq:deltaRhat}), (\ref{eq:deltaFhat}) and (\ref{eq:deltaDFhat}) are abbreviated as
\begin{alignat}{3}
  [\delta_0 \omega] &= [D(\bar{\epsilon}\gamma\psi)], \notag
  \\
  [\delta_0 \hat{\omega}] &= [\bar{\epsilon}\gamma\psi_2] + [F\bar{\epsilon}\gamma\psi], \notag
  \\
  [\delta_0 \hat{R}] &= [R\bar{\epsilon}\gamma \psi] + [D(\bar{\epsilon} \gamma \psi_2)] + [D(F\bar{\epsilon} \gamma \psi)],
  \label{eq:trRhatsimple}
  \\
  [\delta_0 \hat{F}] &= [\bar{\epsilon} \gamma \psi_2], \notag
  \\
  [\delta_0 (D\hat{F})] &= [DF\bar{\epsilon}\gamma\psi] + [D(\bar{\epsilon}\gamma\psi_2)]
  + [FD(\bar{\epsilon}\gamma\psi)]. \notag
\end{alignat}

With respect to the fermionic bilinears of $[\bar{\psi}\gamma\psi]$, we require that two gravitinos are antisymmetric under the exchange in the ansatz. If this is the case, by executing the local supersymmetry transformation on either gravitino, we obtain $[\bar{\epsilon}\gamma\psi_2]$ by partial integrals. Although it is possible to include symmetric terms of $[\bar{\psi}\gamma\psi]$, we drop these out of the ansatz to make it as simple as possible. Fermionic bilinear terms of $[\bar{\psi} \gamma \mathcal{D} \psi_2]$ and $[\overline{\mathcal{D} \psi_2}\gamma \mathcal{D} \psi_2]$ are also dropped out of the ansatz by using partial integrals. (And we also neglect terms which include $[DDF]$ or $[DR]$ in the effective action.)

\subsection{The effective action which generates the variations at $\mathcal{O}(R^0)$} \label{subsec:VarO(R0)}

\begin{figure}[htb]
\begin{center}
\includegraphics[width=13cm]{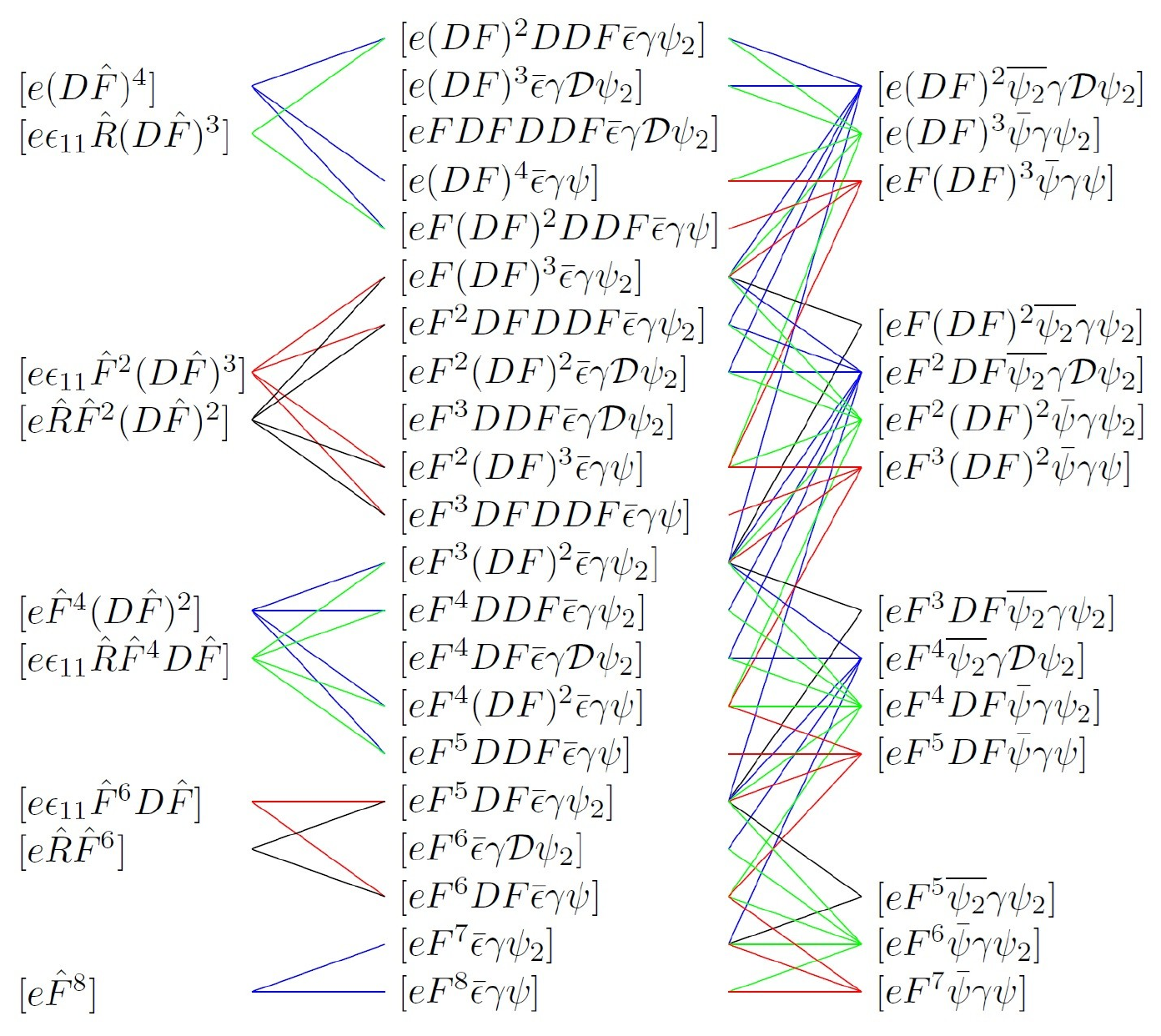}
\caption{Variations at $\mathcal{O}(R^0)$ up to $\mathcal{O}(\psi^2)$ under the local supersymmetry}\label{fig:O(R0)}
\end{center}
\end{figure}

Let us consider the terms in the effective action which generate variations at $\mathcal{O}(R^0)$ up to $\mathcal{O}(\psi^2)$ under the local supersymmetry. Following the ansatz in the previous subsection, the bosonic terms with mass dimension eight at $\mathcal{O}(R^0)$ should be constructed out of $[\hat{F}]$ and $[D\hat{F}]$, and the total number of $[\hat{F}]$ in each term is equal to or more than four. These are classified into
\begin{alignat}{3}
  \mathcal{O}(\hat{F}^4) &: [e(D\hat{F})^4], \notag
  \\
  \mathcal{O}(\hat{F}^5) &: [e\epsilon_{11}\hat{F}^2(D\hat{F})^3], \notag
  \\
  \mathcal{O}(\hat{F}^6) &: [e\hat{F}^4(D\hat{F})^2], \label{eq:bosonR0}
  \\
  \mathcal{O}(\hat{F}^7) &: [e\epsilon_{11}\hat{F}^6D\hat{F}], \notag
  \\
  \mathcal{O}(\hat{F}^8) &: [e\hat{F}^8]. \notag
\end{alignat}
It is also necessary to take into account bosonic terms which linearly depend on the Riemann tensor, and these are written as
\begin{alignat}{3}
  \mathcal{O}(\hat{R}\hat{F}^3) &: [e\epsilon_{11}\hat{R}(D\hat{F})^3], \notag
  \\
  \mathcal{O}(\hat{R}\hat{F}^4) &: [e\hat{R}\hat{F}^2(D\hat{F})^2], \notag
  \\
  \mathcal{O}(\hat{R}\hat{F}^5) &: [e\epsilon_{11}\hat{R}\hat{F}^4D\hat{F}], \label{eq:bosonR}
  \\
  \mathcal{O}(\hat{R}\hat{F}^6) &: [e\hat{R}\hat{F}^6]. \notag
\end{alignat}
Since we examine the cancellation of variations at $\mathcal{O}(R^0)$ in this paper, we only consider the variation of the Riemann tensor in the above. The bosonic terms (\ref{eq:bosonR0}) and (\ref{eq:bosonR}) are written on the left side in Fig.~\ref{fig:O(R0)}.

The fermionic bilinears with mass dimension eight at $\mathcal{O}(R^0)$ should be constructed out of $[F]$, $[DF]$, $[\bar{\psi}\gamma\psi]$, $[\bar{\psi}\gamma\psi_2]$, $[\overline{\psi_2}\gamma\psi_2]$ and $[\overline{\psi_2}\gamma \mathcal{D} \psi_2]$. The total number of $[\hat{F}]$ in each term is equal to or more than two. With this ansatz, the fermionic bilinears are listed as
\begin{alignat}{3}
  \mathcal{O}(F^2) &: 
  [e(DF)^2 \overline{\psi_2}\gamma \mathcal{D}\psi_2], \notag
  \\[0.1cm]
  \mathcal{O}(F^3) &: 
  [e(DF)^3\bar{\psi}\gamma\psi_2], \, 
  [eF(DF)^2\overline{\psi_2}\gamma\psi_2], \,
  [eF^2DF\overline{\psi_2}\gamma\mathcal{D}\psi_2], \notag
  \\[0.1cm]
  \mathcal{O}(F^4) &: 
  [eF(DF)^3\bar{\psi}\gamma\psi], \, 
  [eF^2(DF)^2\bar{\psi}\gamma\psi_2], \, 
  [eF^3DF\overline{\psi_2}\gamma\psi_2], \, 
  [eF^4\overline{\psi_2}\gamma\mathcal{D}\psi_2], \label{eq:fermionR0}
  \\[0.1cm]
  \mathcal{O}(F^5) &: 
  [eF^3(DF)^2\bar{\psi}\gamma\psi], \, 
  [eF^4DF\bar{\psi}\gamma\psi_2], \, 
  [eF^5\overline{\psi_2}\gamma\psi_2], \notag
  \\[0.1cm]
  \mathcal{O}(F^6) &: 
  [eF^5DF\bar{\psi}\gamma\psi], \, 
  [eF^6\bar{\psi}\gamma\psi_2], \notag
  \\[0.1cm]
  \mathcal{O}(F^7) &: 
  [eF^7\bar{\psi}\gamma\psi]. \notag
\end{alignat}
The above fermionic bilinears are written on the right side in the Fig.~\ref{fig:O(R0)}. Note that the derivative $[\mathcal{D}]$ implicitly contains the linear terms on $[F]$.

The variations with mass dimension eight at $\mathcal{O}(R^0)$ up to $\mathcal{O}(\psi^2)$ should be constructed out of $[F]$, $[DF]$, $[DDF]$, $[\bar{\epsilon}\gamma\psi]$, $[\bar{\epsilon}\gamma\psi_2]$ and $[\bar{\epsilon} \gamma\mathcal{D} \psi_2]$. And $[DDF]$ should not be duplicated. The total number of $[F]$ should be equal to or more than two. With this ansatz, the variations are classified into
\begin{alignat}{3}
  \mathcal{O}(F^3) &: 
  [e(DF)^2DDF\bar{\epsilon}\gamma\psi_2], \, 
  [e(DF)^3\bar{\epsilon}\gamma\mathcal{D}\psi_2], \, 
  [eFDFDDF\bar{\epsilon}\gamma\mathcal{D}\psi_2], \notag
  \\[0.1cm]
  \mathcal{O}(F^4) &: 
  [e(DF)^4\bar{\epsilon}\gamma\psi], \, 
  [eF(DF)^2DDF\bar{\epsilon}\gamma\psi], \,
  [eF(DF)^3\bar{\epsilon}\gamma\psi_2], \, 
  [eF^2(DF)DDF\bar{\epsilon}\gamma\psi_2], \notag
  \\
  &\quad 
  [eF^2(DF)^2\bar{\epsilon}\gamma\mathcal{D}\psi_2], \,
  [eF^3DDF\bar{\epsilon}\gamma\mathcal{D}\psi_2], \notag
  \\[0.1cm]
  \mathcal{O}(F^5) &: 
  [eF^2(DF)^3\bar{\epsilon}\gamma\psi], \, 
  [eF^3DFDDF\bar{\epsilon}\gamma\psi], \,
  [eF^3(DF)^2\bar{\epsilon}\gamma\psi_2], \, 
  [eF^4DDF\bar{\epsilon}\gamma\psi_2], \label{eq:varR0}
  \\
  &\quad 
  [eF^4DF\bar{\epsilon}\gamma\mathcal{D}\psi_2], \notag
  \\[0.1cm]
  \mathcal{O}(F^6) &: 
  [eF^4(DF)^2\bar{\epsilon}\gamma\psi], \, 
  [eF^5DDF\bar{\epsilon}\gamma\psi], \,
  [eF^5DF\bar{\epsilon}\gamma\psi_2], \, 
  [eF^6\bar{\epsilon}\gamma\mathcal{D}\psi_2], \notag
  \\[0.1cm]
  \mathcal{O}(F^7) &: 
  [eF^6DF\bar{\epsilon}\gamma\psi], \,
  [eF^7\bar{\epsilon}\gamma\psi_2], \notag
  \\[0.1cm]
  \mathcal{O}(F^8) &: 
  [eF^8\bar{\epsilon}\gamma\psi]. \notag
\end{alignat}
The above variations are shown around the center in the Fig.~\ref{fig:O(R0)}.

So far we have just listed possible terms at $\mathcal{O}(R^0)$ and some of $\mathcal{O}(R)$ up to $\mathcal{O}(\psi^3)$ in the effective action, and possible variations at $\mathcal{O}(R^0)$ up to $\mathcal{O}(\psi^2)$. Now let us consider the variations of the effective action under the local supersymmetry by using the eqs.~(\ref{eq:simplesusytr}), (\ref{eq:trpsi2simple}) and (\ref{eq:trRhatsimple}), and check the cancellation mechanism. 

Up to $\mathcal{O}(\psi^2)$, the variations of the bosonic terms (\ref{eq:bosonR0}) and (\ref{eq:bosonR}) under the local supersymmetry  can be evaluated as follows. First, by using $[\delta_0 e]$ in the eq. (\ref{eq:simplesusytr}) and $[\delta_0(D\hat{F})]$ and $[\delta_0\hat{R}]$ in the eq. (\ref{eq:trRhatsimple}), terms of $\mathcal{O}(\hat{F}^4)$ and $\mathcal{O}(\hat{R}\hat{F}^3)$ are transformed as
\begin{alignat}{3}
  [\delta_0 (e(D\hat{F})^4)] &=  
  [e(DF)^2DDF\bar{\epsilon}\gamma\psi_2]
  + [e(DF)^4\bar{\epsilon}\gamma\psi] 
  + [eF(DF)^2DDF\bar{\epsilon}\gamma\psi], \notag
  \\[0.1cm]
  [\delta_0 (e\epsilon_{11}\hat{R}(D\hat{F})^3)] &=
  [e(DF)^2DDF\bar{\epsilon}\gamma\psi_2] 
  + [eF(DF)^2DDF\bar{\epsilon}\gamma\psi]  
  + [eRDFDDF\bar{\epsilon}\gamma\psi_2] \label{eq:deltaB1B2}
  \\&\quad\,
  + [eDR(DF)^2\bar{\epsilon}\gamma\psi_2]
  + [eR(DF)^3\bar{\epsilon}\gamma\psi]
  + [eRFDFDDF\bar{\epsilon}\gamma\psi] \notag
  \\&\quad\,
  + [eDRF(DF)^2\bar{\epsilon}\gamma\psi]. \notag
\end{alignat}
Note that $\epsilon_{11}^{a_1 \cdots a_{11}}{\bf 1}=-\gamma^{a_1 \cdots a_{11}}$ is used on the right hand side in the second equality. Partial integrals are also executed to obtain the right hand side. Second, terms of $\mathcal{O}(\hat{F}^5)$ and $\mathcal{O}(\hat{R}\hat{F}^4)$ are transformed into
\begin{alignat}{3}
   [\delta_0 (e\epsilon_{11}\hat{F}^2(D\hat{F})^3)] &=  
   [eF(DF)^3\bar{\epsilon}\gamma\psi_2]
   + [eF^2DFDDF\bar{\epsilon}\gamma\psi_2] 
   + [eF^2(DF)^3\bar{\epsilon}\gamma\psi] \notag
   \\&\quad\, 
   + [eF^3DFDDF\bar{\epsilon}\gamma\psi], \notag
   \\[0.1cm]
   [\delta_0 (e\hat{R}\hat{F}^2(D\hat{F})^2)] &= 
   [eF(DF)^3\bar{\epsilon}\gamma\psi_2]
   + [eF^2DFDDF\bar{\epsilon}\gamma\psi_2] 
   + [eF^2(DF)^3\bar{\epsilon}\gamma\psi] \label{eq:deltaB3B4}
   \\&\quad\, 
   + [eF^3DFDDF\bar{\epsilon}\gamma\psi]
   + [eRF(DF)^2\bar{\epsilon}\gamma\psi_2]
   + [eRF^2DDF\bar{\epsilon}\gamma\psi_2] \notag
   \\&\quad\, 
   + [eDRF^2DF\bar{\epsilon}\gamma\psi_2]
   + [eRF^2(DF)^2\bar{\epsilon}\gamma\psi]
   + [eRF^3DDF\bar{\epsilon}\gamma\psi] \notag
   \\&\quad\, 
   + [eDRF^3DF\bar{\epsilon}\gamma\psi]. \notag
\end{alignat}
Third, terms of $\mathcal{O}(\hat{F}^6)$ and $\mathcal{O}(\hat{R}\hat{F}^5)$ are transformed as
\begin{alignat}{3}
   [\delta_0 (e\hat{F}^4(D\hat{F})^2)] &= 
   [eF^3(DF)^2\bar{\epsilon}\gamma\psi_2]
   + [eF^4DDF\bar{\epsilon}\gamma\psi_2] 
   + [eF^4(DF)^2\bar{\epsilon}\gamma\psi] \notag
   \\&\quad\, 
   + [eF^5DDF\bar{\epsilon}\gamma\psi], \notag
   \\[0.1cm]
   [\delta_0 (e\epsilon_{11}\hat{R}\hat{F}^4D\hat{F})] &= 
   [eF^3(DF)^2\bar{\epsilon}\gamma\psi_2]
   + [eF^4DDF\bar{\epsilon}\gamma\psi_2] 
   + [eF^4(DF)^2\bar{\epsilon}\gamma\psi] \label{eq:deltaB5B6}
   \\&\quad\, 
   + [eF^5DDF\bar{\epsilon}\gamma\psi] 
   + [eRF^3DF\bar{\epsilon}\gamma\psi_2]
   + [eDRF^4\bar{\epsilon}\gamma\psi_2] \notag
   \\&\quad 
   + [eRF^4DF\bar{\epsilon}\gamma\psi] 
   + [eDRF^5\bar{\epsilon}\gamma\psi]. \notag
\end{alignat}
Fourth, terms of $\mathcal{O}(\hat{F}^7)$ and $\mathcal{O}(\hat{R}\hat{F}^6)$ are transformed into
\begin{alignat}{3}
   [\delta_0 (e\epsilon_{11}\hat{F}^6D\hat{F})] &= 
   [eF^5DF\bar{\epsilon}\gamma\psi_2] 
   + [eF^6DF\bar{\epsilon}\gamma\psi], \label{eq:deltaB7B8}
   \\[0.1cm]
   [\delta_0 (e\hat{R}\hat{F}^6)] &=
   [eF^5DF\bar{\epsilon}\gamma\psi_2] 
   + [eF^6DF\bar{\epsilon}\gamma\psi]
   + [eRF^5\bar{\epsilon}\gamma\psi_2]
   + [eRF^6\bar{\epsilon}\gamma\psi]. \notag
\end{alignat}
Finally, the term of $\mathcal{O}(\hat{F}^8)$ is transformed as
\begin{alignat}{3}
   [\delta_0 (e\hat{F}^8)] &= 
   [eF^7\bar{\epsilon}\gamma\psi_2] 
   + [eF^8\bar{\epsilon}\gamma\psi]. \label{eq:deltaB9}
\end{alignat}
These transformations are sketched by connecting the bosonic terms and corresponding variations in the Fig~\ref{fig:O(R0)}.

Up to $\mathcal{O}(\psi^2)$, the variations of fermionic bilinears (\ref{eq:fermionR0}) under the local supersymmetry can be evaluated as follows. First, by using $[\delta_0\psi]$ in the eq.~(\ref{eq:simplesusytr}) and $[\delta_0\psi_2]$ of the eq.~(\ref{eq:trpsi2simple}), terms of $\mathcal{O}(F^2)$ and parts of $\mathcal{O}(F^3)$ and $\mathcal{O}(F^4)$ are transformed into 
\begin{alignat}{3}
  [\delta_0 (e(DF)^2 \overline{\psi_2}\gamma \mathcal{D}\psi_2)] &= 
  [e(DF)^2DDF\bar{\epsilon}\gamma\psi_2] 
  + [e(DF)^3\bar{\epsilon}\gamma\mathcal{D}\psi_2] 
  + [eF(DF)^3\bar{\epsilon}\gamma\psi_2] \notag
  \\&\quad\, 
  + [eF^2DFDDF\bar{\epsilon}\gamma\psi_2] 
  + [eF^2(DF)^2\bar{\epsilon}\gamma\mathcal{D}\psi_2]
  + [eF^3(DF)^2\bar{\epsilon}\gamma\psi_2] \notag
  \\&\quad\, 
  + [eRDFDDF\bar{\epsilon}\gamma\psi_2] 
  + [eR(DF)^2\bar{\epsilon}\gamma\mathcal{D}\psi_2] 
  + [eRF(DF)^2\bar{\epsilon}\gamma\psi_2], \notag
  \\[0.1cm]
  [\delta_0 (e(DF)^3\bar{\psi}\gamma\psi_2)] &= 
  [e(DF)^2DDF\bar{\epsilon}\gamma\psi_2]
  + [e(DF)^3\bar{\epsilon}\gamma\mathcal{D}\psi_2]
  + [e(DF)^4\bar{\epsilon}\gamma\psi] \label{eq:deltaF1F2F3}
  \\&\quad\, 
  + [eF(DF)^3\bar{\epsilon}\gamma\psi_2]
  + [eF^2(DF)^3\bar{\epsilon}\gamma\psi]
  + [eR(DF)^3\bar{\epsilon}\gamma\psi], \notag
  \\[0.1cm]
  [\delta_0 (eF(DF)^3\bar{\psi}\gamma\psi)] &= 
  [e(DF)^4\bar{\epsilon}\gamma\psi]
  + [eF(DF)^2DDF\bar{\epsilon}\gamma\psi]
  + [eF(DF)^3\bar{\epsilon}\gamma\psi_2] \notag
  \\&\quad\, 
  + [eF^2(DF)^3\bar{\epsilon}\gamma\psi]. \notag
\end{alignat}
Note that partial integrals are executed to obtain the right hand side, and the covariant derivative on the Majorana gravitino is combined with $F$ to make $\mathcal{D}$ defined in the eq. (\ref{eq:curlD}). Note also that we restrict the ranks of the gamma matrices of the fermionic bilinear terms $[eF(DF)^3\bar{\psi}\gamma\psi]$ so that they only contain $[\psi_2]$ after the variations. That is, allowed ranks of the gamma matrices are $1$, $5$ and $9$ for $[eF(DF)^3\bar{\psi}\gamma\psi]$. Second, parts of $\mathcal{O}(F^3)$, $\mathcal{O}(F^4)$ and $\mathcal{O}(F^5)$ are transformed as
\begin{alignat}{3}
  [\delta_0 (eF(DF)^2\overline{\psi_2}\gamma\psi_2)] &= 
  [eF(DF)^3\bar{\epsilon}\gamma\psi_2]
  + [eF^3(DF)^2\bar{\epsilon}\gamma\psi_2]
  + [eRF(DF)^2\bar{\epsilon}\gamma\psi_2], \notag
  \\[0.1cm]
  [\delta_0 (eF^2DF\overline{\psi_2}\gamma\mathcal{D}\psi_2)] &= 
  [eF(DF)^3\bar{\epsilon}\gamma\psi_2] 
  + [eF^2DFDDF\bar{\epsilon}\gamma\psi_2]
  + [eF^2(DF)^2\bar{\epsilon}\gamma\mathcal{D}\psi_2] \notag
  \\&\quad\, 
  + [eF^3(DF)^2\bar{\epsilon}\gamma\psi_2]
  + [eF^4DDF\bar{\epsilon}\gamma\psi_2] 
  + [eF^4DF\bar{\epsilon}\gamma\mathcal{D}\psi_2]\notag
  \\&\quad\, 
  + [eF^5DF\bar{\epsilon}\gamma\psi_2] 
  + [eRF(DF)^2\bar{\epsilon}\gamma\psi_2]
  + [eRF^2DDF\bar{\epsilon}\gamma\psi_2]  \notag
  \\&\quad\, 
  + [eRF^2DF\bar{\epsilon}\gamma\mathcal{D}\psi_2]
  + [eRF^3DF\bar{\epsilon}\gamma\psi_2], \label{eq:deltaF4F5F6F7}
  \\[0.1cm]
  [\delta_0 (eF^2(DF)^2\bar{\psi}\gamma\psi_2)] &=
  [eF(DF)^3\bar{\epsilon}\gamma\psi_2] 
  + [eF^2DFDDF\bar{\epsilon}\gamma\psi_2] 
  + [eF^2(DF)^2\bar{\epsilon}\gamma\mathcal{D}\psi_2] \notag
  \\&\quad\, 
  + [eF^2(DF)^3\bar{\epsilon}\gamma\psi]
  + [eF^3(DF)^2\bar{\epsilon}\gamma\psi_2]
  + [eF^4(DF)^2\bar{\epsilon}\gamma\psi] \notag
  \\&\quad\, 
  + [eRF^2(DF)^2\bar{\epsilon}\gamma\psi], \notag
  \\[0.1cm]
  [\delta_0 (eF^3(DF)^2\bar{\psi}\gamma\psi)] &=
  [eF^2(DF)^3\bar{\epsilon}\gamma\psi] 
  + [eF^3DFDDF\bar{\epsilon}\gamma\psi]
  + [eF^3(DF)^2\bar{\epsilon}\gamma\psi_2] \notag
  \\&\quad\, 
  + [eF^4(DF)^2\bar{\epsilon}\gamma\psi]. \notag
\end{alignat}
Possible ranks of the gamma matrices are $2$, $6$ and $10$ for $[eF^3(DF)^2\bar{\psi}\gamma\psi]$. Third, parts of $\mathcal{O}(F^4)$, $\mathcal{O}(F^5)$ and $\mathcal{O}(F^6)$ are transformed as
\begin{alignat}{3}
  [\delta_0 (eF^3DF\overline{\psi_2}\gamma\psi_2)] &=
  [eF^3(DF)^2\bar{\epsilon}\gamma\psi_2]
  + [eF^5DF\bar{\epsilon}\gamma\psi_2]
  + [eRF^3DF\bar{\epsilon}\gamma\psi_2], \notag
  \\[0.1cm]
  [\delta_0 (eF^4\overline{\psi_2}\gamma\mathcal{D}\psi_2)] &=
  [eF^3(DF)^2\bar{\epsilon}\gamma\psi_2]
  + [eF^4DF\bar{\epsilon}\gamma\mathcal{D}\psi_2]
  + [eF^5DF\bar{\epsilon}\gamma\psi_2] \notag
  \\&\quad\, 
  + [eF^6\bar{\epsilon}\gamma\mathcal{D}\psi_2]
  + [eF^7\bar{\epsilon}\gamma\psi_2]
  + [eRF^3DF\bar{\epsilon}\gamma\psi_2] \notag
  \\&\quad\, 
  + [eRF^4\bar{\epsilon}\gamma\mathcal{D}\psi_2]
  + [eRF^5\bar{\epsilon}\gamma\psi_2], \label{eq:deltaF8F9F10F11}
  \\[0.1cm]
  [\delta_0 (eF^4DF\bar{\psi}\gamma\psi_2)] &=
  [eF^3(DF)^2\bar{\epsilon}\gamma\psi_2] 
  + [eF^4DDF\bar{\epsilon}\gamma\psi_2]
  + [eF^4DF\bar{\epsilon}\gamma\mathcal{D}\psi_2] \notag
  \\&\quad\, 
  + [eF^4(DF)^2\bar{\epsilon}\gamma\psi]
  + [eF^5DF\bar{\epsilon}\gamma\psi_2]
  + [eF^6DF\bar{\epsilon}\gamma\psi]
  + [eRF^4DF\bar{\epsilon}\gamma\psi], \notag
  \\[0.1cm]
  [\delta_0 ([eF^5DF\bar{\psi}\gamma\psi])] &=
  [eF^4(DF)^2\bar{\epsilon}\gamma\psi]
  + [eF^5DDF\bar{\epsilon}\gamma\psi]
  + [eF^5DF\bar{\epsilon}\gamma\psi_2] 
  + [eF^6DF\bar{\epsilon}\gamma\psi]. \notag
\end{alignat}
Possible ranks of the gamma matrices are $1$, $5$ and $9$ for $[eF^5DF\bar{\psi}\gamma\psi]]$. Finally, parts of $\mathcal{O}(F^5)$, $\mathcal{O}(F^6)$ and $\mathcal{O}(F^7)$ are transformed into
\begin{alignat}{3}
  [\delta_0 (eF^5\overline{\psi_2}\gamma\psi_2)] &=
  [eF^5DF\bar{\epsilon}\gamma\psi_2]
  + [eF^7\bar{\epsilon}\gamma\psi_2]
  + [eRF^5\bar{\epsilon}\gamma\psi_2], \notag
  \\[0.1cm]
  [\delta_0 (eF^6\bar{\psi}\gamma\psi_2)] &=
  [eF^5DF\bar{\epsilon}\gamma\psi_2]
  + [eF^6\bar{\epsilon}\gamma\mathcal{D}\psi_2]
  + [eF^6DF\bar{\epsilon}\gamma\psi]
  + [eF^7\bar{\epsilon}\gamma\psi_2] \label{eq:deltaF12F13F14}
  \\&\quad\, 
  + [eF^8\bar{\epsilon}\gamma\psi] 
  + [eRF^6\bar{\epsilon}\gamma\psi], \notag
  \\[0.1cm]
  [\delta_0 (eF^7\bar{\psi}\gamma\psi)] &=
  [eF^6DF\bar{\epsilon}\gamma\psi] 
  + [eF^7\bar{\epsilon}\gamma\psi_2]
  + [eF^8\bar{\epsilon}\gamma\psi]. \notag
\end{alignat}
Possible ranks of the gamma matrices are $2$, $6$ and $10$ for $[eF^7\bar{\psi}\gamma\psi]$. The results of the transformations under the local supersymmetry are sketched by connecting the fermionic bilinears and corresponding variations in the Fig~\ref{fig:O(R0)}.

As is clear from the result of the above variations, after we complete the cancellation mechanism in the Fig.~\ref{fig:O(R0)}, there remain variations at $\mathcal{O}(R)$ up to $\mathcal{O}(\psi^2)$. Thus it is inevitable to consider the cancellation of variations at $\mathcal{O}(R)$ up to $\mathcal{O}(\psi^2)$ in the next stage. And this procedure lasts until we finish the cancellation of variations at $\mathcal{O}(R^4)$. Although, we do not classify possible terms beyond $\mathcal{O}(R)$ in the effective action, and terms beyond $\mathcal{O}(R^0)$ in the variations, it is straightforward to write down these terms by following the ansatz given in the subsection \ref{subsec:Ansatz}.

\begin{figure}[htb]
\begin{center}
\includegraphics[width=15.5cm]{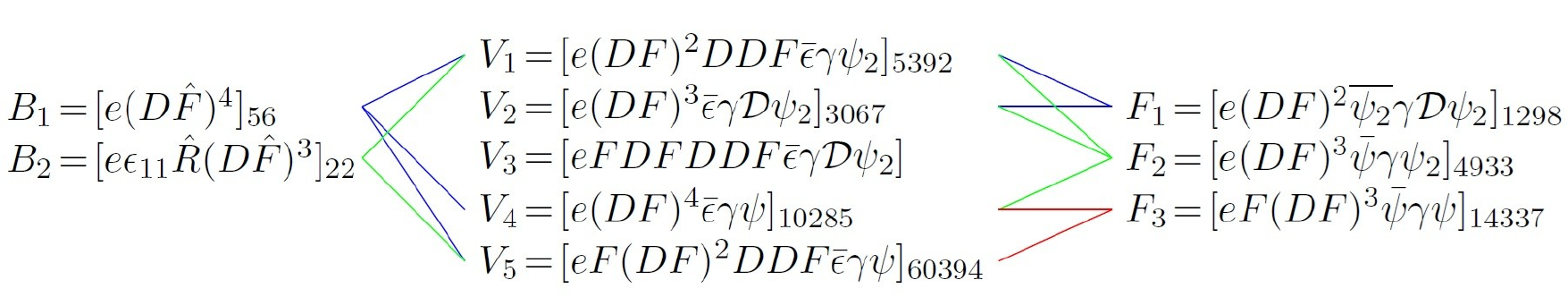}
\caption{$[e(D\hat{F})^4]$ and related parts}\label{fig:O(R0)part}
\end{center}
\end{figure}

It is also clear that,  even at $\mathcal{O}(R^0)$, we should deal with a lot of terms in order to check the local supersymmetry. Thus we restrict our calculations to the upper part of the Fig.~\ref{fig:O(R0)}, which is directly related to $[e(D\hat{F})^4]$ by the local supersymmetry. These are extracted in Fig.~\ref{fig:O(R0)part}, and we assign $B_1$, $B_2$ for the bosonic terms in the effective action, $F_1$, $F_2$, $F_3$ for the fermionic bilinears, and $V_1,\cdots, V_5$ for the variations. A number of independent terms for each class is shown in the subscript. Details are explained in the next section.\\

\section{Classifications of Terms in the Effective Action and the Variations} \label{sec:ActionVariation}

In this paper, we consider the cancellation under the local supersymmetry within the Fig.~\ref{fig:O(R0)part}. Note that we skip a classification of $V_3$, because it is not entered in the cancellation mechanism. In order to execute this task, it is necessary to figure out a basis of terms in each class. For generic higher derivative terms, we construct the basis in each class by following steps below. 
\noindent
\begin{enumerate}
\item First, we generate all possible adjacency matrices in each class, which specify numbers of contracted indices by the matrix notation. As explained below, some of the adjacency matrices are mapped into each other by changing the ordering of identical tensors, so we should consider a conjugacy class of these matrices. The representative matrix is denoted by ADJ.

\item Second, we create an index list which corresponds to the ADJ. At this stage, we impose symmetric or anti-symmetric properties of $\hat{F}_{abcd}$, $\hat{R}_{abcd}$, $\psi_{ab}$ and $\gamma_{a_1\cdots a_n}$ to remove duplication of terms in each class. For example, terms which contain $F_{acde}F_{bcde}\psi_{ab}$ should vanish. The index list is denoted as IND. 

\item Third, for the terms created by the IND, we consider Bianchi identities up to $\mathcal{O}(\psi^2)$: 
\begin{alignat}{3}
  D_{[e} F_{abcd]}&=0, \quad D_{[e} R_{ab]cd}=0, \quad
  \mathcal{D}_{[a} \psi_{bc]} = \tfrac{1}{4} R_{ij[ab} \gamma^{ij} \psi_{c]} 
  + DF_{[ab} \psi_{c]} + F^2_{[ab} \psi_{c]}. \label{eq:BianchiDP}
\end{alignat}
The last one relates terms in different classes. For the variations which partially contain $D_aD_bF_{cdef}$, we also use the commutation relation of
\begin{alignat}{3}
  [D_a,D_b] F_{cdef} = R_c{}^g{}_{ab} F_{gdef} + R_d{}^g{}_{ab} F_{cgef} 
  + R_e{}^g{}_{ab} F_{cdgf} + R_f{}^g{}_{ab} F_{cdeg}. \label{eq:comD} 
\end{alignat}

\item Finally, dimension dependent identities should be taken into account, which occur when a number of different indices in the IND is greater than $d$, the number of the spacetime dimensions. For each IND, it is possible to antisymmetrize any $d+1$ indices, and it should be zero in $d$ dimensions.  
\end{enumerate}

After four steps in the above, we obtain independent terms for each basis. We often denote the term in the basis by TNS. In order to obtain the basis in the Fig.~\ref{fig:O(R0)part}, we should deal with more than millions of terms,  so we use the Mathematica codes to classify these terms. Tips for the Mathematica codes are given in the appendix \ref{app:Code}. In each subsection below, we show results of the basis in each class. The number of the terms for the basis is shown in the subscript in the Fig.~\ref{fig:O(R0)part}.

\subsection{$B_1=[e(D\hat{F})^4]_{56}$} \label{subsec:B1}

\begin{table}[htb]
  \centering
  \begin{tabular}{l|c}
    \#ADJ & 146 \\
    \#IND & 137 \\
    \#Bianchi id. & 81 \\
    \#TNS & 56 \\
    \#TNS reduced by EOM & 24
  \end{tabular}
  \caption{Basis for $B_1 = [e(D\hat{F})^4]$.} \label{tb:B1}
\end{table}

Let us begin with bosonic terms in the class of $B_1=[e(D\hat{F})^4]$. Although we use $[\hat{F}]$ to represent $B_1$, this means that we construct the basis of $[e(DF)^4]$ and replace $F_{abcd}$ with the supercovariant expression $\hat{F}_{abcd}$ in the end. The result is summarized in Table~\ref{tb:B1}, and we will find 56 independent terms for the basis in $B_1$.

As an example of the tensor structure in $B_1$, let us take our choice like
\begin{alignat}{3}
  \text{TNS} = eD_{f_1}F_{f_2 f_3 f_4 f_5}D_{f_2}F_{f_1 f_6 f_7 f_8}
  D_{f_3}F_{f_6 f_7 f_9 f_{10}}D_{f_8}F_{f_4 f_5 f_9 f_{10}}, \label{eq:TNSDF4}
\end{alignat}
and an index list of the TNS is extracted as
\begin{alignat}{3}
  &\text{IND} = \big\{\{\!f_1\!\},\{\!f_2,\!f_3,\!f_4,\!f_5\!\},\{\!f_2\!\},
  \{\!f_1,\!f_6,\!f_7,\!f_8\!\}, \{\!f_3\!\},\{\!f_6,\!f_7,\!f_9,\!f_{10}\!\},\{\!f_8\!\},
  \{\!f_4,\!f_5,\!f_9,\!f_{10}\!\}\big\}. \label{eq:INDDF4}
\end{alignat}
A rank of the IND, which represents a set of ranks of aligned tensors, is given by
\begin{alignat}{3}
  \text{RNK}=\{1,4,1,4,1,4,1,4\}. \label{eq:RNKDF4}
\end{alignat}
We define an adjacency matrix of the IND so that $(i,j)$ component of the matrix is equal to a number of intersection between $i$th tensor and $j$th one. Then a representative matrix of the adjacency matrices for the IND is represented as
\begin{alignat}{3}
  \text{ADJ} &= 
  \begin{pmatrix}
    0 & 0 & 0 & 1 & 0 & 0 & 0 & 0 \\
    0 & 0 & 1 & 0 & 1 & 0 & 0 & 2 \\
    0 & 1 & 0 & 0 & 0 & 0 & 0 & 0 \\
    1 & 0 & 0 & 0 & 0 & 2 & 1 & 0 \\
    0 & 1 & 0 & 0 & 0 & 0 & 0 & 0 \\
    0 & 0 & 0 & 2 & 0 & 0 & 0 & 2 \\
    0 & 0 & 0 & 1 & 0 & 0 & 0 & 0 \\
    0 & 2 & 0 & 0 & 0 & 2 & 0 & 0 
  \end{pmatrix}. \label{eq:ADJDF4}
\end{alignat}
All diagonal components are zero since $F_{abcd}$ is the antisymmetric tensor. And the sum of each row or each column is equal to the corresponding rank of the tensor. Since the order of four $[DF]$s is exchangeable in the tensor (\ref{eq:TNSDF4}), the corresponding adjacency matrix is not unique in general. Thus we consider a conjugacy class of the adjacency matrices which is obtained by identifying permutations of four $[DF]$s. We call this representative matrix ADJ. 

Now we reverse the above procedure to obtain TNSs for the basis. By using the Mathematica code explained in the appendix \ref{app:Code}, it is possible to generate  146 ADJs for $B_1$. Next we create INDs from each ADJs uniquely. Notice that this is not one to one correspondence since we should take into account that $F_{abcd}$ is completely antisymmetric. By using the Mathematica code, we find that there are 137 INDs for $B_1$. 

In this procedure, however, the Bianchi identity in the eq.~(\ref{eq:BianchiDP}) has not been used. There are 81 Bianchi identities among 137 INDs. Thus we obtain the basis of 56 TNSs for $B_1$. Among these, we explicitly show 24 terms which do not include $D^a \hat{F}_{abcd}$ partially.
\begin{alignat}{3}
  B_1[1] &= e D_{f_1} \hat{F}_{f_2 f_3 f_4 f_5} D_{f_2} \hat{F}_{f_1 f_6 f_7 f_8} 
  D_{f_3} \hat{F}_{f_6 f_7 f_9 f_{10}} D_{f_8} \hat{F}_{f_4 f_5 f_9 f_{10}} , \notag 
  \\
  B_1[2] &= e D_{f_1} \hat{F}_{f_2 f_3 f_4 f_5} D_{f_2} \hat{F}_{f_1 f_6 f_7 f_8} 
  D_{f_3} \hat{F}_{f_4 f_6 f_9 f_{10}} D_{f_7} \hat{F}_{f_5 f_8 f_9 f_{10}} , \notag 
  \\ 
  B_1[3] &= e D_{f_1} \hat{F}_{f_2 f_3 f_4 f_5} D_{f_2} \hat{F}_{f_1 f_3 f_6 f_7} 
  D_{f_4} \hat{F}_{f_6 f_8 f_9 f_{10}} D_{f_7} \hat{F}_{f_5 f_8 f_9 f_{10}} , \notag 
  \\ 
  B_1[4] &= e D_{f_1} \hat{F}_{f_2 f_3 f_4 f_5} D_{f_2} \hat{F}_{f_1 f_3 f_6 f_7} 
  D_{f_4} \hat{F}_{f_5 f_8 f_9 f_{10}} D_{f_6} \hat{F}_{f_7 f_8f_9 f_{10}} , \notag 
  \\ 
  B_1[5] &= e D_{f_1} \hat{F}_{f_2 f_3 f_4 f_5} D_{f_1} \hat{F}_{f_6 f_7 f_8 f_9} 
  D_{f_2} \hat{F}_{f_6 f_7 f_8 f_{10}} \hat{F}_{f_9} \hat{F}_{f_3 f_4 f_5 f_{10}} , \notag 
  \\ 
  B_1[6] &= e D_{f_1} \hat{F}_{f_2 f_3 f_4 f_5} D_{f_1} \hat{F}_{f_6 f_7 f_8 f_9} 
  D_{f_2} \hat{F}_{f_3 f_6 f_7 f_{10}} D_{f_8} \hat{F}_{f_4 f_5 f_9 f_{10}} , \notag
  \\ 
  B_1[7] &= e D_{f_1} \hat{F}_{f_2 f_3 f_4 f_5} D_{f_1} \hat{F}_{f_6 f_7 f_8 f_9} 
  D_{f_2} \hat{F}_{f_3 f_6 f_7 f_{10}} D_{f_4} \hat{F}_{f_5 f_8 f_9 f_{10}} , \notag
  \\
  B_1[8] &= e D_{f_1} \hat{F}_{f_2 f_3 f_4 f_5} D_{f_1} \hat{F}_{f_6 f_7 f_8 f_9} 
  D_{f_2} \hat{F}_{f_3 f_4 f_6 f_{10}} D_{f_7} \hat{F}_{f_5 f_8 f_9 f_{10}} , \notag 
  \\
  B_1[9] &= e D_{f_1} \hat{F}_{f_2 f_3 f_4 f_5} D_{f_1} \hat{F}_{f_6 f_7 f_8 f_9} 
  D_{f_2} \hat{F}_{f_3 f_4 f_6 f_{10}} D_{f_5} \hat{F}_{f_7 f_8 f_9 f_{10}} , \notag 
  \\
  B_1[10] &= e D_{f_1} \hat{F}_{f_2 f_3 f_4 f_5} D_{f_1} \hat{F}_{f_6 f_7 f_8 f_9} 
  D_{f_2} \hat{F}_{f_3 f_4 f_5 f_{10}} D_{f_6} \hat{F}_{f_7 f_8 f_9 f_{10}} , \notag
  \\
  B_1[11] &= e D_{f_1} \hat{F}_{f_2 f_3 f_4 f_5} D_{f_1} \hat{F}_{f_2 f_6 f_7 f_8} 
  D_{f_9} \hat{F}_{f_3 f_4 f_6 f_{10}} D_{f_9} \hat{F}_{f_5 f_7 f_8 f_{10}} , \notag
  \\
  B_1[12] &= e D_{f_1} \hat{F}_{f_2 f_3 f_4 f_5} D_{f_1} \hat{F}_{f_2 f_6 f_7 f_8} 
  D_{f_9} \hat{F}_{f_3 f_4 f_5 f_{10}} D_{f_9} \hat{F}_{f_6 f_7 f_8 f_{10}} , \label{eq:B1}
  \\ 
  B_1[13] &= e D_{f_1} \hat{F}_{f_2 f_3 f_4 f_5} D_{f_1} \hat{F}_{f_2 f_6 f_7 f_8} 
  D_{f_3} \hat{F}_{f_6 f_7 f_9 f_{10}} D_{f_8} \hat{F}_{f_4 f_5 f_9 f_{10}} , \notag 
  \\ 
  B_1[14] &= e D_{f_1} \hat{F}_{f_2 f_3 f_4 f_5} D_{f_1} \hat{F}_{f_2 f_6 f_7 f_8} 
  D_{f_3} \hat{F}_{f_4 f_6 f_9 f_{10}} D_{f_7} \hat{F}_{f_5 f_8 f_9 f_{10}} , \notag 
  \\ 
  B_1[15] &= e D_{f_1} \hat{F}_{f_2 f_3 f_4 f_5} D_{f_1} \hat{F}_{f_2 f_6 f_7 f_8} 
  D_{f_3} \hat{F}_{f_4 f_6 f_9 f_{10}} D_{f_5} \hat{F}_{f_7 f_8 f_9 f_{10}} , \notag 
  \\ 
  B_1[16] &= e D_{f_1} \hat{F}_{f_2 f_3 f_4 f_5} D_{f_1} \hat{F}_{f_2 f_6 f_7 f_8} 
  D_{f_3} \hat{F}_{f_4 f_5 f_9 f_{10}} D_{f_6} \hat{F}_{f_7 f_8 f_9 f_{10}} , \notag 
  \\ 
  B_1[17] &= e D_{f_1} \hat{F}_{f_2 f_3 f_4 f_5} D_{f_1} \hat{F}_{f_2 f_3 f_6 f_7} 
  D_{f_8} \hat{F}_{f_4 f_6 f_9 f_{10}} D_{f_8} \hat{F}_{f_5 f_7 f_9 f_{10}} , \notag 
  \\ 
  B_1[18] &= e D_{f_1} \hat{F}_{f_2 f_3 f_4 f_5} D_{f_1} \hat{F}_{f_2 f_3 f_6 f_7} 
  D_{f_8} \hat{F}_{f_4 f_5 f_9 f_{10}} D_{f_8} \hat{F}_{f_6 f_7 f_9 f_{10}} , \notag 
  \\ 
  B_1[19] &= e D_{f_1} \hat{F}_{f_2 f_3 f_4 f_5} D_{f_1} \hat{F}_{f_2 f_3 f_6 f_7} 
  D_{f_4} \hat{F}_{f_6 f_8 f_9 f_{10}} D_{f_7} \hat{F}_{f_5 f_8 f_9 f_{10}} , \notag 
  \\ 
  B_1[20] &= e D_{f_1} \hat{F}_{f_2 f_3 f_4 f_5} D_{f_1} \hat{F}_{f_2 f_3 f_6 f_7} 
  D_{f_4} \hat{F}_{f_6 f_8 f_9 f_{10}} D_{f_5} \hat{F}_{f_7 f_8 f_9 f_{10}} , \notag 
  \\ 
  B_1[21] &= e D_{f_1} \hat{F}_{f_2 f_3 f_4 f_5} D_{f_1} \hat{F}_{f_2 f_3 f_6 f_7} 
  D_{f_4} \hat{F}_{f_5 f_8 f_9 f_{10}} D_{f_6} \hat{F}_{f_7 f_8 f_9 f_{10}} , \notag 
  \\ 
  B_1[22] &= e D_{f_1} \hat{F}_{f_2 f_3 f_4 f_5} D_{f_1} \hat{F}_{f_2 f_3 f_4 f_6} 
  D_{f_7} \hat{F}_{f_5 f_8 f_9 f_{10}} D_{f_7} \hat{F}_{f_6 f_8 f_9 f_{10}} , \notag 
  \\ 
  B_1[23] &= e D_{f_1} \hat{F}_{f_2 f_3 f_4 f_5} D_{f_1} \hat{F}_{f_2 f_3 f_4 f_6} 
  D_{f_5} \hat{F}_{f_7 f_8 f_9 f_{10}} D_{f_6} \hat{F}_{f_7 f_8 f_9 f_{10}} , \notag 
  \\ 
  B_1[24] &= e D_{f_1} \hat{F}_{f_2 f_3 f_4 f_5} D_{f_1} \hat{F}_{f_2 f_3 f_4 f_5} 
  D_{f_6} \hat{F}_{f_7 f_8 f_9 f_{10}} D_{f_6} \hat{F}_{f_7 f_8 f_9 f_{10}} . \notag
\end{alignat}
Here $F_{abcd}$ is replaced with $\hat{F}_{abcd}$ in the end. There are remaining 32 terms which partially include $D^a \hat{F}_{abcd}$. These can be mapped to other terms by using the equations of motion for the eleven dimensional supergravity.
\begin{alignat}{3}
  B_1[25] &= e D_{f_1} \hat{F}_{f_1 f_2 f_3 f_4} D_{f_5} \hat{F}_{f_2 f_6 f_7 f_8} 
  D_{f_6} \hat{F}_{f_3 f_5 f_9 f_{10}} D_{f_7} \hat{F}_{f_4 f_8 f_9 f_{10}}, \notag 
  \\ 
  B_1[26] &= e D_{f_1} \hat{F}_{f_1 f_2 f_3 f_4} D_{f_5} \hat{F}_{f_2 f_6 f_7 f_8} 
  D_{f_5} \hat{F}_{f_3 f_6 f_9 f_{10}} D_{f_7} \hat{F}_{f_4 f_8 f_9 f_{10}}, \notag 
  \\ 
  B_1[27] &= e D_{f_1} \hat{F}_{f_1 f_2 f_3 f_4} D_{f_5} \hat{F}_{f_2 f_5 f_6 f_7} 
  D_{f_8} \hat{F}_{f_3 f_6 f_9 f_{10}} D_{f_8} \hat{F}_{f_4 f_7 f_9 f_{10}}, \notag 
  \\ 
  B_1[28] &= e D_{f_1} \hat{F}_{f_1 f_2 f_3 f_4} D_{f_5} \hat{F}_{f_2 f_5 f_6 f_7} 
  D_{f_8} \hat{F}_{f_3 f_6 f_8 f_9} D_{f_{10}} \hat{F}_{f_4 f_7 f_9 f_{10}}, \notag 
  \\ 
  B_1[29] &= e D_{f_1} \hat{F}_{f_1 f_2 f_3 f_4} D_{f_5} \hat{F}_{f_2 f_3 f_6 f_7} 
  D_{f_8} \hat{F}_{f_4 f_8 f_9 f_{10}} D_{f_5} \hat{F}_{f_6 f_7 f_9 f_{10}}, \notag 
  \\ 
  B_1[30] &= e D_{f_1} \hat{F}_{f_1 f_2 f_3 f_4} D_{f_5} \hat{F}_{f_2 f_3 f_6 f_7}
  D_{f_6} \hat{F}_{f_4 f_8 f_9 f_{10}} D_{f_5} \hat{F}_{f_7 f_8 f_9 f_{10}}, \notag 
  \\ 
  B_1[31] &= e D_{f_1} \hat{F}_{f_1 f_2 f_3 f_4} D_{f_5} \hat{F}_{f_2 f_3 f_6 f_7} 
  D_{f_8} \hat{F}_{f_4 f_5 f_9 f_{10}} D_{f_8} \hat{F}_{f_6 f_7 f_9 f_{10}}, \notag 
  \\ 
  B_1[32] &= e D_{f_1} \hat{F}_{f_1 f_2 f_3 f_4} D_{f_5} \hat{F}_{f_2 f_3 f_6 f_7} 
  D_{f_5} \hat{F}_{f_4 f_8 f_9 f_{10}} D_{f_6} \hat{F}_{f_7 f_8 f_9 f_{10}}, \notag 
  \\ 
  B_1[33] &= e D_{f_1} \hat{F}_{f_1 f_2 f_3 f_4} D_{f_5} \hat{F}_{f_2 f_3 f_6 f_7} 
  D_{f_5} \hat{F}_{f_4 f_6 f_8 f_9} D_{f_{10}} \hat{F}_{f_7 f_8 f_9 f_{10}}, \notag 
  \\ 
  B_1[34] &= e D_{f_1} \hat{F}_{f_1 f_2 f_3 f_4} D_{f_5} \hat{F}_{f_2 f_3 f_5 f_6} 
  D_{f_7} \hat{F}_{f_4 f_8 f_9 f_{10}} D_{f_7} \hat{F}_{f_6 f_8 f_9 f_{10}}, \notag 
  \\ 
  B_1[35] &= e D_{f_1} \hat{F}_{f_1 f_2 f_3 f_4} D_{f_5} \hat{F}_{f_2 f_3 f_5 f_6} 
  D_{f_7} \hat{F}_{f_4 f_7 f_8 f_9} D_{f_{10}} \hat{F}_{f_6 f_8 f_9 f_{10}}, \notag 
  \\
  B_1[36] &= e D_{f_1} \hat{F}_{f_1 f_2 f_3 f_4} D_{f_5} \hat{F}_{f_2 f_3 f_4 f_6} 
  D_{f_7} \hat{F}_{f_5 f_8 f_9 f_{10}} D_{f_7} \hat{F}_{f_6 f_8 f_9 f_{10}}, \notag 
  \\ 
  B_1[37] &= e D_{f_1} \hat{F}_{f_1 f_2 f_3 f_4} D_{f_5} \hat{F}_{f_2 f_3 f_4 f_6} 
  D_{f_5} \hat{F}_{f_7 f_8 f_9 f_{10}} D_{f_6} \hat{F}_{f_7 f_8 f_9 f_{10}}, \notag 
  \\ 
  B_1[38] &= e D_{f_1} \hat{F}_{f_1 f_2 f_3 f_4} D_{f_5} \hat{F}_{f_2 f_3 f_4 f_6} 
  D_{f_5} \hat{F}_{f_6 f_7 f_8 f_9} D_{f_{10}} \hat{F}_{f_7 f_8 f_9 f_{10}}, \notag 
  \\ 
  B_1[39] &= e D_{f_1} \hat{F}_{f_1 f_2 f_3 f_4} D_{f_5} \hat{F}_{f_2 f_3 f_4 f_5} 
  D_{f_6} \hat{F}_{f_7 f_8 f_9 f_{10}} D_{f_6} \hat{F}_{f_7 f_8 f_9 f_{10}}, \notag 
  \\
  B_1[40] &= e D_{f_1} \hat{F}_{f_1 f_2 f_3 f_4} D_{f_5} \hat{F}_{f_2 f_3 f_4 f_5} 
  D_{f_6} \hat{F}_{f_6 f_7 f_8 f_9} D_{f_{10}} \hat{F}_{f_7 f_8 f_9 f_{10}}, \label{eq:B1_2}
  \\
  B_1[41] &= e D_{f_1} \hat{F}_{f_1 f_2 f_3 f_4} D_{f_2} \hat{F}_{f_5 f_6 f_7 f_8} 
  D_{f_5} \hat{F}_{f_3 f_6 f_7 f_9} D_{f_{10}} \hat{F}_{f_4 f_8 f_9 f_{10}}, \notag
  \\
  B_1[42] &= e D_{f_1} \hat{F}_{f_1 f_2 f_3 f_4} D_{f_2} \hat{F}_{f_5 f_6 f_7 f_8} 
  D_{f_9} \hat{F}_{f_3 f_4 f_9 f_{10}} D_{f_5} \hat{F}_{f_6 f_7 f_8 f_{10}}, \notag 
  \\
  B_1[43] &= e D_{f_1} \hat{F}_{f_1 f_2 f_3 f_4} D_{f_2} \hat{F}_{f_5 f_6 f_7 f_8} 
  D_{f_9} \hat{F}_{f_3 f_4 f_5 f_{10}} D_{f_9} \hat{F}_{f_6 f_7 f_8 f_{10}}, \notag
  \\
  B_1[44] &= e D_{f_1} \hat{F}_{f_1 f_2 f_3 f_4} D_{f_2} \hat{F}_{f_5 f_6 f_7 f_8} 
  D_{f_9} \hat{F}_{f_3 f_4 f_5 f_9} D_{f_{10}} \hat{F}_{f_6 f_7 f_8 f_{10}}, \notag 
  \\
  B_1[45] &= e D_{f_1} \hat{F}_{f_1 f_2 f_3 f_4} D_{f_2} \hat{F}_{f_5 f_6 f_7 f_8} 
  D_{f_5} \hat{F}_{f_3 f_4 f_9 f_{10}} D_{f_6} \hat{F}_{f_7 f_8 f_9 f_{10}}, \notag 
  \\ 
  B_1[46] &= e D_{f_1} \hat{F}_{f_1 f_2 f_3 f_4} D_{f_2} \hat{F}_{f_5 f_6 f_7 f_8} 
  D_{f_5} \hat{F}_{f_3 f_4 f_6 f_9} D_{f_{10}} \hat{F}_{f_7 f_8 f_9 f_{10}}, \notag 
  \\ 
  B_1[47] &= e D_{f_1} \hat{F}_{f_1 f_2 f_3 f_4} D_{f_2} \hat{F}_{f_5 f_6 f_7 f_8} 
  D_{f_3} \hat{F}_{f_5 f_6 f_9 f_{10}} D_{f_7} \hat{F}_{f_4 f_8 f_9 f_{10}}, \notag 
  \\ 
  B_1[48] &= e D_{f_1} \hat{F}_{f_1 f_2 f_3 f_4} D_{f_2} \hat{F}_{f_5 f_6 f_7 f_8} 
  D_{f_3} \hat{F}_{f_5 f_6 f_7 f_9} D_{f_{10}} \hat{F}_{f_4 f_8 f_9 f_{10}}, \notag 
  \\ 
  B_1[49] &= e D_{f_1} \hat{F}_{f_1 f_2 f_3 f_4} D_{f_2} \hat{F}_{f_3 f_5 f_6 f_7} 
  D_{f_8} \hat{F}_{f_4 f_8 f_9 f_{10}} D_{f_5} \hat{F}_{f_6 f_7 f_9 f_{10}}, \notag 
  \\ 
  B_1[50] &= e D_{f_1} \hat{F}_{f_1 f_2 f_3 f_4} D_{f_2} \hat{F}_{f_3 f_5 f_6 f_7} 
  D_{f_8} \hat{F}_{f_4 f_5 f_9 f_{10}} D_{f_8} \hat{F}_{f_6 f_7 f_9 f_{10}}, \notag 
  \\ 
  B_1[51] &= e D_{f_1} \hat{F}_{f_1 f_2 f_3 f_4} D_{f_2} \hat{F}_{f_3 f_5 f_6 f_7} 
  D_{f_8} \hat{F}_{f_4 f_5 f_8 f_9} D_{f_{10}} \hat{F}_{f_6 f_7 f_9 f_{10}}, \notag 
  \\ 
  B_1[52] &= e D_{f_1} \hat{F}_{f_1 f_2 f_3 f_4} D_{f_2} \hat{F}_{f_3 f_5 f_6 f_7} 
  D_{f_5} \hat{F}_{f_4 f_8 f_9 f_{10}} D_{f_6} \hat{F}_{f_7 f_8 f_9 f_{10}}, \notag 
  \\ 
  B_1[53] &= e D_{f_1} \hat{F}_{f_1 f_2 f_3 f_4} D_{f_2} \hat{F}_{f_3 f_5 f_6 f_7} 
  D_{f_5} \hat{F}_{f_4 f_6 f_8 f_9} D_{f_{10}} \hat{F}_{f_7 f_8 f_9 f_{10}}, \notag 
  \\ 
  B_1[54] &= e D_{f_1} \hat{F}_{f_1 f_2 f_3 f_4} D_{f_2} \hat{F}_{f_3 f_5 f_6 f_7} 
  D_{f_4} \hat{F}_{f_5 f_8 f_9 f_{10}} D_{f_6} \hat{F}_{f_7 f_8 f_9 f_{10}}, \notag 
  \\ 
  B_1[55] &= e D_{f_1} \hat{F}_{f_1 f_2 f_3 f_4} D_{f_2} \hat{F}_{f_3 f_5 f_6 f_7} 
  D_{f_4} \hat{F}_{f_5 f_6 f_8 f_9} D_{f_{10}} \hat{F}_{f_7 f_8 f_9 f_{10}}, \notag 
  \\ 
  B_1[56] &= e D_{f_1} \hat{F}_{f_1 f_2 f_3 f_4} D_{f_2} \hat{F}_{f_3 f_4 f_5 f_6} 
  D_{f_5} \hat{F}_{f_6 f_7 f_8 f_9} D_{f_{10}} \hat{F}_{f_7 f_8 f_9 f_{10}}. \notag
\end{alignat}
The summary of the basis in $B_1$ is shown in the Table~\ref{tb:B1}.

\subsection{$B_2=[e\epsilon_{11}\hat{R}(D\hat{F})^3]_{22}$} \label{subsec:B2}

\begin{table}[hb]
  \centering
  \begin{tabular}{l|c}
    \#ADJ & 890 \\
    \#IND & 743 \\
    \#Bianchi id. & 681 \\
    \#Dimension dep. id. & 603 \\
    \#TNS & 22 \\
    \#TNS reduced by EOM & 10
  \end{tabular}
  \caption{Basis for $B_2 = [e\epsilon_{11}\hat{R}(D\hat{F})^3]$.} \label{tb:B2}
\end{table}

Let us classify terms in the class of $B_2=[e\epsilon_{11}\hat{R}(D\hat{F})^3]$. Although we use $[\hat{R}]$ and $[\hat{F}]$ for $B_2$, this means that we construct the basis of $[e\epsilon_{11}R(DF)^3]$ and replace $R_{abcd}$ and $F_{abcd}$ with the supercovariant expressions $\hat{R}_{abcd}$ and $\hat{F}_{abcd}$ in the end. The result is summarized in Table~\ref{tb:B2}, and we will find 22 independent terms for the basis in $B_2$.

As an example of the tensor structure in $B_2$, we select
\begin{alignat}{3}
  \text{TNS} = e\epsilon^{11}_{f_3 f_4 f_6 f_7 f_9 f_{10} f_{11} f_{12} f_{13} f_{14} f_{15}}
  R_{f_1 f_2 f_3 f_4}D_{f_5}F_{f_1 f_2 f_6 f_7}D_{f_5}F_{f_8 f_9 f_{10} f_{11}}
  D_{f_8}F_{f_{12} f_{13} f_{14} f_{15}}. \label{eq:TNSepR(DF)3}
\end{alignat}
An index list of the above TNS is given by
\begin{alignat}{3}
  &\text{IND} = \big\{\{\!f_1,\!f_2\!\},\{\!f_3,\!f_4\!\},\{\!f_5\!\},\{\!f_1,\!f_2,\!f_6,\!f_7\!\}, 
  \{\!f_5\!\},\{\!f_8,\!f_9,\!f_{10},\!f_{11}\!\},\{\!f_8\!\},\{\!f_{12},\!f_{13},\!f_{14},\!f_{15}\!\}, \notag
  \\
  &\qquad\qquad
  \{\!f_3, \!f_4, \!f_6, \!f_7, \!f_9, \!f_{10}, \!f_{11}, \!f_{12}, \!f_{13}, \!f_{14}, \!f_{15}\!\}\big\}, \label{eq:INDepR(DF)3}
\end{alignat}
and a rank of the IND, which represents a set of ranks of aligned tensors, is written by
\begin{alignat}{3}
  \text{RNK}=\{2,2,1,4,1,4,1,4,11\}. \label{eq:RNKepR(DF)3}
\end{alignat}
Notice that the indices of $\epsilon_{11}$ are put at the end of the list to make the code run faster. The indices of the Riemann tensor $R_{f_1 f_2 f_3 f_4}$ is divided into two antisymmetric parts $\{\{\!f_1,\!f_2\!\},\{\!f_3,\!f_4\!\}\}$ to make the diagonal elements of the adjacency matrix to be zero. The indices of the Ricci tensor $R_{f_1f_2}$ and those of the scalar curvature $R$ can be expressed as $\{\{\!f_1,\!f_3\!\},\{\!f_2,\!f_3\!\}\}$ and $\{\{\!f_1,\!f_2\!\},\{\!f_1,\!f_2\!\}\}$, respectively. A representative matrix of adjacency matrices for the IND is expressed as
\begin{alignat}{3}
  \text{ADJ} &= 
  \begin{pmatrix}
    0 & 0 & 0 & 2 & 0 & 0 & 0 & 0 & 0 \\
    0 & 0 & 0 & 0 & 0 & 0 & 0 & 0 & 2 \\
    0 & 0 & 0 & 0 & 1 & 0 & 0 & 0 & 0 \\
    2 & 0 & 0 & 0 & 0 & 0 & 0 & 0 & 2 \\
    0 & 0 & 1 & 0 & 0 & 0 & 0 & 0 & 0 \\
    0 & 0 & 0 & 0 & 0 & 0 & 1 & 0 & 3 \\
    0 & 0 & 0 & 0 & 0 & 1 & 0 & 0 & 0 \\
    0 & 0 & 0 & 0 & 0 & 0 & 0 & 0 & 4 \\
    0 & 2 & 0 & 2 & 0 & 3 & 0 & 4 & 0 
  \end{pmatrix}. \label{eq:ADJepR(DF)3}
\end{alignat}
The sum of each row or each column is equal to the corresponding rank of the tensor. Since the order of three $[DF]$s can be exchanged in the tensor (\ref{eq:TNSepR(DF)3}), we consider a conjugacy class of the adjacency matrices, which is obtained by identifying  permutations of three $[DF]$s. We call this representative matrix ADJ. 

By using the Mathematica code explained in the appendix \ref{app:Code}, we find that there are 890 ADJs for $B_2$. Then it is possible to generate INDs from ADJs uniquely. Notice that this is not one to one correspondence since we should take into account symmetric or antisymmetric properties of $R_{abcd}$, $F_{abcd}$ and $\epsilon^{11}_{a_1 \cdots a_{11}}$. By using the Mathematica code, we find that there are 743 INDs for $B_2$. 

In this procedure, however, the Bianchi identities in the eq.~(\ref{eq:BianchiDP}) have not been used. There are 681 Bianchi identities among 743 terms. Furthermore, we should take into account the dimension dependent identities which arise by antisymmetrizing 12 indices among 15 ones. There are 603 dimension dependent identities among 743 terms, but some of them become trivial by imposing the Bianchi identities. 

Carefully solving the overlap of the Bianchi identities and the dimension dependent ones, we obtain 22 TNSs for $B_2$. Among these, there exist 10 terms which do not include parts like $D^a \hat{F}_{abcd}$.
\begin{alignat}{3}
  B_2[1] &\!=\! e \epsilon^{11}_{f_3 f_4 f_7 f_8 f_9 f_{10} f_{11} f_{12} f_{13} f_{14} f_{15}}
  \hat{R}_{f_1 f_2 f_3 f_4} D_{f_5} \hat{F}_{f_1 f_2 f_6 f_7} 
  D_{f_5} \hat{F}_{f_8 f_9 f_{10} f_{11}} D_{f_6} \hat{F}_{f_{12} f_{13} f_{14} f_{15}}, \notag 
  \\ 
  B_2[2] &\!=\! e \epsilon^{11}_{f_3 f_4 f_6 f_7 f_8 f_{10} f_{11} f_{12} f_{13} f_{14} f_{15}} 
  \hat{R}_{f_1 f_2 f_3 f_4} D_{f_1} \hat{F}_{f_5 f_6 f_7 f_8} 
  D_{f_9} \hat{F}_{f_2 f_5 f_{10} f_{11}} D_{f_9} \hat{F}_{f_{12} f_{13} f_{14} f_{15}}, \notag 
  \\ 
  B_2[3] &\!=\! e \epsilon^{11}_{f_3 f_4 f_7 f_8 f_9 f_{10} f_{11} f_{12} f_{13} f_{14} f_{15}} 
  \hat{R}_{f_1 f_2 f_3 f_4} D_{f_1} \hat{F}_{f_5 f_6 f_7 f_8} 
  D_{f_5} \hat{F}_{f_2 f_9 f_{10} f_{11}} D_{f_6} \hat{F}_{f_{12} f_{13} f_{14} f_{15}}, \notag 
  \\ 
  B_2[4] &\!=\! e \epsilon^{11}_{f_4 f_5 f_6 f_7 f_8 f_{10} f_{11} f_{12} f_{13} f_{14} f_{15}} 
  \hat{R}_{f_1 f_2 f_3 f_4} D_{f_1} \hat{F}_{f_5 f_6 f_7 f_8} 
  D_{f_9} \hat{F}_{f_2 f_3 f_{10} f_{11}} D_{f_9} \hat{F}_{f_{12} f_{13} f_{14} f_{15}}, \notag 
  \\
  B_2[5] &\!=\! e \epsilon^{11}_{f_4 f_6 f_7 f_8 f_9 f_{10} f_{11} f_{12} f_{13} f_{14} f_{15}} 
  \hat{R}_{f_1 f_2 f_3 f_4} D_{f_1} \hat{F}_{f_3 f_5 f_6 f_7} 
  D_{f_2} \hat{F}_{f_8 f_9 f_{10} f_{11}} D_{f_5} \hat{F}_{f_{12} f_{13} f_{14} f_{15}}, 
  \label{eq:B2} 
  \\ 
  B_2[6] &\!=\! e \epsilon^{11}_{f_5 f_6 f_7 f_8 f_9 f_{10} f_{11} f_{12} f_{13} f_{14} f_{15}} 
  \hat{R}_{f_1 f_2 f_3 f_4} D_{f_1} \hat{F}_{f_3 f_5 f_6 f_7} 
  D_{f_2} \hat{F}_{f_8 f_9 f_{10} f_{11}} D_{f_4} \hat{F}_{f_{12} f_{13} f_{14} f_{15}}, \notag 
  \\ 
  B_2[7] &\!=\! e \epsilon^{11}_{f_4 f_6 f_7 f_8 f_9 f_{10} f_{11} f_{12} f_{13} f_{14} f_{15}} 
  \hat{R}_{f_1 f_2 f_3 f_4} D_{f_1} \hat{F}_{f_2 f_5 f_6 f_7} 
  D_{f_3} \hat{F}_{f_8 f_9 f_{10} f_{11}} D_{f_5} \hat{F}_{f_{12} f_{13} f_{14} f_{15}}, \notag 
  \\ 
  B_2[8] &\!=\! e \epsilon^{11}_{f_4 f_5 f_6 f_8 f_9 f_{10} f_{11} f_{12} f_{13} f_{14} f_{15}} 
  \hat{R}_{f_1 f_2 f_3 f_4} D_{f_1} \hat{F}_{f_2 f_3 f_5 f_6} 
  D_{f_7} \hat{F}_{f_8 f_9 f_{10} f_{11}} D_{f_7} \hat{F}_{f_{12} f_{13} f_{14} f_{15}}, \notag 
  \\ 
  B_2[9] &\!=\! e \epsilon^{11}_{f_3 f_6 f_7 f_8 f_9 f_{10} f_{11} f_{12} f_{13} f_{14} f_{15}} 
  \hat{R}_{f_1 f_2 f_1 f_3} D_{f_4} \hat{F}_{f_2 f_5 f_6 f_7} 
  D_{f_4} \hat{F}_{f_8 f_9 f_{10} f_{11}} D_{f_5} \hat{F}_{f_{12} f_{13} f_{14} f_{15}}, \notag
  \\ 
  B_2[10] &\!=\! e \epsilon^{11}_{f_4 f_5 f_6 f_8 f_9 f_{10} f_{11} f_{12} f_{13} f_{14} f_{15}} 
  \hat{R}_{f_1 f_2 f_1 f_3} D_{f_2} \hat{F}_{f_3 f_4 f_5 f_6} 
  D_{f_7} \hat{F}_{f_8 f_9 f_{10} f_{11}} D_{f_7} \hat{F}_{f_{12} f_{13} f_{14} f_{15}}. \notag
\end{alignat}
On the other hand, there remain 12 terms which partially include $D^a \hat{F}_{abcd}$. These will be mapped to other terms by using the equations of motion for the eleven dimensional supergravity.
\begin{alignat}{3}
  B_2[11] &\!=\! e \epsilon^{11}_{f_3 f_4 f_6 f_7 f_8 f_9 f_{10} f_{11} f_{13} f_{14} f_{15}}
  \hat{R}_{f_1 f_2 f_3 f_4} D_{f_5} \hat{F}_{f_1 f_2 f_6 f_7} 
  D_{f_5} \hat{F}_{f_8 f_9 f_{10} f_{11}} D_{f_{12}} \hat{F}_{f_{12} f_{13} f_{14} f_{15}}, \notag 
  \\
  B_2[12] &\!=\! e \epsilon^{11}_{f_3 f_4 f_6 f_8 f_9 f_{10} f_{11} f_{12} f_{13} f_{14} f_{15}} 
  \hat{R}_{f_1 f_2 f_3 f_4} D_{f_5} \hat{F}_{f_1 f_2 f_5 f_6} 
  D_{f_7} \hat{F}_{f_8 f_9 f_{10} f_{11}} D_{f_7} \hat{F}_{f_{12} f_{13} f_{14} f_{15}}, \notag
  \\
  B_2[13] &\!=\! e \epsilon^{11}_{f_3 f_4 f_5 f_6 f_7 f_8 f_{11} f_{12} f_{13} f_{14} f_{15}} 
  \hat{R}_{f_1 f_2 f_3 f_4} D_{f_1} \hat{F}_{f_5 f_6 f_7 f_8} 
  D_{f_9} \hat{F}_{f_2 f_9 f_{10} f_{11}} D_{f_{10}} \hat{F}_{f_{12} f_{13} f_{14} f_{15}}, 
  \notag  
  \\
  B_2[14] &\!=\! e \epsilon^{11}_{f_3 f_4 f_6 f_7 f_8 f_9 f_{10} f_{11} f_{13} f_{14} f_{15}} 
  \hat{R}_{f_1 f_2 f_3 f_4} D_{f_1} \hat{F}_{f_5 f_6 f_7 f_8} 
  D_{f_5} \hat{F}_{f_2 f_9 f_{10} f_{11}} D_{f_{12}} \hat{F}_{f_{12} f_{13} f_{14} f_{15}}, \notag  
  \\ 
  B_2[15] &\!=\! e \epsilon^{11}_{f_3 f_4 f_6 f_7 f_8 f_9 f_{10} f_{11} f_{13} f_{14} f_{15}} 
  \hat{R}_{f_1 f_2 f_3 f_4} D_{f_1} \hat{F}_{f_5 f_6 f_7 f_8} 
  D_{f_2} \hat{F}_{f_5 f_9 f_{10} f_{11}} D_{f_{12}} \hat{F}_{f_{12} f_{13} f_{14} f_{15}}, \notag 
  \\ 
  B_2[16] &\!=\! e \epsilon^{11}_{f_4 f_5 f_6 f_7 f_8 f_9 f_{10} f_{11} f_{13} f_{14} f_{15}} 
  \hat{R}_{f_1 f_2 f_3 f_4} D_{f_1} \hat{F}_{f_3 f_5 f_6 f_7} 
  D_{f_2} \hat{F}_{f_8 f_9 f_{10} f_{11}} D_{f_{12}} \hat{F}_{f_{12} f_{13} f_{14} f_{15}}, \label{eq:B2_2} 
  \\ 
  B_2[17] &\!=\! e \epsilon^{11}_{f_3 f_4 f_6 f_7 f_8 f_9 f_{10} f_{11} f_{13} f_{14} f_{15}} 
  \hat{R}_{f_1 f_2 f_3 f_4} D_{f_1} \hat{F}_{f_2 f_5 f_6 f_7} 
  D_{f_5} \hat{F}_{f_8 f_9 f_{10} f_{11}} D_{f_{12}} \hat{F}_{f_{12} f_{13} f_{14} f_{15}}, \notag 
  \\ 
  B_2[18] &\!=\! e \epsilon^{11}_{f_4 f_5 f_6 f_7 f_8 f_9 f_{10} f_{11} f_{13} f_{14} f_{15}} 
  \hat{R}_{f_1 f_2 f_3 f_4} D_{f_1} \hat{F}_{f_2 f_5 f_6 f_7} 
  D_{f_3} \hat{F}_{f_8 f_9 f_{10} f_{11}} D_{f_{12}} \hat{F}_{f_{12} f_{13} f_{14} f_{15}}, \notag 
  \\ 
  B_2[19] &\!=\! e \epsilon^{11}_{f_3 f_5 f_6 f_8 f_9 f_{10} f_{11} f_{12} f_{13} f_{14} f_{15}} 
  \hat{R}_{f_1 f_2 f_1 f_3} D_{f_4} \hat{F}_{f_2 f_4 f_5 f_6} 
  D_{f_7} \hat{F}_{f_8 f_9 f_{10} f_{11}} D_{f_7} \hat{F}_{f_{12} f_{13} f_{14} f_{15}}, \notag
  \\ 
  B_2[20] &\!=\! e \epsilon^{11}_{f_3 f_5 f_6 f_7 f_8 f_9 f_{10} f_{11} f_{13} f_{14} f_{15}} 
  \hat{R}_{f_1 f_2 f_1 f_3} D_{f_2} \hat{F}_{f_4 f_5 f_6 f_7} 
  D_{f_4} \hat{F}_{f_8 f_9 f_{10} f_{11}} D_{f_{12}} \hat{F}_{f_{12} f_{13} f_{14} f_{15}}, \notag 
  \\ 
  B_2[21] &\!=\! e \epsilon^{11}_{f_4 f_5 f_6 f_7 f_8 f_9 f_{10} f_{11} f_{13} f_{14} f_{15}} 
  \hat{R}_{f_1 f_2 f_1 f_3} D_{f_2} \hat{F}_{f_4 f_5 f_6 f_7} 
  D_{f_3} \hat{F}_{f_8 f_9 f_{10} f_{11}} D_{f_{12}} \hat{F}_{f_{12} f_{13} f_{14} f_{15}}, \notag 
  \\ 
  B_2[22] &\!=\! e \epsilon^{11}_{f_4 f_5 f_6 f_8 f_9 f_{10} f_{11} f_{12} f_{13} f_{14} f_{15}} 
  \hat{R}_{f_1 f_2 f_1 f_2} D_{f_3} \hat{F}_{f_3 f_4 f_5 f_6} 
  D_{f_7} \hat{F}_{f_8 f_9 f_{10} f_{11}} D_{f_7} \hat{F}_{f_{12} f_{13} f_{14} f_{15}}. \notag
\end{alignat}
Note that some of $B_2[11]$, $\cdots$, $B_2[22]$ may be related to terms which do not include $D^a \hat{F}_{abcd}$ by using the dimension dependent identities. In order to obtain the basis of (\ref{eq:B2}) and (\ref{eq:B2_2}), we selected terms which include $D^a \hat{F}_{abcd}$ as many as possible. The summary of the basis in $B_2$ is shown in the Table~\ref{tb:B2}.

\subsection{$F_1=[e(DF)^2\overline{\psi_2}\gamma\mathcal{D}\psi_2]_{1298}$} \label{subsec:F1}

\begin{table}[hb]
  \centering
  \begin{tabular}{l|cccccc|c}
    LEG & 1 & 3 & 5 & 7 & 9 & 11 & total \\ \hline
    \#ADJ & 380 & 1119 & 1318 & 897 & 376 & 94 & 4184 \\
    \#IND & 372 & 1093 & 1296 & 868 & 360 & 83 & 4072 \\
    \#Bianchi id. & 230 & 681 & 862 & 626 & 297 & 76 & 2772 \\
    \#Dim. dep. id. & 0 & 0 & 0 & 0 & 16 & 46 & 62 \\
    \#TNS & 142 & 412 & 434 & 242 & 63 & 5 & 1298 \\
    \#TNS reduced by EOM & 36 & 59 & 40 & 8 & 0 & 0 & 143
  \end{tabular}
  \caption{Basis for $F_1 = [e(DF)^2\overline{\psi_2}\gamma\mathcal{D}\psi_2]$.} \label{tb:F1}
\end{table}

Let us consider terms in the class of $F_1=[e(DF)^2\overline{\psi_2}\gamma\mathcal{D}\psi_2]$. The result is summarized in Table \ref{tb:F1}, and we will find 1298 independent terms for the basis in $F_1$.

As an example of the tensor structure in $F_1$, we choose
\begin{alignat}{3}
  \text{TNS} = eD_{f_1}F_{f_2 f_3 f_4 f_5}D_{f_6}F_{f_2 f_7 f_8 f_9}
  \bar{\psi}_{f_3 f_7}\gamma_{f_1 f_5 f_6}\mathcal{D}_{f_4}\psi_{f_8 f_9}. \label{eq:TNS(DF)2P2DP2}
\end{alignat}
An index list of the TNS is expressed as
\begin{alignat}{3}
  &\text{IND} = \big\{\{\!f_1\!\},\{\!f_2,\!f_3,\!f_4,\!f_5\!\},\{\!f_6\!\},\{\!f_2,\!f_7,\!f_8,\!f_9\!\},\{\!f_3,\!f_7\!\},
  \{\!f_4\!\},\{\!f_8,\!f_9\!\},\{\!f_1,\!f_5,\!f_6\!\}\big\}, \label{eq:IND(DF)2P2DP2}
\end{alignat}
and a rank of the IND, which represents a set of ranks of aligned tensors, is written by
\begin{alignat}{3}
  \text{RNK}=\{1,4,1,4,2,1,2,\text{LEG}\}, \qquad \text{LEG}=3. \label{eq:RNK(DF)2P2DP2}
\end{alignat}
Here LEG is the rank of the gamma matrix in the TNS, and those indices are put at the end of the list to make the code run faster. Then a representative matrix of the adjacency matrices for the above IND is written by
\begin{alignat}{3}
  \text{ADJ} &= 
  \begin{pmatrix}
    0 & 0 & 0 & 0 & 0 & 0 & 0 & 1 \\
    0 & 0 & 0 & 1 & 1 & 1 & 0 & 1 \\
    0 & 0 & 0 & 0 & 0 & 0 & 0 & 1 \\
    0 & 1 & 0 & 0 & 1 & 0 & 2 & 0 \\
    0 & 1 & 0 & 1 & 0 & 0 & 0 & 0 \\
    0 & 1 & 0 & 0 & 0 & 0 & 0 & 0 \\
    0 & 0 & 0 & 2 & 0 & 0 & 0 & 0 \\
    1 & 1 & 1 & 0 & 0 & 0 & 0 & 0 
  \end{pmatrix}. \label{eq:ADJ(DF)2P2DP2}
\end{alignat}
The sum of each row or each column is equal to the corresponding rank of the tensor. Since the order of two $[DF]$s can be exchanged in the tensor (\ref{eq:TNS(DF)2P2DP2}), we consider a conjugacy class of the adjacency matrices which is obtained by identifying  permutations of two $[DF]$s. We call this representative matrix ADJ. 

By using the Mathematica code explained in the appendix \ref{app:Code}, we find that there are 4184 ADJs for $F_1$. Then it is possible to generate INDs from ADJs uniquely. Notice that this is not one to one correspondence since we should take into account antisymmetric properties of $F_{abcd}$,  $\psi_{ab}$ and $\gamma_{a_1\cdots a_n}$. By using the Mathematica code, we find that there are 4072 INDs for $F_1$. 

In this procedure, however, the Bianchi identities in the eq.~(\ref{eq:BianchiDP}) have not been used. There are 2772 Bianchi identities among 4072 INDs. Furthermore, we should take into account the dimension dependent identities for LEG $=9,11$, which arise by antisymmetrization of 12 indices among $(15+\text{LEG})/2$ ones. There are 62 dimension dependent identities, but some of them become trivial by imposing the Bianchi identities. 

Carefully solving the overlap of the Bianchi identities and the dimension dependent ones, we obtain 1298 TNSs for the basis in $F_1$. We denote each term in the basis as $F_1[m,i]$, where $m$ represents the value of LEG and $i$ runs from 1 to \#TNS for each $m$. Among these, we explicitly show 5 terms for $m=11$.
\begin{alignat}{3}
  F_1[11,1] &= e D_{f_1} F_{f_2 f_3 f_4 f_5} D_{f_6} F_{f_7 f_8 f_9 f_{10}} 
  \bar{\psi}_{f_1 f_{11}} \gamma_{f_2 f_3 f_4 f_5 f_7 f_8 f_9 f_{10} f_{11} f_{12} f_{13}} 
  D_{f_6} \psi_{f_{12} f_{13}}, \notag 
  \\ 
  F_1[11,2] &= e D_{f_1} F_{f_2 f_3 f_4 f_5} D_{f_6} F_{f_1 f_7 f_8 f_9} 
  \bar{\psi}_{f_{10} f_{11}} \gamma_{f_2 f_3 f_4 f_5 f_7 f_8 f_9 f_{10} f_{11} f_{12} f_{13}} 
  D_{f_6} \psi_{f_{12} f_{13}}, \notag 
  \\ 
  F_1[11,3] &= e D_{f_1} F_{f_2 f_3 f_4 f_5} D_{f_1} F_{f_6 f_7 f_8 f_9} 
  \bar{\psi}_{f_{10} f_{11}} \gamma_{f_2 f_3 f_4 f_5 f_6 f_7 f_8 f_9 f_{11} f_{12} f_{13}} 
  D_{f_{10}} \psi_{f_{12} f_{13}}, \label{eq:F1[11]} 
  \\ 
  F_1[11,4] &= e D_{f_1} F_{f_2 f_3 f_4 f_5} D_{f_1} F_{f_6 f_7 f_8 f_9} 
  \bar{\psi}_{f_{10} f_{11}} \gamma_{f_3 f_4 f_5 f_6 f_7 f_8 f_9 f_{10} f_{11} f_{12} f_{13}} 
  D_{f_2} \psi_{f_{12} f_{13}}, \notag 
  \\ 
  F_1[11,5] &= e D_{f_1} F_{f_1 f_2 f_3 f_4} D_{f_5} F_{f_6 f_7 f_8 f_9} 
  \bar{\psi}_{f_{10} f_{11}} \gamma_{f_2 f_3 f_4 f_6 f_7 f_8 f_9 f_{10} f_{11} f_{12} f_{13}} 
  D_{f_5} \psi_{f_{12} f_{13}}. \notag
\end{alignat}
These are mapped to other terms by using the equations of motion for the eleven dimensional supergravity. The summary of the basis in $F_1$ is shown in the Table~\ref{tb:F1}.

\subsection{$F_2=[e(DF)^3\bar{\psi}\gamma\psi_2]_{4933}$} \label{subsec:F2}

\begin{table}[htb]
  \centering
  \begin{tabular}{l|cccccc|c}
    LEG & 0 & 2 & 4 & 6 & 8 & 10 & total \\ \hline
    \#ADJ & 211 & 1873 & 3885 & 3939 & 2459 & 1063 & 13430 \\
    \#IND & 194 & 1809 & 3774 & 3813 & 2345 & 981 & 12916 \\
    \#Bianchi id. & 107 & 989 & 2129 & 2328 & 1585 & 741 & 7879 \\
    \#Dim. dep. id. & 0 & 0 & 0 & 24 & 209 & 444 & 677 \\
    \#TNS & 87 & 820 & 1645 & 1484 & 733 & 164 & 4933 \\
    \#TNS reduced by EOM & 31 & 285 & 504 & 400 & 155 & 0 & 1375
  \end{tabular}
  \caption{Basis for $F_2 = [e(DF)^3\bar{\psi}\gamma\psi_2]$.} \label{tb:F2}
\end{table}

Let us consider terms in the class of $F_2=[e(DF)^3\bar{\psi}\gamma\psi_2]$. The result is shown in Table~\ref{tb:F2}, and we obtain 4933 independent terms for the basis in $F_2$. 

An example of the tensor structure in this class is written by
\begin{alignat}{3}
  \text{TNS} = e D_{f_1} F_{f_2 f_3 f_4 f_5} D_{f_6} F_{f_2 f_7 f_8 f_9} 
  D_{f_{10}} F_{f_3 f_7 f_{11} f_{12}} \bar{\psi}_{f_4} 
  \gamma_{f_1 f_5 f_6 f_9 f_{10} f_{12}} \psi_{f_8 f_{11}}. \label{eq:TNS(DF)3PP2}
\end{alignat}
From this TNS, we extract an index list like
\begin{alignat}{3}
  &\text{IND} = 
  \big\{\{\!f_1\!\},\{\!f_2,\!f_3,\!f_4,\!f_5\!\},\{\!f_6\!\},\{\!f_2,\!f_7,\!f_8,\!f_9\!\},
  \{\!f_{10}\!\},\{\!f_3,\!f_7,\!f_{11},\!f_{12}\!\}, \notag
  \\
  &\qquad\qquad
  \{\!f_4\!\},\{\!f_8,\!f_{11}\!\},\{\!f_1,\!f_5,\!f_6,\!f_9,\!f_{10},\!f_{12}\!\}\big\}, \label{eq:IND(DF)3PP2}
\end{alignat}
and a rank of the IND, which represents a set of ranks of aligned tensors, is written by
\begin{alignat}{3}
  \text{RNK}=\{1,4,1,4,1,4,1,2,\text{LEG}\}, \qquad \text{LEG}=6. \label{eq:RNK(DF)3PP2}
\end{alignat}
Here LEG is the rank of the gamma matrix in the TNS, and those indices are put at the end of the list to make the code run faster. Then a representative matrix of the adjacency matrices for the above IND is written by
\begin{alignat}{3}
  \text{ADJ} &= 
  \begin{pmatrix}
    0 & 0 & 0 & 0 & 0 & 0 & 0 & 0 & 1 \\
    0 & 0 & 0 & 1 & 0 & 1 & 1 & 0 & 1 \\
    0 & 0 & 0 & 0 & 0 & 0 & 0 & 0 & 1 \\
    0 & 1 & 0 & 0 & 0 & 1 & 0 & 1 & 1 \\
    0 & 0 & 0 & 0 & 0 & 0 & 0 & 0 & 1 \\
    0 & 1 & 0 & 1 & 0 & 0 & 0 & 1 & 1 \\
    0 & 1 & 0 & 0 & 0 & 0 & 0 & 0 & 0 \\
    0 & 0 & 0 & 1 & 0 & 1 & 0 & 0 & 0 \\
    1 & 1 & 1 & 1 & 1 & 1 & 0 & 0 & 0 
  \end{pmatrix}. \label{eq:ADJ(DF)3PP2}
\end{alignat}
The sum of each row or each column is equal to the corresponding rank of the tensor. Since the order of three $[DF]$s can be exchanged in the tensor (\ref{eq:TNS(DF)3PP2}), we consider a conjugacy class of the adjacency matrices which is obtained by identifying the permutations of three $[DF]$s. We call this representative matrix ADJ. 

By using the Mathematica code explained in the appendix \ref{app:Code}, we find that there are 13430 ADJs for $F_2$. Then it is possible to generate INDs from ADJs uniquely. Notice that this is not one to one correspondence since we should take into account antisymmetric properties of $F_{abcd}$,  $\psi_{ab}$ and $\gamma_{a_1\cdots a_n}$. By using the Mathematica code, we find that there are 12916 INDs for $F_2$. 

In this procedure, however, the Bianchi identities in the eq.~(\ref{eq:BianchiDP}) has not been used. There are 7879 Bianchi identities among 12916 INDs. Furthermore, we should take into account the dimension dependent identities for LEG $=6,8,10$, which arise by antisymmetrization of 12 indices among $(18+\text{LEG})/2$ ones. There are 677 dimension dependent identities, but some of them become trivial by imposing the Bianchi identities. 

Carefully solving the overlap of the Bianchi identities and the dimension dependent ones, we obtain 4933 TNSs for the basis in $F_2$. We denote each term in the basis as $F_2[m,i]$, where $m$ represents
the value of LEG and $i$ runs from 1 to \#TNS for each $m$. The summary of the basis in $F_2$ is shown in the Table~\ref{tb:F2}.

\subsection{$F_3=[eF(DF)^3 \bar{\psi} \gamma \psi]_{14337}$} \label{subsec:F3}

\begin{table}[htb]
  \centering
  \begin{tabular}{l|cccccc|c}
    LEG & 1 & 3 & 5 & 7 & 9 & 11 & total \\ \hline
    \#ADJ & 7324 & (29043) & 45332 & (40612) & 23828 & (9890) & 76484 \\
    \#IND & 3774 & (13390) & 23797 & (18351) & 12659 & (4148) & 40230 \\
    \#Bianchi id. & 2129 & (7396) & 13839 & (11187) & 8454 & (3004) & 24422 \\
    \#Dim. dep. id. & 0 &  & 619 & & 5240 & & 5859 \\
    \#TNS & 1645 & 0 & 9856 & 0 & 2836 & 0 & 14337 \\
    \#TNS reduced by EOM & 748 & 0 & 4996 & 0 & 1451 & 0 & 7195 
  \end{tabular}
  \caption{Basis for $F_3 = [eF(DF)^3 \bar{\psi} \gamma \psi]$.} \label{tb:F3}
\end{table}

Let us consider terms in the class of $F_3=[eF(DF)^3 \bar{\psi} \gamma \psi]$. The result is shown in Table~\ref{tb:F3}, and we will have 14337 independent terms for the basis in $F_3$. 

An example of the tensor structure in this class is written by
\begin{alignat}{3}
  \text{TNS} = e F_{f_1 f_2 f_3 f_4} D_{f_5} F_{f_6 f_7 f_8 f_9} 
  D_{f_6} F_{f_5 f_{10} f_{11} f_{12}} D_{f_7} F_{f_8 f_{10} f_{11} f_{13}}
  \psi_{f_9}\gamma_{f_1 f_2 f_3 f_4 f_{13}}\psi_{f_{12}}. \label{eq:TNSF(DF)3PP}
\end{alignat}
From this we extract an index list of the TNS like
\begin{alignat}{3}
  &\text{IND} = \big\{\{\!f_1,\!f_2,\!f_3,\!f_4\!\},\{\!f_5\!\},\{\!f_6,\!f_7,\!f_8,\!f_9\!\},
  \{\!f_6\!\},\{\!f_5,\!f_{10},\!f_{11},\!f_{12}\!\}, \notag
  \\
  &\qquad\qquad
  \{\!f_7\!\},\{\!f_8,\!f_{10},\!f_{11},\!f_{13}\!\},\{\!f_9\!\},\{\!f_{12}\!\},\{\!f_1,\!f_2,\!f_3,\!f_4,\!f_{13}\!\}\big\}. \label{eq:INDF(DF)3PP}
\end{alignat}
A rank of the IND, which represents a set of ranks of aligned tensors, is written by
\begin{alignat}{3}
  \text{RNK}=\{4,1,4,1,4,1,4,1,1,\text{LEG}\}, \qquad \text{LEG}=5. \label{eq:RNKF(DF)3PP}
\end{alignat}
Here LEG is the rank of the gamma matrix in the TNS, and those indices are put at the end of the list to make the code run faster. Then a representative matrix of the adjacency matrices for the above IND is written by
\begin{alignat}{3}
  \text{ADJ} &= 
  \begin{pmatrix}
    0 & 0 & 0 & 0 & 0 & 0 & 0 & 0 & 0 & 4 \\
    0 & 0 & 0 & 0 & 1 & 0 & 0 & 0 & 0 & 0 \\
    0 & 0 & 0 & 1 & 0 & 1 & 1 & 1 & 0 & 0 \\
    0 & 0 & 1 & 0 & 0 & 0 & 0 & 0 & 0 & 0 \\
    0 & 1 & 0 & 0 & 0 & 0 & 2 & 0 & 1 & 0 \\
    0 & 0 & 1 & 0 & 0 & 0 & 0 & 0 & 0 & 0 \\
    0 & 0 & 1 & 0 & 2 & 0 & 0 & 0 & 0 & 1 \\
    0 & 0 & 1 & 0 & 0 & 0 & 0 & 0 & 0 & 0 \\
    0 & 0 & 0 & 0 & 1 & 0 & 0 & 0 & 0 & 0 \\
    4 & 0 & 0 & 0 & 0 & 0 & 1 & 0 & 0 & 0
  \end{pmatrix}. \label{eq:ADJF(DF)3PP}
\end{alignat}
The sum of each row or each column is equal to the corresponding rank of the tensor. Since the order of three $[DF]$s can be exchanged in the tensor (\ref{eq:TNSF(DF)3PP}), we consider a conjugacy class of the adjacent matrices which is obtained by identifying the permutations of three $[DF]$s. We call this representative matrix ADJ. 

By using the Mathematica code explained in the appendix \ref{app:Code}, we find that there are 76484 ADJs for $F_3$ with LEG $=1,5,9$. And it is possible to generate INDs from ADJs uniquely. Notice that this is not one to one correspondence since we should take into account antisymmetric properties of $F_{abcd}$ and $\gamma_{a_1\cdots a_n}$. We also consider two gravitinos are antisymmetric under the exchange for LEG $=1,5,9$ and symmetric for LEG $=3,7,11$, because they are Majorana spinors. In this paper, we only consider LEG $=1,5,9$ to make the ansatz simple. By using the Mathematica code, we find that there are 40230 INDs for $F_3$. 

In this procedure, however, the Bianchi identities in the eq.~(\ref{eq:BianchiDP}) has not been used. There are 24422 Bianchi identities among 40230 INDs. Furthermore, we should take into account the dimension dependent identities for LEG $=5,9$, which arise by antisymmetrization of 12 indices among $(21+\text{LEG})/2$ ones. There are 5859 dimension dependent identities, but some of them become trivial by imposing the Bianchi identities. 

Carefully solving the overlap of the Bianchi identities and the dimension dependent ones, we obtain 14337 TNSs for the basis in $F_3$. We denote each term in the basis as $F_3[m,i]$, where $m$ represents the value of LEG and $i$ runs from 1 to \#TNS for each $m$. The summary of the basis in $F_3$ is shown in the Table~\ref{tb:F3}.

\subsection{$V_1=[e(DF)^2DDF\bar{\epsilon}\gamma\psi_2]_{5392}$} \label{subsec:V1}

\begin{table}[htb]
  \centering
  \begin{tabular}{l|cccccc|c}
    LEG & 0 & 2 & 4 & 6 & 8 & 10 & total \\ \hline
    \#ADJ & 604 & 5491 & 11431 & 11570 & 7168 & 3040 & 39304 \\
    \#IND & 587 & 5427 & 11320 & 11444 & 7054 & 2958 & 38790 \\
    \#Bianchi id. etc. & 495 & 4482 & 9491 & 9838 & 6290 & 2729 & 33325 \\
    \#Dim. dep. id. & 0 & 0 & 0 & 81 & 632 & 1350 & 2063 \\
    \#TNS & 92 & 945 & 1829 & 1606 & 752 & 168 & 5392
  \end{tabular}
  \caption{Basis for $V_1 = [e(DF)^2DDF\bar{\epsilon}\gamma\psi_2]$.} \label{tb:V1}
\end{table}

Let us classify variations in the class of $V_1=[e(DF)^2DDF\bar{\epsilon}\gamma\psi_2]$. The result is summarized in Table.~\ref{tb:V1}, and the number of the independent terms becomes 5392 for the basis in $V_1$.

As an example of the tensor in this class, we choose
\begin{alignat}{3}
  \text{TNS} = eD_{f_1}F_{f_2 f_3 f_4 f_5}D_{f_6}F_{f_7 f_8 f_9 f_{10}}
  D_{f_{11}}D_{f_{11}}F_{f_2 f_3 f_7 f_{12}}
  \bar{\epsilon}\gamma_{f_4 f_5 f_8 f_9 f_{10} f_{12}}\psi_{f_1 f_6}. \label{eq:TNS(DF)2DDFP2}
\end{alignat}
From the TNS, we extract an index list like
\begin{alignat}{3}
  &\text{IND} = 
  \big\{\{\!f_1\!\},\{\!f_2,\!f_3,\!f_4,\!f_5\!\},\{\!f_6\!\},\{\!f_7,\!f_8,\!f_9,\!f_{10}\!\},
  \{\!f_{11}\!\},\{\!f_{11}\!\},\{\!f_2,\!f_3,\!f_7,\!f_{12}\!\}, \notag
  \\
  &\qquad\qquad
  \{\!f_1,\!f_6\!\},\{\!f_4,\!f_5,\!f_8,\!f_9,\!f_{10},\!f_{12}\!\}\big\}, \label{eq:IND(DF)2DDFP2}
\end{alignat}
and the rank of the IND is written by
\begin{alignat}{3}
  \text{RNK}=\{1,4,1,4,1,1,4,2,\text{LEG}\}, \qquad \text{LEG}=6. \label{eq:RNK(DF)2DDFP2}
\end{alignat}
Here LEG is the rank of the gamma matrix in the TNS, and those indices are put at the end of the list to make the code run faster. Then a representative matrix of the adjacency matrices of the above IND is expressed as
\begin{alignat}{3}
  \text{ADJ} &= 
  \begin{pmatrix}
    0 & 0 & 0 & 0 & 0 & 0 & 0 & 1 & 0 \\
    0 & 0 & 0 & 0 & 0 & 0 & 2 & 0 & 2 \\
    0 & 0 & 0 & 0 & 0 & 0 & 0 & 1 & 0 \\
    0 & 0 & 0 & 0 & 0 & 0 & 1 & 0 & 3 \\
    0 & 0 & 0 & 0 & 0 & 1 & 0 & 0 & 0 \\
    0 & 0 & 0 & 0 & 1 & 0 & 0 & 0 & 0 \\
    0 & 2 & 0 & 1 & 0 & 0 & 0 & 0 & 1 \\
    1 & 0 & 1 & 0 & 0 & 0 & 0 & 0 & 0 \\
    0 & 2 & 0 & 3 & 0 & 0 & 1 & 0 & 0 
  \end{pmatrix}. \label{eq:ADJ(DF)2DDFP2}
\end{alignat}
The sum of each row or each column is equal to the corresponding rank of the tensor. Since the order of two $[DF]$s can be exchanged in the tensor (\ref{eq:TNS(DF)2DDFP2}), we consider a conjugacy class of the adjacency matrices which is obtained by identifying the permutations of  two $[DF]$s. We call this representative matrix ADJ. 

By using the Mathematica code explained in the appendix \ref{app:Code}, we find that there are 39304 ADJs for $V_1$. Then it is possible to generate INDs from ADJs uniquely. Notice that this is not one to one correspondence since we should take into account antisymmetric properties of $F_{abcd}$,  $\psi_{ab}$ and $\gamma_{a_1\cdots a_n}$. By using the Mathematica code, we find that there are 38790 INDs for $V_1$. 

In this procedure, however, the Bianchi identities in the eq.~(\ref{eq:BianchiDP}) have not been used. We should also take into account the commutation relations in the eq.~(\ref{eq:comD}), which generate terms in other classes not shown in the Fig.~\ref{fig:O(R0)part}. There are 33325 Bianchi identities and the commutation relations among 38790 INDs. Furthermore, it is necessary to consider the dimension dependent identities for LEG $=6,8,10$ which arise by antisymmetrization of 12 indices among $(18+\text{LEG})/2$ ones. There are 2063 dimension dependent identities, but some of them become trivial by imposing the Bianchi identities and the commutation relations. 

Carefully solving the overlap of the Bianchi identities, the commutation relations and the dimension dependent identities, we obtain 5392 TNSs for the basis in $V_1$. We denote each term in the basis as $V_1[n,j]$, where $n$ represents the value of LEG and $j$ runs from 1 to \#TNS for each $n$. The summary of the basis in $V_1$ is shown in the Table~\ref{tb:V1}.

\subsection{$V_2=[e(DF)^3\bar{\epsilon}\gamma\mathcal{D}\psi_2]_{3067}$} \label{subsec:V2}

\begin{table}[hb]
  \centering
  \begin{tabular}{l|cccccc|c}
    LEG & 0 & 2 & 4 & 6 & 8 & 10 & total \\ \hline
    \#ADJ & 211 & 1873 & 3885 & 3939 & 2459 & 1063 & 13430 \\
    \#IND & 194 & 1809 & 3774 & 3813 & 2345 & 981 & 12916 \\
    \#Bianchi id. & 145 & 1287 & 2724 & 2909 & 1887 & 850 & 9802 \\
    \#Dim. dep. id. & 0 & 0 & 0 & 24 & 209 & 444 & 677 \\
    \#TNS & 49 & 522 & 1050 & 904 & 447 & 95 & 3067 
  \end{tabular}
  \caption{Basis for $V_2 = [e(DF)^3\bar{\epsilon}\gamma\mathcal{D}\psi_2]$.} \label{tb:V2}
\end{table}

Let us consider variations in the class of $V_2=[e(DF)^3\bar{\epsilon}\gamma\mathcal{D}\psi_2]$.  The result is summarized in Table.~\ref{tb:V2}, and the number of the independent terms is 3067 for the basis in $V_2$. 

As an example of the tensor in this class, we select
\begin{alignat}{3}
  \text{TNS} = e D_{f_1} F_{f_2 f_3 f_4 f_5} D_{f_6} F_{f_2 f_7 f_8 f_9}
  D_{f_{10}} F_{f_{11} f_{12} f_{13} f_{14}}
  \bar{\epsilon} \gamma_{f_1 f_4 f_5 f_6 f_7 f_8 f_9 f_{10} f_{13} f_{14}}
  \mathcal{D}_{f_{11}} \psi_{f_3 f_{12}}. \label{eq:TNS(DF)3DP2}
\end{alignat}
From the TNS, we extract an index list like
\begin{alignat}{3}
  &\text{IND} = 
  \big\{\{\!f_1\!\},\{\!f_2,\!f_3,\!f_4,\!f_5\!\},\{\!f_6\!\},\{\!f_2,\!f_7,\!f_8,\!f_9\!\},
  \{\!f_{10}\!\},\{\!f_{11},\!f_{12},\!f_{13},\!f_{14}\!\}, \notag
  \\
  &\qquad\qquad
  \{\!f_{11}\!\},\{\!f_3,\!f_{12}\!\},\{\!f_1,\!f_4,\!f_5,\!f_6,\!f_7,\!f_8,\!f_9,\!f_{10},\!f_{13},\!f_{14}\!\}\big\}, \label{eq:IND(DF)3DP2}
\end{alignat}
and the rank of the IND is written by
\begin{alignat}{3}
  \text{RNK}=\{1,4,1,4,1,4,1,2,\text{LEG}\}, \qquad \text{LEG}=10. \label{eq:RNK(DF)3DP2}
\end{alignat}
Here LEG is the rank of the gamma matrix in the TNS, and those indices are put at the end of the list to make the code run faster. Then a representative matrix of the adjacency matrices of the above IND is written by
\begin{alignat}{3}
  \text{ADJ} &= 
  \begin{pmatrix}
    0 & 0 & 0 & 0 & 0 & 0 & 0 & 0 & 1 \\
    0 & 0 & 0 & 1 & 0 & 0 & 0 & 1 & 2 \\
    0 & 0 & 0 & 0 & 0 & 0 & 0 & 0 & 1 \\
    0 & 1 & 0 & 0 & 0 & 0 & 0 & 0 & 3 \\
    0 & 0 & 0 & 0 & 0 & 0 & 0 & 0 & 1 \\
    0 & 0 & 0 & 0 & 0 & 0 & 1 & 1 & 2 \\
    0 & 0 & 0 & 0 & 0 & 1 & 0 & 0 & 0 \\
    0 & 1 & 0 & 0 & 0 & 1 & 0 & 0 & 0 \\
    1 & 2 & 1 & 3 & 1 & 2 & 0 & 0 & 0 
  \end{pmatrix}. \label{eq:ADJ(DF)3DP2}
\end{alignat}
The sum of each row or each column is equal to the corresponding rank of the tensor. Since the order of three $[DF]$s can be exchanged in the tensor (\ref{eq:TNS(DF)3DP2}), we should consider a conjugacy class of the adjacent matrices which is obtained by identifying the permutations of three $[DF]$s. We call this representative matrix ADJ. 

By using the Mathematica code explained in the appendix \ref{app:Code}, we find that there are 13430 ADJs for $V_2$. Then it is possible to generate INDs from ADJs uniquely. Notice that this is not one to one correspondence since we should take into account antisymmetric properties of $F_{abcd}$,  $\psi_{ab}$ and $\gamma_{a_1\cdots a_n}$. By using the Mathematica code, we find that there are 12916 INDs for $V_2$. 

In this procedure, however, the Bianchi identities in the eq.~(\ref{eq:BianchiDP}) have not been used. There are 9802 Bianchi identities among 12916 INDs. It should be noted that the Bianchi identities for $\mathcal{D}\psi_2$ give relations among terms in $[e(DF)^3\bar{\epsilon}\gamma\mathcal{D}\psi_2]$, $[e(DF)^4\bar{\epsilon}\gamma\psi]$, $[eF^2(DF)^3\bar{\epsilon}\gamma\psi]$ and $[eR(DF)^3\bar{\epsilon}\gamma\psi]$. So we should deal with these relations carefully when we consider cancellations of the variations in each class, and it will be done in section~\ref{sec:Cancel}. Furthermore, we should take into account the dimension dependent identities  for LEG $=6,8,10$, which arise by antisymmetrization of 12 indices among $(18+\text{LEG})/2$ ones. There are 677 dimension dependent identities, but some of them become trivial by imposing the Bianchi identities. 

Carefully solving the overlap of the Bianchi identities and the dimension dependent ones, we obtain 3067 TNSs for the basis in $V_2$. We denote each term in the basis as $V_2[n,j]$, where $n$ represents the value of LEG and $j$ runs from 1 to \#TNS for each $n$. The summary of the basis in $V_2$ is shown in the Table~\ref{tb:V2}.

\subsection{$V_4=[e(DF)^4\bar{\epsilon}\gamma\psi]_{10285}$} \label{subsec:V4}

\begin{table}[hb]
  \centering
  \begin{tabular}{l|cccccc|c}
    LEG & 1 & 3 & 5 & 7 & 9 & 11 & total \\ \hline
    \#ADJ & 1909 & 7440 & 11587 & 10411 & 6164 & 2605 & 40116 \\
    \#IND & 1839 & 7067 & 11208 & 9890 & 5807 & 2318 & 38129 \\
    \#Bianchi id. & 1194 & 4684 & 7594 & 7124 & 4471 & 1922 & 26989 \\
    \#Dim. dep. id. & 0 & 12 & 239 & 1133 & 2310 & 2009 & 5703 \\
    \#TNS & 645 & 2383 & 3595 & 2619 & 967 & 76 & 10285
  \end{tabular}
  \caption{Basis for $V_4 = [e(DF)^4\bar{\epsilon}\gamma\psi]$.} \label{tb:V4}
\end{table}

Let us consider variations in the class of $V_4=[e(DF)^4\bar{\epsilon}\gamma\psi]$. The result is summarized in Table.~\ref{tb:V4}, and the number of the independent terms is 10285 for the basis in $V_4$.
 
As an example of the tensor in this class, we make a choice of 
\begin{alignat}{3}
  \text{TNS} = e D_{f_1} F_{f_2 f_3 f_4 f_5} D_{f_6} F_{f_1 f_7 f_8 f_9} 
  D_{f_2} F_{f_7 f_{10} f_{11} f_{12}} D_{f_8} F_{f_3 f_4 f_{10} f_{13}}
  \bar{\epsilon}\gamma_{f_5 f_9 f_{11} f_{12} f_{13}}\psi_{f_6}. \label{eq:TNS(DF)4P}
\end{alignat}
From this TNS, we extract an index list like
\begin{alignat}{3}
  &\text{IND} = 
  \big\{\{\!f_1\!\},\{\!f_2,\!f_3,\!f_4,\!f_5\!\},\{\!f_6\!\},\{\!f_1,\!f_7,\!f_8,\!f_9\!\},
  \{\!f_2\!\},\{\!f_7,\!f_{10},\!f_{11},\!f_{12}\!\}, \notag
  \\
  &\qquad\qquad
  \{\!f_8\!\},\{\!f_3,\!f_4,\!f_{10},\!f_{13}\!\},\{\!f_6\!\},\{\!f_5,\!f_9,\!f_{11},\!f_{12},\!f_{13}\!\}\big\}, \label{eq:IND(DF)4P}
\end{alignat}
and the rank of this IND is written by
\begin{alignat}{3}
  \text{RNK}=\{1,4,1,4,1,4,1,4,1,\text{LEG}\}, \qquad \text{LEG}=5. \label{eq:RNK(DF)4P}
\end{alignat}
Here LEG is the rank of the gamma matrix in the TNS, and these indices are put at the end of the list to make the code run faster. Then a representative matrix of the adjacency matrices of the IND is written by
\begin{alignat}{3}
  \text{ADJ} &= 
  \begin{pmatrix}
    0 & 0 & 0 & 1 & 0 & 0 & 0 & 0 & 0 & 0 \\
    0 & 0 & 0 & 0 & 1 & 0 & 0 & 2 & 0 & 1 \\
    0 & 0 & 0 & 0 & 0 & 0 & 0 & 0 & 1 & 0 \\
    1 & 0 & 0 & 0 & 0 & 1 & 1 & 0 & 0 & 1 \\
    0 & 1 & 0 & 0 & 0 & 0 & 0 & 0 & 0 & 0 \\
    0 & 0 & 0 & 1 & 0 & 0 & 0 & 1 & 0 & 2 \\
    0 & 0 & 0 & 1 & 0 & 0 & 0 & 0 & 0 & 0 \\
    0 & 2 & 0 & 0 & 0 & 1 & 0 & 0 & 0 & 1 \\
    0 & 0 & 1 & 0 & 0 & 0 & 0 & 0 & 0 & 0 \\
    0 & 1 & 0 & 1 & 0 & 2 & 0 & 1 & 0 & 0
  \end{pmatrix}. \label{eq:ADJ(DF)4P}
\end{alignat}
The sum of each row or each column is equal to the corresponding rank of the tensor. Since the order of four $[DF]$s can be exchanged in the tensor (\ref{eq:TNS(DF)4P}), we consider a conjugacy class of the adjacency matrices which is obtained by identifying the permutations of four $[DF]$s. We call this representative matrix ADJ. 

By using the Mathematica code explained in the appendix \ref{app:Code}, we find that there are 40116 ADJs for $V_4$. Then it is possible to generate INDs from ADJs uniquely. Notice that this is not one to one correspondence since we should take into account antisymmetric properties of $F_{abcd}$ and $\gamma_{a_1\cdots a_n}$. By using the Mathematica code, we find that there are 38129 INDs for $V_4$. 

In this procedure, however, the Bianchi identities in the eq.~(\ref{eq:BianchiDP}) have not been used. There are 26989 Bianchi identities among 38129 INDs. Furthermore, we should take into account the dimension dependent identities for LEG $=3,5,7,9,11$, which arise by antisymmetrization of 12 indices among $(21+\text{LEG})/2$ ones. There are 5703 dimension dependent identities, but some of them become trivial by imposing the Bianchi identities. 

Carefully solving the overlap of the Bianchi identities and the dimension dependent ones, we obtain 10285 TNSs for the basis in $V_4$. We denote each term in the basis as $V_4[n,j]$, where $n$ represents the value of LEG and $j$ runs from 1 to \#TNS for each $n$. The summary of the basis in $V_4$ is shown in the Table~\ref{tb:V4}.

\subsection{$V_5=[eF(DF)^2DDF\bar{\epsilon}\gamma\psi]_{60394}$} \label{subsec:V5}

\begin{table}[h]
  \centering
  \begin{tabular}{l|cccccc|c}
    LEG & 1 & 3 & 5 & 7 & 9 & 11 & total \\ \hline
    \#ADJ & 21635 & 86165 & 134539 & 120261 & 70201 & 28820 & 461621 \\
    \#IND & 21495 & 85652 & 133794 & 119450 & 69487 & 28349 & 458227 \\
    \#Bianchi id. etc. & 17886 & 71422 & 112953 & 103026 & 61651 & 25884 & 392822 \\
    \#Dim. dep. id. & 0 & 192 & 2907 & 13806 & 27878 & 24510 & 69293 \\
    \#TNS & 3609 & 14230 & 20772 & 15647 & 5673 & 463 & 60394 
  \end{tabular}
  \caption{Basis for $V_5 = [eF(DF)^2DDF\bar{\epsilon}\gamma\psi]$.} \label{tb:V5}
\end{table}

Let us investigate variations in the class of $V_5=[eF(DF)^2DDF\bar{\epsilon}\gamma\psi]$. The result is summarized in Table.~\ref{tb:V5}, and the number of the independent terms is 60394 for the basis in $V_5$. 

As an example of the tensor in this class, we choose
\begin{alignat}{3}
  \text{TNS} = e F_{f_1 f_2 f_3 f_4} D_{f_5} F_{f_6 f_7 f_8 f_9} 
  D_{f_{10}} F_{f_6 f_7 f_{11} f_{12}} D_{f_8} D_{f_{13}} F_{f_5 f_{10} f_{11} f_{13}}
  \bar{\epsilon} \gamma_{f_1 f_2 f_3 f_4 f_{12}} \psi_{f_9}. \label{eq:TNSF(DF)2DDFP}
\end{alignat}
From this TNS, we extract an index list like
\begin{alignat}{3}
  &\text{IND} = 
  \big\{\{\!f_1,\!f_2,\!f_3,\!f_4\!\},\{\!f_5\!\},\{\!f_6,\!f_7,\!f_8,\!f_9\!\},
  \{\!f_{10}\!\},\{\!f_6,\!f_7,\!f_{11},\!f_{12}\!\}, \notag
  \\
  &\qquad\qquad
  \{\!f_8\!\},\{\!f_{13}\!\},\{\!f_5,\!f_{10},\!f_{11},\!f_{13}\!\},
  \{\!f_9\!\},\{\!f_1,\!f_2,\!f_3,\!f_4,\!f_{12}\!\}\big\}, \label{eq:INDF(DF)2DDFP}
\end{alignat}
and the rank of the IND is written by
\begin{alignat}{3}
  \text{RNK}=\{4,1,4,1,4,1,1,4,1,\text{LEG}\}, \qquad \text{LEG}=5. \label{eq:RNKF(DF)2DDFP}
\end{alignat}
Here LEG is the rank of the gamma matrix in the TNS, and those indices are put at the end of the list to make the code run faster. Then a representative matrix of the adjacency matrices of the IND is written by
\begin{alignat}{3}
  \text{ADJ} &= 
  \begin{pmatrix}
    0 & 0 & 0 & 0 & 0 & 0 & 0 & 0 & 0 & 4 \\
    0 & 0 & 0 & 0 & 0 & 0 & 0 & 1 & 0 & 0 \\
    0 & 0 & 0 & 0 & 2 & 1 & 0 & 0 & 1 & 0 \\
    0 & 0 & 0 & 0 & 0 & 0 & 0 & 1 & 0 & 0 \\
    0 & 0 & 2 & 0 & 0 & 0 & 0 & 1 & 0 & 1 \\
    0 & 0 & 1 & 0 & 0 & 0 & 0 & 0 & 0 & 0 \\
    0 & 0 & 0 & 0 & 0 & 0 & 0 & 1 & 0 & 0 \\
    0 & 1 & 0 & 1 & 1 & 0 & 1 & 0 & 0 & 0 \\
    0 & 0 & 1 & 0 & 0 & 0 & 0 & 0 & 0 & 0 \\
    4 & 0 & 0 & 0 & 1 & 0 & 0 & 0 & 0 & 0
  \end{pmatrix}. \label{eq:ADJF(DF)2DDFP}
\end{alignat}
The sum of each row or each column is equal to the corresponding rank of the tensor. Since the order of two $[DF]$s can be exchanged in the tensor (\ref{eq:TNSF(DF)2DDFP}), we consider a conjugacy class of the adjacency matrices which is obtained by identifying the permutations of two $[DF]$s. We call this representative matrix ADJ. 

By using the Mathematica code explained in the appendix \ref{app:Code}, we find that there are 461621 ADJs for $V_5$. Then it is possible to generate INDs from ADJs uniquely. Notice that this is not one to one correspondence since we should take into account antisymmetric properties of $F_{abcd}$ and $\gamma_{a_1\cdots a_n}$. By using the Mathematica code, we find that there are 458227 INDs for $V_5$. 

In this procedure, however, the Bianchi identities in the eq.~(\ref{eq:BianchiDP}) have not been used. We should also take into account the commutation relations in the eq.~(\ref{eq:comD}), which generate terms in other classes not shown in the Fig.~\ref{fig:O(R0)part}. There are 392822 Bianchi identities and commutation relations among 458227 INDs. Furthermore, we should take into account the dimension dependent identities for LEG $=3,5,7,9,11$, which arise by antisymmetrization of 12 indices among $(21+\text{LEG})/2$ ones. There are 69293 dimension dependent identities, but some of them become trivial by imposing the Bianchi identities and the commutation relations. 

Carefully solving the overlap of the Bianchi identities, the commutation relations and the dimension dependent identities, we obtain 60394 TNSs for the basis in $V_5$. We denote each term in the basis as $V_5[n,j]$, where $n$ represents the value of LEG and $j$ runs from 1 to \#TNS for each $n$. The summary of the basis in $V_5$ is shown in the Table~\ref{tb:V5}.

\section{Variations under the Local Supersymmetry} \label{sec:SusyVariation}

In the Fig.~\ref{fig:O(R0)part}, the variations of the effective action and the cancellation mechanism under the local supersymmetry are shown. The sketch of the transformations for the bosonic terms is given by the eq.~(\ref{eq:deltaB1B2}), and that for the fermionic bilinears is written in the eq.~(\ref{eq:deltaF1F2F3}) by using the abbreviated notation. In this section, we calculate details of these variations for $B_1$, $B_2$, $F_1$, $F_2$ and $F_3$ in the effective action.

\subsection{$\delta_0 B_1=\delta_0 [e(D\hat{F})^4]_{56}$} \label{subsec:deltaB1}

Let us begin with the bosonic part of the effective action. The terms in the class of $B_1$ are expressed as $eD_e\hat{F}_{abcd}X^{eabcd}$, where $X^{eabcd}$ consists of $[(D\hat{F})^3]$. By using the eq.~(\ref{eq:susytr}) and the eq.~(\ref{eq:deltaDFhat}), the variation of $eD_e\hat{F}_{abcd}X^{eabcd}$ up to $\mathcal{O}(\psi^2)$ is given by
\begin{alignat}{3}
  &\delta_0 (eD_e\hat{F}_{abcd}X^{eabcd}) \notag
  \\
  &= eD_eF_{abcd}X^{eabcd}\bar{\epsilon}\gamma^f\psi_f 
  + e\delta_0 (D_e\hat{F}_{abcd})X^{eabcd} + eD_eF_{abcd} \delta_0  X^{eabcd} \notag
  \\
  &= eD_eF_{abcd}X^{eabcd}\bar{\epsilon}\gamma^f\psi_f + eD_eF_{abcd} \delta_0  X^{eabcd} 
  - eD_fF_{abcd}X^{eabcd}\bar{\epsilon}\gamma^f\psi_e \notag
  \\&\quad\,
  + eD_eX^{eabcd} \bar{\epsilon}(\gamma_{ab} \psi_{cd} + \gamma_{ac} \psi_{db} + \gamma_{ad} \psi_{bc} 
  + \gamma_{cd} \psi_{ab} + \gamma_{db} \psi_{ac} + \gamma_{bc} \psi_{ad}) \notag
  \\&\quad\,
  + eD_e(F^f{}_{bcd}X^{eabcd}) \bar{\epsilon} \gamma_{[a}\psi_{f]} 
  + eD_a(F^f{}_{bcd}X^{eabcd}) \bar{\epsilon} \gamma_{(f}\psi_{e)} 
  - eD_f(F^f{}_{bcd}X^{eabcd}) \bar{\epsilon} \gamma_{(a}\psi_{e)} \notag
  \\&\quad\,
  - eD_e(F^f{}_{cda}X^{eabcd}) \bar{\epsilon} \gamma_{[b}\psi_{f]} 
  - eD_b(F^f{}_{cda}X^{eabcd}) \bar{\epsilon} \gamma_{(f}\psi_{e)} 
  + eD_f(F^f{}_{cda}X^{eabcd}) \bar{\epsilon} \gamma_{(b}\psi_{e)} \notag
  \\&\quad\,
  + eD_e(F^f{}_{dab} X^{eabcd}) \bar{\epsilon} \gamma_{[c}\psi_{f]} 
  + eD_c(F^f{}_{dab} X^{eabcd}) \bar{\epsilon} \gamma_{(f}\psi_{e)} 
  - eD_f(F^f{}_{dab} X^{eabcd}) \bar{\epsilon} \gamma_{(c}\psi_{e)} \notag
  \\&\quad\,
  - eD_e(F^f{}_{abc} X^{eabcd}) \bar{\epsilon} \gamma_{[d}\psi_{f]}
  - eD_d(F^f{}_{abc} X^{eabcd}) \bar{\epsilon} \gamma_{(f}\psi_{e)}
  + eD_f(F^f{}_{abc} X^{eabcd}) \bar{\epsilon} \gamma_{(d}\psi_{e)} \notag
  \\
  &=  eD_eX^{eabcd} \bar{\epsilon}(\gamma_{ab} \psi_{cd} + \gamma_{ac} \psi_{db} + \gamma_{ad} \psi_{bc} 
  + \gamma_{cd} \psi_{ab} + \gamma_{db} \psi_{ac} + \gamma_{bc} \psi_{ad}) \notag
  \\&\quad\,
  + eD_eF_{abcd}X^{eabcd}\bar{\epsilon}\gamma^f\psi_f
  - eD_fF_{abcd}X^{eabcd}\bar{\epsilon}\gamma^f\psi_e \notag
  \\&\quad\,
  + eD_eF^f{}_{bcd}X^{eabcd} \bar{\epsilon} \gamma_{[a}\psi_{f]} 
  + eD_aF^f{}_{bcd}X^{eabcd} \bar{\epsilon} \gamma_{(f}\psi_{e)} 
  - eD_fF^f{}_{bcd}X^{eabcd} \bar{\epsilon} \gamma_{(a}\psi_{e)} \notag
  \\&\quad\,
  - eD_eF^f{}_{cda}X^{eabcd} \bar{\epsilon} \gamma_{[b}\psi_{f]} 
  - eD_bF^f{}_{cda}X^{eabcd} \bar{\epsilon} \gamma_{(f}\psi_{e)} 
  + eD_fF^f{}_{cda}X^{eabcd} \bar{\epsilon} \gamma_{(b}\psi_{e)} \notag
  \\&\quad\,
  + eD_eF^f{}_{dab} X^{eabcd} \bar{\epsilon} \gamma_{[c}\psi_{f]} 
  + eD_cF^f{}_{dab} X^{eabcd} \bar{\epsilon} \gamma_{(f}\psi_{e)} 
  - eD_fF^f{}_{dab} X^{eabcd} \bar{\epsilon} \gamma_{(c}\psi_{e)} \notag
  \\&\quad\,
  - eD_eF^f{}_{abc} X^{eabcd} \bar{\epsilon} \gamma_{[d}\psi_{f]}
  - eD_dF^f{}_{abc} X^{eabcd} \bar{\epsilon} \gamma_{(f}\psi_{e)}
  + eD_fF^f{}_{abc} X^{eabcd} \bar{\epsilon} \gamma_{(d}\psi_{e)} \label{eq:deltaB1}
  \\&\quad\,
  + eF^f{}_{bcd}D_eX^{eabcd} \bar{\epsilon} \gamma_{[a}\psi_{f]} 
  + eF^f{}_{bcd}D_aX^{eabcd} \bar{\epsilon} \gamma_{(f}\psi_{e)} 
  - eF^f{}_{bcd}D_fX^{eabcd} \bar{\epsilon} \gamma_{(a}\psi_{e)} \notag
  \\&\quad\,
  - eF^f{}_{cda}D_eX^{eabcd} \bar{\epsilon} \gamma_{[b}\psi_{f]} 
  - eF^f{}_{cda}D_bX^{eabcd} \bar{\epsilon} \gamma_{(f}\psi_{e)} 
  + eF^f{}_{cda}D_fX^{eabcd} \bar{\epsilon} \gamma_{(b}\psi_{e)} \notag
  \\&\quad\,
  + eF^f{}_{dab} D_eX^{eabcd} \bar{\epsilon} \gamma_{[c}\psi_{f]} 
  + eF^f{}_{dab} D_cX^{eabcd} \bar{\epsilon} \gamma_{(f}\psi_{e)} 
  - eF^f{}_{dab} D_fX^{eabcd} \bar{\epsilon} \gamma_{(c}\psi_{e)} \notag
  \\&\quad\,
  - eF^f{}_{abc} D_eX^{eabcd} \bar{\epsilon} \gamma_{[d}\psi_{f]}
  - eF^f{}_{abc} D_dX^{eabcd} \bar{\epsilon} \gamma_{(f}\psi_{e)}
  + eF^f{}_{abc} D_fX^{eabcd} \bar{\epsilon} \gamma_{(d}\psi_{e)} \notag
  \\&\quad\,
  + eD_eF_{abcd}\delta_0 X^{eabcd}. \notag
\end{alignat}
Note that $eD_eF_{abcd}\delta_0 X^{eabcd}$ contain similar terms, except for the term $(\delta_0 e)D_e F_{abcd}X^{eabcd}=eD_eF_{abcd}X^{eabcd}\bar{\epsilon}\gamma^f\psi_f$. The terms in the variation are expanded by the bases in $V_1$, $V_4$ and $V_5$. We will express these variations as follows.
\begin{alignat}{3}
  \delta_0 B_1[i] &=
  \sum_{n,j} M_{B_1[i] V_1[n,j]} V_1[n,j] + \sum_{n,j} M_{B_1[i] V_4[n,j]} V_4[n,j] 
  + \sum_{n,j} M_{B_1[i] V_5[n,j]} V_5[n,j] \notag
  \\
  &= 
  \begin{pmatrix}
    M_{B_1[i] V_1[n,j]} & 0 & M_{B_1[i] V_4[n,j]} & M_{B_1[i] V_5[n,j]}
  \end{pmatrix}
  \begin{pmatrix}
    V_1[n,j] \\ V_2[n,j] \\ V_4[n,j] \\ V_5[n,j]
  \end{pmatrix}. \label{eq:delB1}
\end{alignat}
Here $n$ runs possible values of LEG for each class, and $j=1,\cdots,\#\text{IND}$ for fixed $n$. The summations of $n$ and $j$ are understood in the second equality. We also abbreviate the above expression by using the matrix notation as
\begin{alignat}{3}
  \delta_0 B_1 &=
  \begin{pmatrix}
    M_{B_1V_1} & 0 & M_{B_1 V_4} & M_{B_1 V_5}
  \end{pmatrix}
  \begin{pmatrix}
    V_1 \\ V_2 \\ V_4 \\ V_5
  \end{pmatrix}. \label{eq:delB1_2}
\end{alignat}
The eq.~(\ref{eq:delB1}) corresponds to the $i$-th component of the eq.~(\ref{eq:delB1_2}).

\subsection{$\delta_0 B_2=\delta_0 [e\epsilon_{11}\hat{R}(D\hat{F})^3]_{22}$} \label{subsec:deltaB2}

The terms in the class of $B_2$ are expressed as $e\hat{R}_{abcd}X^{abcd}$, where $X^{abcd}$ consists of $[\epsilon_{11}(D\hat{F})^3]$. Since we examine the cancellation of the variations at $\mathcal{O}(R^0)$, only the variation of the Riemann tensor should be taken into account at this order. By using the transformation of the eq.~(\ref{eq:deltaRhat}), the variation of $e\hat{R}_{abcd}X^{abcd}$ up to $\mathcal{O}(\psi^2)$ at $\mathcal{O}(R^0)$ is given by
\begin{alignat}{3}
  &\delta_0 (e\hat{R}_{abcd}X^{abcd}) \notag
  \\
  &= eR_{abcd}X^{abcd} \bar{\epsilon}\gamma^f\psi_f + e\delta_0 \hat{R}_{abcd}X^{abcd} 
  + eR_{abcd}\delta_0 X^{abcd} \notag
  \\
  &= e D_{a} X^{abcd} (\tfrac{1}{4} \bar{\epsilon} \gamma_b \psi_{cd} 
  + \tfrac{1}{4} \bar{\epsilon} \gamma_c \psi_{bd} 
  + \tfrac{1}{4} \bar{\epsilon} \gamma_d \psi_{cb}
  + \tfrac{1}{12} F_{cdij} \bar{\epsilon} \gamma^{ij} \psi_{b}
  + \tfrac{1}{288} F^{ijkl} \bar{\epsilon} \gamma_{cdijkl} \psi_{b}) \notag
  \\
  &\quad
  + e D_{b} X^{abcd} (\tfrac{1}{4} \bar{\epsilon} \gamma_a \psi_{dc} 
  + \tfrac{1}{4} \bar{\epsilon} \gamma_c \psi_{da} 
  + \tfrac{1}{4} \bar{\epsilon} \gamma_d \psi_{ac}
  - \tfrac{1}{12} F_{cdij} \bar{\epsilon} \gamma^{ij} \psi_{a}
  - \tfrac{1}{288} F^{ijkl} \bar{\epsilon} \gamma_{cdijkl} \psi_{a}) \label{eq:deltaRX}
  \\
  &\quad
  + e D_{c} X^{abcd} (\tfrac{1}{4} \bar{\epsilon} \gamma_d \psi_{ab} 
  + \tfrac{1}{4} \bar{\epsilon} \gamma_a \psi_{db} 
  + \tfrac{1}{4} \bar{\epsilon} \gamma_b \psi_{ad}
  + \tfrac{1}{12} F_{abij} \bar{\epsilon} \gamma^{ij} \psi_{d}
  + \tfrac{1}{288} F^{ijkl} \bar{\epsilon} \gamma_{abijkl} \psi_{d})\notag
  \\
  &\quad
  + e D_{d} X^{abcd} (\tfrac{1}{4} \bar{\epsilon} \gamma_c \psi_{ba} 
  + \tfrac{1}{4} \bar{\epsilon} \gamma_a \psi_{bc} 
  + \tfrac{1}{4} \bar{\epsilon} \gamma_b \psi_{ca}
  - \tfrac{1}{12} F_{abij} \bar{\epsilon} \gamma^{ij} \psi_{c}
  - \tfrac{1}{288} F^{ijkl} \bar{\epsilon} \gamma_{abijkl} \psi_{c}) \notag
  \\
  &\quad
  - \tfrac{1}{2} e X^{abcd} (R_{ebcd} \bar{\epsilon} \gamma^e \psi_a 
  + R_{aecd} \bar{\epsilon} \gamma^e \psi_b 
  + R_{abed} \bar{\epsilon} \gamma^e \psi_c 
  + R_{abce} \bar{\epsilon} \gamma^e \psi_d) \notag
  \\
  &\quad
  + eR_{abcd}X^{abcd} \bar{\epsilon}\gamma^f\psi_f + eR_{abcd}\delta_0 X^{abcd}. \notag
\end{alignat}
The last two lines in the above are irrelevant in this paper. Note that $\delta_0 \epsilon^{11}_{a_1\cdots a_{11}} = 0$, $D_a\epsilon^{11}_{a_1\cdots a_{11}} = 0$ and $\bar{\epsilon}\epsilon^{11}_{a_1\cdots a_{11}}\psi_b = -\bar{\epsilon}\gamma_{a_1\cdots a_{11}}\psi_b$. By substituting $X^{abcd}=\epsilon^{11}_{g_1\cdots g_{11}}Y^{g_1\cdots g_{11}abcd}$ into the above, we obtain
\begin{alignat}{3}
  &\delta_0 (e\epsilon^{11}_{g_1\cdots g_{11}}\hat{R}_{abcd}Y^{g_1\cdots g_{11}abcd}) \notag
  \\
  &= - \tfrac{1}{4} e D_{a} Y^{g_1\cdots g_{11}abcd} 
  (\bar{\epsilon}\gamma_{g_1\cdots g_{11}}\gamma_b \psi_{cd} 
  + \bar{\epsilon}\gamma_{g_1\cdots g_{11}}\gamma_c \psi_{bd} 
  + \bar{\epsilon}\gamma_{g_1\cdots g_{11}}\gamma_d \psi_{cb}) \notag
  \\&\quad
  - \tfrac{1}{4} e D_{b}Y^{g_1\cdots g_{11}abcd} 
  (\bar{\epsilon}\gamma_{g_1\cdots g_{11}}\gamma_a \psi_{dc} 
  + \bar{\epsilon}\gamma_{g_1\cdots g_{11}}\gamma_c \psi_{da} 
  + \bar{\epsilon}\gamma_{g_1\cdots g_{11}}\gamma_d \psi_{ac}) \notag
  \\&\quad
  - \tfrac{1}{4} e D_{c} Y^{g_1\cdots g_{11}abcd} 
  (\bar{\epsilon}\gamma_{g_1\cdots g_{11}}\gamma_d \psi_{ab} 
  + \bar{\epsilon}\gamma_{g_1\cdots g_{11}}\gamma_a \psi_{db} 
  + \bar{\epsilon}\gamma_{g_1\cdots g_{11}}\gamma_b \psi_{ad}) \notag
  \\&\quad
  - \tfrac{1}{4} e D_{d} Y^{g_1\cdots g_{11}abcd} 
  (\bar{\epsilon}\gamma_{g_1\cdots g_{11}}\gamma_c \psi_{ba} 
  + \bar{\epsilon}\gamma_{g_1\cdots g_{11}}\gamma_a \psi_{bc} 
  + \bar{\epsilon}\gamma_{g_1\cdots g_{11}}\gamma_b \psi_{ca}) \label{eq:deltaB2}
  \\
  &\quad
  - \tfrac{1}{12} e F_{cdij} D_{a} Y^{g_1\cdots g_{11}abcd} 
  \bar{\epsilon}\gamma_{g_1\cdots g_{11}}\gamma^{ij} \psi_{b}
  - \tfrac{1}{288} e F^{ijkl} D_{a} Y^{g_1\cdots g_{11}abcd} 
  \bar{\epsilon}\gamma_{g_1\cdots g_{11}}\gamma_{cdijkl} \psi_{b} \notag
  \\&\quad
  + \tfrac{1}{12} e F_{cdij} D_{b}Y^{g_1\cdots g_{11}abcd} 
  \bar{\epsilon}\gamma_{g_1\cdots g_{11}}\gamma^{ij} \psi_{a}
  + \tfrac{1}{288} e F^{ijkl} D_{b}Y^{g_1\cdots g_{11}abcd} 
  \bar{\epsilon}\gamma_{g_1\cdots g_{11}}\gamma_{cdijkl} \psi_{a} \notag
  \\&\quad
  - \tfrac{1}{12} e F_{abij} D_{c} Y^{g_1\cdots g_{11}abcd} 
  \bar{\epsilon}\gamma_{g_1\cdots g_{11}}\gamma^{ij} \psi_{d}
  - \tfrac{1}{288} e F^{ijkl} D_{c} Y^{g_1\cdots g_{11}abcd} 
  \bar{\epsilon}\gamma_{g_1\cdots g_{11}}\gamma_{abijkl} \psi_{d} \notag
  \\&\quad
  + \tfrac{1}{12} e F_{abij} D_{d} Y^{g_1\cdots g_{11}abcd} 
  \bar{\epsilon}\gamma_{g_1\cdots g_{11}}\gamma^{ij} \psi_{c}
  + \tfrac{1}{288} e F^{ijkl} D_{d} Y^{g_1\cdots g_{11}abcd} 
  \bar{\epsilon}\gamma_{g_1\cdots g_{11}}\gamma_{abijkl} \psi_{c} \notag
  \\
  &\quad
  + \tfrac{1}{2} e Y^{g_1\cdots g_{11}abcd} (R_{ebcd} 
  \bar{\epsilon}\gamma_{g_1\cdots g_{11}}\gamma^e \psi_a 
  + R_{aecd} \bar{\epsilon}\gamma_{g_1\cdots g_{11}}\gamma^e \psi_b \notag
  \\&\qquad\qquad\qquad\qquad\quad
  + R_{abed} \bar{\epsilon}\gamma_{g_1\cdots g_{11}}\gamma^e \psi_c 
  + R_{abce} \bar{\epsilon}\gamma_{g_1\cdots g_{11}}\gamma^e \psi_d) \notag
  \\
  &\quad
  - e R_{abcd}Y^{g_1\cdots g_{11}abcd} \bar{\epsilon}\gamma_{g_1\cdots g_{11}}\gamma^f\psi_f 
  + e\epsilon^{11}_{g_1\cdots g_{11}}R_{abcd}\delta_0 Y^{g_1\cdots g_{11}abcd}. \notag
\end{alignat}
If we neglect the last three lines, which are linear with respect to the Riemann tensor, the terms in the variation are expanded by the bases in $V_1$ and $V_5$. We will express these variations as follows.
\begin{alignat}{3}
  \delta_0 B_2[i] &=
  \sum_{n,j} M_{B_2[i] V_1[n,j]} V_1[n,j] + \sum_{n,j} M_{B_2[i] V_5[n,j]} V_5[n,j] \notag
  \\
  &= 
  \begin{pmatrix}
    M_{B_2[i] V_1[n,j]} & 0 & 0 & M_{B_2[i] V_5[n,j]}
  \end{pmatrix}
  \begin{pmatrix}
    V_1[n,j] \\ V_2[n,j] \\ V_4[n,j] \\ V_5[n,j]
  \end{pmatrix}. \label{eq:delB2}
\end{alignat}
Here $n$ runs possible values of LEG for each class, and $j=1,\cdots,\#\text{IND}$ for fixed $n$. The summations of $n$ and $j$ are understood in the last line. We also abbreviate the above by using the matrix notation as
\begin{alignat}{3}
  \delta_0 B_2 &=
  \begin{pmatrix}
    M_{B_2 V_1} & 0 & 0 & M_{B_2 V_5}
  \end{pmatrix}
  \begin{pmatrix}
    V_1 \\ V_2 \\ V_4 \\ V_5
  \end{pmatrix}. \label{eq:delB2_2}
\end{alignat}
The eq.~(\ref{eq:delB2}) corresponds to the $i$-th component of the eq.~(\ref{eq:delB2_2}).

\subsection{$\delta_0 F_1=\delta_0 [e(DF)^2\overline{\psi_2}\gamma\mathcal{D}\psi_2]$} \label{subsec:deltaF1}

The terms in the class of $F_1$ are written as $eX\overline{\psi_{ab}}\gamma^{(n)}\mathcal{D}_e\psi_{cd}$, where $[X]$ consists of $[(DF)^2]$ and $[\gamma^{(n)}]$ is a gamma matrix with odd  rank of $n$. Of course, $X$ and $\gamma^{(n)}$ have indices, but we abbreviate them for simplicity. Since we examine the cancellation of the variations at $\mathcal{O}(\psi)$, only the variation of the gravitino should be taken into account at this order. By using the transformation of the eq.~(\ref{eq:comm}), the variation of $eX\overline{\psi_{ab}}\gamma^{(n)}\mathcal{D}_c\psi_{de}$ up to $\mathcal{O}(\psi^2)$ is given by
\begin{alignat}{3}
  &\delta_0 (eX\bar{\psi}_{ab}\gamma^{(n)}\mathcal{D}_e\psi_{cd}) \notag
  \\
  &= eX\overline{\delta_0 \psi_{ab}}\gamma^{(n)}\mathcal{D}_e\psi_{cd}
  + (-1)^{\frac{n(n+1)}{2}} e X
  \overline{\mathcal{D}_e\delta_0 \psi_{cd}}\gamma^{(n)}\psi_{ab} \notag
  \\
  &= eX\overline{\delta_0 \psi_{ab}}\gamma^{(n)}\mathcal{D}_e\psi_{cd}
  - (-1)^{\frac{n(n+1)}{2}} e X
  \overline{\delta_0 \psi_{cd}}\gamma^{(n)}\mathcal{D}_e\psi_{ab} \notag
  \\&\quad\,
  - (-1)^{\frac{n(n+1)}{2}} e D_eX \overline{\delta_0 \psi_{cd}}\gamma^{(n)}\psi_{ab}
  + (-1)^{\frac{n(n+1)}{2}} e X 
  \overline{\delta_0 \psi_{cd}}(\overline{F_e}\gamma^{(n)}+\gamma^{(n)}F_e)\psi_{ab} \notag
  \\
  &= - 2(-1)^{\frac{n(n+1)}{2}} e D_eX
  \bar{\epsilon}\overline{DF_{cd}}\gamma^{(n)}\psi_{ab} \notag
  \\&\quad\,
  + 2 eX \bar{\epsilon}\overline{DF_{ab}}\gamma^{(n)}\mathcal{D}_e\psi_{cd}
  - 2(-1)^{\frac{n(n+1)}{2}} eX 
  \bar{\epsilon}\overline{DF_{cd}}\gamma^{(n)}\mathcal{D}_e\psi_{ab} \notag
  \\&\quad\,
  + 2(-1)^{\frac{n(n+1)}{2}} eX \bar{\epsilon}
  \overline{DF_{cd}}(\overline{F_e}\gamma^{(n)}+\gamma^{(n)}F_e)\psi_{ab} \notag
  \\&\quad\,
  - 2(-1)^{\frac{n(n+1)}{2}} eD_eX 
  \bar{\epsilon}\overline{F^2_{cd}}\gamma^{(n)}\psi_{ab} \label{eq:deltaF1}
  \\&\quad\,
  + 2 eX \bar{\epsilon}\overline{F^2_{ab}}\gamma^{(n)}\mathcal{D}_e\psi_{cd}
  - 2(-1)^{\frac{n(n+1)}{2}} eX
  \bar{\epsilon}\overline{F^2_{cd}}\gamma^{(n)}\mathcal{D}_e\psi_{ab} \notag
  \\&\quad\,
  + 2(-1)^{\frac{n(n+1)}{2}} eX
  \bar{\epsilon}\overline{F^2_{cd}}(\overline{F_e}\gamma^{(n)}+\gamma^{(n)}F_e)\psi_{ab} \notag
  \\&\quad\,
  + \tfrac{1}{2}(-1)^{\frac{n(n+1)}{2}} e R_{fgcd} D_eX
  \bar{\epsilon}\gamma^{fg}\gamma^{(n)}\psi_{ab} \notag
  \\&\quad\,
  - \tfrac{1}{2} e R_{fgab} X \bar{\epsilon}\gamma^{fg}\gamma^{(n)}\mathcal{D}_e\psi_{cd}
  + \tfrac{1}{2} (-1)^{\frac{n(n+1)}{2}} e R_{fgcd} X
  \bar{\epsilon}\gamma^{fg}\gamma^{(n)}\mathcal{D}_e\psi_{ab} \notag
  \\&\quad\,
  - \tfrac{1}{2}(-1)^{\frac{n(n+1)}{2}} e R_{fgcd}X 
  \bar{\epsilon}\gamma^{fg}(\overline{F_e}\gamma^{(n)}+\gamma^{(n)}F_e)\psi_{ab}. \notag
\end{alignat}
If we neglect last seven lines, which do not appear in the Fig.~\ref{fig:O(R0)part}, the terms in the variation are classified by the bases in $V_1$ and $V_2$. We will express these variations as follows.
\begin{alignat}{3}
  \delta_0 F_1[m,i] &=
  \sum_{n,j} M_{F_1[m,i] V_1[n,j]} V_1[n,j] + \sum_{n,j} M_{F_1[m,i] V_2[n,j]} V_2[n,j] \notag
  \\
  &= 
  \begin{pmatrix}
    M_{F_1[m,i] V_1[n,j]} & M_{F_1[m,i] V_2[n,j]} & 0 & 0
  \end{pmatrix}
  \begin{pmatrix}
    V_1[n,j] \\ V_2[n,j] \\ V_4[n,j] \\ V_5[n,j]
  \end{pmatrix}. \label{eq:delF1}
\end{alignat}
Here $n$ runs possible values of LEG for each class, and $j=1,\cdots,\#\text{IND}$ for fixed $n$. The summations of $n$ and $j$ are understood in the last line. We also abbreviate the above by using the matrix notation as
\begin{alignat}{3}
  \delta_0 F_1 &=
  \begin{pmatrix}
    M_{F_1 V_1} & M_{F_1 V_2} & 0 & 0
  \end{pmatrix}
  \begin{pmatrix}
    V_1 \\ V_2 \\ V_4 \\ V_5
  \end{pmatrix}. \label{eq:delF1_2}
\end{alignat}
The eq.~(\ref{eq:delF1}) corresponds to the $(m,i)$-th component of the eq.~(\ref{eq:delF1_2}).

\subsection{$\delta_0 F_2=\delta_0 [e(DF)^3\bar{\psi}\gamma\psi_2]$} \label{subsec:deltaF2}

The terms in the class of $F_2$ are written by $eX\overline{\psi_c}\gamma^{(n)}\psi_{ab}$, where $[X]$ consists of $[(DF)^3]$ and $[\gamma^{(n)}]$ is a gamma matrix with even rank of $n$. Of course, $X$ and $\gamma^{(n)}$ have indices but we abbreviate them for simplicity. Since we examine the cancellation of the variations at $\mathcal{O}(\psi)$, only the variation of the gravitino should be taken into account at this order. By using the transformations of the eq.~(\ref{eq:susytr}) and the  eq.~(\ref{eq:comm}), the variation of $eX\overline{\psi_c}\gamma^{(n)}\psi_{ab}$ up to $\mathcal{O}(\psi^2)$ is given by
\begin{alignat}{3}
  &\delta_0 (eX\overline{\psi_c}\gamma^{(n)}\psi_{ab}) \notag
  \\
  &= 2eX\overline{\mathcal{D}_c \epsilon}\gamma^{(n)}\psi_{ab}
  + (-1)^{\frac{n(n+1)}{2}}eX\overline{\delta_0 \psi_{ab}}\gamma^{(n)}\psi_c \notag
  \\
  &= - 2eD_cX\bar{\epsilon}\gamma^{(n)}\psi_{ab} \notag
  \\&\quad\,
  - 2eX\bar{\epsilon}\gamma^{(n)}\mathcal{D}_c\psi_{ab} \notag
  \\&\quad\,
  + 2(-1)^{\frac{n(n+1)}{2}}eX\bar{\epsilon}\overline{DF_{ab}}\gamma^{(n)}\psi_c \label{eq:deltaF2}
  \\&\quad\,
  + 2eX\bar{\epsilon}(\overline{F_c}\gamma^{(n)}+\gamma^{(n)}F_c)\psi_{ab} \notag
  \\&\quad\,
  + 2(-1)^{\frac{n(n+1)}{2}}eX\bar{\epsilon}\overline{F^2_{ab}}\gamma^{(n)}\psi_c \notag
  \\&\quad\,
  - \tfrac{1}{2}(-1)^{\frac{n(n+1)}{2}}eR_{deab}X\bar{\epsilon}\gamma^{de}\gamma^{(n)}\psi_c. \notag
\end{alignat}
If we neglect last three lines, which do not appear in the Fig.~\ref{fig:O(R0)part}, the terms in the variation are expanded by the bases in $V_1$, $V_2$ and $V_4$. We will express these variations as follows.
\begin{alignat}{3}
  \delta_0 F_2[m,i] &=
  \sum_{n,j} M_{F_2[m,i] V_1[n,j]} V_1[n,j] + \sum_{n,j} M_{F_2[m,i] V_2[n,j]} V_2[n,j] 
  + \sum_{n,j} M_{F_2[m,i] V_4[n,j]} V_4[n,j] \notag
  \\
  &= 
  \begin{pmatrix}
    M_{F_2[m,i] V_1[n,j]} & M_{F_2[m,i] V_2[n,j]} & M_{F_2[m,i] V_4[n,j]} & 0 
  \end{pmatrix}
  \begin{pmatrix}
    V_1[n,j] \\ V_2[n,j] \\ V_4[n,j] \\ V_5[n,j]
  \end{pmatrix}. \label{eq:delF2}
\end{alignat}
Here $n$ runs possible values of LEG for each class, and $j=1,\cdots,\#\text{IND}$ for fixed $n$. The summations of $n$ and $j$ are understood in the last line. We also abbreviate the above by using the matrix notation as
\begin{alignat}{3}
  \delta_0 F_2 &=
  \begin{pmatrix}
    M_{F_2 V_1} & M_{F_2 V_2} & M_{F_2 V_4} & 0
  \end{pmatrix}
  \begin{pmatrix}
    V_1 \\ V_2 \\ V_4 \\ V_5
  \end{pmatrix}. \label{eq:delF2_2}
\end{alignat}
The eq.~(\ref{eq:delF2}) corresponds to the $(m,i)$-th component of the eq.~(\ref{eq:delF2_2}).

\subsection{$\delta_0 F_3=\delta_0 [eF(DF)^3 \bar{\psi} \gamma \psi]$} \label{subsec:deltaF3}

The terms in the class of $F_3$ are expressed as $eX\overline{\psi_a}\gamma^{(n)}\psi_b$, where $[X]$ consists of $[F(DF)^3]$ and $[\gamma^{(n)}]$ is a gamma matrix with rank of $n=1,5,9$. Of course, $X$ and $\gamma^{(n)}$ have indices but we abbreviate them for simplicity. Since we examine the cancellation of the variations at $\mathcal{O}(\psi)$, only the variation of the gravitino should be taken into account at this order. By using the transformation of the eq.~(\ref{eq:susytr}), the variation of $eX\overline{\psi_a}\gamma^{(n)}\psi_b$ up to $\mathcal{O}(\psi^2)$ is given by
\begin{alignat}{3}
  &\delta_0 (eX\overline{\psi_a}\gamma^{(n)}\psi_b) \notag
  \\
  &= 2eX\overline{\mathcal{D}_a \epsilon}\gamma^{(n)}\psi_b
  - 2eX\overline{\mathcal{D}_b \epsilon}\gamma^{(n)}\psi_a \notag
  \\
  &= - 2eD_aX\bar{\epsilon}\gamma^{(n)}\psi_b 
  + 2eD_bX\bar{\epsilon}\gamma^{(n)}\psi_a \label{eq:deltaF3}
  \\&\quad\,
  - 2eX\bar{\epsilon}\gamma^{(n)}\psi_{ab} \notag
  \\&\quad\,
  + 2eX\bar{\epsilon}(\overline{F_a}\gamma^{(n)}+\gamma^{(n)}F_a)\psi_b 
  - 2eX\bar{\epsilon}(\overline{F_b}\gamma^{(n)}+\gamma^{(n)}F_b)\psi_a. \notag
\end{alignat}
If we neglect last two lines, which do not appear in the Fig.~\ref{fig:O(R0)part}, the terms in the variation are expanded by the bases in $V_4$ and $V_5$. We will express these variations as follows.
\begin{alignat}{3}
  \delta_0 F_3[m,i] &=
  \sum_{n,j} M_{F_3[m,i] V_4[n,j]} V_4[n,j] + \sum_{n,j} M_{F_3[m,i] V_5[n,j]} V_5[n,j] \notag
  \\
  &= 
  \begin{pmatrix}
    0 & 0 & M_{F_3[m,i] V_4[n,j]} & M_{F_3[m,i] V_5[n,j]}
  \end{pmatrix}
  \begin{pmatrix}
    V_1[n,j] \\ V_2[n,j] \\ V_4[n,j] \\ V_5[n,j]
  \end{pmatrix}. \label{eq:delF3}
\end{alignat}
Here $n$ runs possible values of LEG for each class, and $j=1,\cdots,\#\text{IND}$ for fixed $n$. The summations of $n$ and $j$ are understood in the last line. We also abbreviate the above by using the matrix notation as
\begin{alignat}{3}
  \delta_0 F_3 &=
  \begin{pmatrix}
    0 & 0 & M_{F_3 V_4} & M_{F_3 V_5}
  \end{pmatrix}
  \begin{pmatrix}
    V_1 \\ V_2 \\ V_4 \\ V_5
  \end{pmatrix}. \label{eq:delF3_2}
\end{alignat}
The eq.~(\ref{eq:delF3}) corresponds to the $(m,i)$-th component of the eq.~(\ref{eq:delF3_2}).

\section{Cancellation of Variations under the Local Supersymmetry} \label{sec:Cancel}

Finally, we figure out the structure of the effective action within the Fig.~\ref{fig:O(R0)part} in this subsection. First we write down a generic form of the effective action which is a linear combination of possible terms with arbitrary coefficients. Then we consider the variations of the generic action and examine the cancellation to determine the coefficients in the effective action. While examining the cancellation, we should take into account the Bianchi identities, the commutation relations and the dimension dependent identities for the variations. It is also important to consider the field redefinition ambiguities and corrections to the local supersymmetry transformations.

\subsection{Variation of the Generic Action} \label{subsec:sumv}

Let us investigate the variations of the generic action under the local supersymmetry. In the section~\ref{sec:ActionVariation}, we have classified terms in the action and the variations. The bosonic terms in the generic action are expanded by $B_1[i]$ and $B_2[i]$, where $i=1,\cdots,\#\text{TNS}$ for each class. The gravitino bilinears are expanded by $F_1[m,i]$, $F_2[m,i]$, $F_3[m,i]$, where $m$ takes possible values in LEG for each class, and $i=1,\cdots,\#\text{TNS}$ for each $m$. The numbers of $\text{TNS}$s are summarized in the Tables~\ref{tb:B1}-\ref{tb:F3}. The generic Lagrangian density within the Fig.~\ref{fig:O(R0)part} are expressed by a linear combination of all possible terms, and explicitly written as
\begin{alignat}{3}
  \mathcal{L}_1 &= \! \sum_{i} \! b_1[i] B_1[i] \!+\! \sum_{i} \! b_2[i] B_2[i] 
  \!+\! \sum_{m,i} \! f_1[m,\! i] F_1[m,\! i] \!+\! \sum_{m,i} \! f_2[m,\! i] F_2[m,\! i]
  \!+\! \sum_{m,i} \! f_3[m,\! i] F_3[m,\! i] \notag
  \\
  &= \begin{pmatrix}
    b_1[i] \!&\! b_2[i] \!&\! f_1[m,i] \!&\! f_2[m,i] \!&\! f_3[m,i]
  \end{pmatrix}
  \begin{pmatrix}
    B_1[i] \\ B_2[i] \\ F_1[m,i] \\ F_2[m,i] \\ F_3[m,i]
  \end{pmatrix}. \label{eq:Lag}
\end{alignat}
The summations of $m$ and $i$ are understood in the last line. The coefficients of $b_1[i]$, $b_2[i], f_1[m,i], f_2[m,i]$ and $f_3[m,i]$ will be determined by imposing the local supersymmetry, and the number of the coefficients is $56+22+1298+4933+14337=20646$ in total. For simplicity, we abbreviate the eq.~(\ref{eq:Lag}) by using the matrix notation as
\begin{alignat}{3}
  \mathcal{L}_1 
  &= \begin{pmatrix}
    b_1 \!&\! b_2 \!&\! f_1 \!&\! f_2 \!&\! f_3
  \end{pmatrix}
  \begin{pmatrix}
    B_1 \\ B_2 \\ F_1 \\ F_2 \\ F_3
  \end{pmatrix}. \label{eq:Lag2}
\end{alignat}

Now we consider the local supersymmetry transformation of the above generic Lagrangian density. The transformations of the terms in $B_1$, $B_2$, $F_1$, $F_2$ and $F_3$ are evaluated by the eqs. (\ref{eq:delB1_2}), (\ref{eq:delB2_2}), (\ref{eq:delF1_2}), (\ref{eq:delF2_2}) and (\ref{eq:delF3_2}), respectively, so the variation of the Lagrangian density (\ref{eq:Lag2}) is expanded by the bases of $V_{1}[n,j]$, $V_{2}[n,j]$, $V_{4}[n,j]$ and $V_{5}[n,j]$ as 
\begin{alignat}{3}
  \delta_0 \mathcal{L}_1 
  &= \begin{pmatrix}
    b_1 \!&\! b_2 \!&\! f_1 \!&\! f_2 \!&\! f_3
  \end{pmatrix}
  \begin{pmatrix}
    \delta_0 B_1 \\ \delta_0 B_2 \\ \delta_0 F_1 \\ \delta_0 F_2 \\ \delta_0 F_3
  \end{pmatrix} \notag
  \\
  &=
  \begin{pmatrix}
    b_1 \!&\! b_2 \!&\! f_1 \!&\! f_2 \!&\! f_3
  \end{pmatrix} 
  \begin{pmatrix}
    M_{B_1 V_1} & 0 & M_{B_1 V_4} & M_{B_1 V_5} \\
    M_{B_2 V_1} & 0 & 0 & M_{B_2 V_5} \\
    M_{F_1 V_1} & M_{F_1 V_2} & 0 & 0 \\
    M_{F_2 V_1} & M_{F_2 V_2} & M_{F_2 V_4} & 0 \\
    0 & 0 & M_{F_3 V_4} & M_{F_3 V_5}
  \end{pmatrix} 
  \begin{pmatrix}
    V_1 \\ V_2 \\ V_4 \\ V_5 
  \end{pmatrix}. \label{eq:Mmatrix}
\end{alignat}
The number of variations, which is the sum of $\#\text{IND}$s in the Tables \ref{tb:V1}-\ref{tb:V5}, is $38790+12916+38129+458227=548062$ in total. We call $5 \times 4$ block matrix $M$, which is $20646\times 548062$ matrix.

\subsection{Solve the Bianchi identities, the commutation relations and the dimension dependent identities} \label{subsec:BiaDim}

Here we reduce the number of independent terms in the variations by imposing the Bianchi identities (\ref{eq:BianchiDP}), the commutation relations of two derivatives (\ref{eq:comD}) and the dimension dependent identities. 

Note that the Bianchi identities which arise from $D_{[e}F_{abcd]}=0$ and the dimension dependent identities do not mix terms among $V_1$, $V_2$, $V_4$ and $V_5$. On the other hand, the Bianchi identities which come from $\mathcal{D}_{[a} \psi_{bc]} = \tfrac{1}{4} R_{ij[ab} \gamma^{ij} \psi_{c]} + DF_{[ab} \psi_{c]} + F^2_{[ab} \psi_{c]}$ generate identities which mix terms in $V_2$ and $V_4$ like
\begin{alignat}{3}
  e X\bar{\epsilon}\gamma^{(n)}\mathcal{D}_{c}\psi_{ab}
  &= - e X\bar{\epsilon}\gamma^{(n)}\mathcal{D}_{a}\psi_{bc}
  - e X\bar{\epsilon}\gamma^{(n)}\mathcal{D}_{b}\psi_{ca} \label{eq:V2V4}
  \\&\quad\,
  + 3 eX\bar{\epsilon}\gamma^{(n)}DF_{[ab}\psi_{c]}
  +3  eX\bar{\epsilon}\gamma^{(n)}F^2_{[ab}\psi_{c]}
  + \tfrac{3}{4} eR_{ij[ab}X\bar{\epsilon}\gamma^{(n)}\gamma^{ij}\psi_{c]}. \notag
\end{alignat}
Here the indices of $X=(DF)^3$ and ones of the gamma matrix $\gamma^{(n)}$ are abbreviated. If we neglect the last two terms, which are irrelevant for the cancellation within the Fig.~\ref{fig:O(R0)part}, the above equation implies some terms in $V_{2}$ are related to some terms in $V_{4}$. 

Similarly, the commutation relation of two derivatives in the eq.~(\ref{eq:comD}) relates some terms in $V_1$ or $V_5$ to those which are not listed in the Fig.~\ref{fig:O(R0)part}. For example, terms in $V_1$ are written as
\begin{alignat}{3}
  e X D_a D_b F_{cdef} \bar{\epsilon} \gamma^{(n)} \psi_2
  &= e X D_b D_a F_{cdef} \bar{\epsilon} \gamma^{(n)} \psi_2  \label{eq:V1V1}
  \\&\quad\,
  + e R_c{}^g{}_{ab} F_{gdef} X \bar{\epsilon} \gamma^{(n)} \psi_2
  + e R_d{}^g{}_{ab} F_{cgef} X \bar{\epsilon} \gamma^{(n)} \psi_2 \notag
  \\&\quad\,
  + e R_e{}^g{}_{ab} F_{cdgf} X \bar{\epsilon} \gamma^{(n)} \psi_2
  + e R_f{}^g{}_{ab} F_{cdeg} X \bar{\epsilon} \gamma^{(n)} \psi_2, \notag
\end{alignat}
where the indices of $X=(DF)^2$ and ones of the gamma matrix $\gamma^{(n)}$ are abbreviated. As a result, the variations satisfy following relations within the Fig.~\ref{fig:O(R0)part}.
\begin{alignat}{3}
  0 &=
  \begin{pmatrix}
    B_{V_1 V_1} & 0 & 0 & 0 \\
    0 & B_{V_2 V_2}  & B_{V_2 V_4} & 0 \\
    0 & 0 & B_{V_4 V_4} & 0 \\
    0 & 0 & 0 & B_{V_5 V_5}
  \end{pmatrix} 
  \begin{pmatrix}
    V_1 \\ V_2 \\ V_4 \\ V_5 
  \end{pmatrix}. \label{eq:Bmatrix}
\end{alignat}
The number of rows of the matrix, which is given by the sum of $\#\text{Bianchi id. (etc.)}$ in the Tables \ref{tb:V1}-\ref{tb:V5}, is $33325+9802+26989+392822=462938$ in total.

Next we consider the dimension dependent identities. These only relate terms within each variation and are expressed as
\begin{alignat}{3}
  0 &=
  \begin{pmatrix}
    D_{V_1 V_1} & 0 & 0 & 0 \\
    0 & D_{V_2 V_2}  & 0 & 0 \\
    0 & 0 & D_{V_4 V_4} & 0 \\
    0 & 0 & 0 & D_{V_5 V_5}
  \end{pmatrix} 
  \begin{pmatrix}
    V_1 \\ V_2 \\ V_4 \\ V_5 
  \end{pmatrix}. \label{eq:Dmatrix}
\end{alignat}
The number of rows of the matrix, which is given by the sum of $\#\text{Dim. dep. id.}$ in the Tables \ref{tb:V1}-\ref{tb:V5},  is $2063+677+5703+69293=77736$ in total. 

By solving the Bianchi identities, the commutation relations and the dimension dependent identities, we obtain independent terms of $\#\text{TNS}_{V_1}=5392$ for $V_1$, $\#\text{TNS}_{V_2}=3067$ for $V_2$, $\#\text{TNS}_{V_4}=10285$ for $V_4$ and $\#\text{TNS}_{V_5}=60394$ for $V_5$. We express the reduction of the variations by using a projection matrix $P$ as
\begin{alignat}{3}
  \begin{pmatrix}
    V_1 \\ V_2 \\ V_4 \\ V_5 
  \end{pmatrix} &= P
  \begin{pmatrix}
    V_1 \\ V_2 \\ V_4 \\ V_5 
  \end{pmatrix}
  =
  \begin{pmatrix}
    P_{V_1 V_1} & 0 & 0 & 0 \\
    0 & P_{V_2 V_2}  & P_{V_2 V_4} & 0 \\
    0 & 0 & P_{V_4 V_4} & 0 \\
    0 & 0 & 0 & P_{V_5 V_5}
  \end{pmatrix} 
  \begin{pmatrix}
    V_1 \\ V_2 \\ V_4 \\ V_5 
  \end{pmatrix}. \label{eq:Pmatrix}
\end{alignat}
Since $P$ is the projection matrix, we require
\begin{alignat}{3}
  &P_{V_v V_v}^2 = P_{V_v V_v} =
  \begin{pmatrix}
    0 & A_v \\
    0 & {\bf 1}_{\#\text{TNS}_{V_v}}
  \end{pmatrix}, \quad
  (v=1,2,4,5), \notag
  \\
  &P_{V_2 V_2} P_{V_2 V_4} + P_{V_2 V_4} P_{V_4 V_4} = P_{V_2 V_4} =
  \begin{pmatrix}
    0 & B \\
    0 & 0
  \end{pmatrix}, \label{eq:proj}
\end{alignat}
The matrix expressions in the above are true when the independent terms of each $V_v$ are moved to lower rows and dependent ones to upper rows. $A_v$ is $(\#\text{IND}_{V_v}-\#\text{TNS}_{V_v}) \times \#\text{TNS}_{V_v}$ matrix and $B$ is $(\#\text{IND}_{V_2}-\#\text{TNS}_{V_2}) \times \#\text{TNS}_{V_4}$ matrix.

\subsection{The field redefinitions} \label{subsec:fieldredef}

As mentioned in the section \ref{sec:ActionVariation}, we reduce the number of terms in the Lagrangian density $\mathcal{L}_1$ by using the field redefinition ambiguities. In order to investigate these ambiguities, it should be noted that  $F_{ijkl}F^{ijkl}$, $F_{aijk} F_{bijk}$, $\gamma^{ab} \psi_{ab}$, $\gamma^{b} \psi_{ab}$ and $D_d F^{dabc}$ are written in terms of $R_{ab}$, $R$, $\epsilon_{11}^{abcijklmnop} F_{ijkl} F_{mnop}$ with the equations of motion (\ref{eq:EoM}) for the eleven dimensional supergravity  as
\begin{alignat}{3}
  F_{ijkl}F^{ijkl} &= 144 R + 16 E(e)^a{}_a, \notag
  \\
  F_{aijk} F_b{}^{ijk} &= 12 R_{ab} + 12 \eta_{ab} R - 6 E(e)_{ab} 
  + 2 \eta_{ab} E(e)^i{}_i, \notag
  \\
  \gamma^{ab} \psi_{ab} &= - \tfrac{1}{9} \gamma_a E(\psi)^a, \label{eq:EOM2}
  \\
  \gamma^{b} \psi_{ab} &= \tfrac{4}{9} E(\psi)_a - \tfrac{1}{18} \gamma_{ab} E(\psi)^b, \notag
  \\
  D_d F^{dabc} &= \tfrac{1}{8 \cdot 144} \epsilon_{11}^{abcijklmnop} F_{ijkl} F_{mnop} 
  + 6 E(A)^{abc}. \notag
\end{alignat}
Then terms in the effective action which contain the left hand sides of the above are removed by using the field redefinitions. Details are explained below.

First, let us redefine the vielbein as
\begin{alignat}{3}
  e^\mu{}_a \;\longrightarrow\; e^\mu{}_a - 16 e^\mu{}_a X^{(e)} 
  + 6 X^{(e) \mu}{}_a - 2 e^\mu{}_a X^{(e) i}{}_i, \label{eq:redefvl}
\end{alignat}
where $X^{(e)}$ and $X(e)_{ab}$ represent $\ell_p^6$ times mass dimension six terms. Then the Lagrangian density $\mathcal{L} = \mathcal{L}_0 + \ell_p^6 \mathcal{L}_1$ up to the next leading order changes as
\begin{alignat}{3}
  \mathcal{L} \;\longrightarrow\; 
  \mathcal{L} &= \mathcal{L}_0 + \ell_p^6 \mathcal{L}_1 - 16 e E(e)^a{}_a X^{(e)}
  - e \big( - 6 E(e)^a{}_\mu + 2 e^a{}_\mu E(e)^i{}_i \big) X^{(e) \mu}{}_a, \label{eq:redefLvl}
  \\
  &= \mathcal{L}_0 + \ell_p^6 \mathcal{L}_1 - e \big( F_{ijkl}F^{ijkl} \!-\! 144 R \big) X^{(e)} 
  - e \big( F_{aijk} F_b{}^{ijk} \!-\! 12 R_{ab} \!-\! 12 \eta_{ab} R \big) X^{(e) ab}. \notag
\end{alignat}
Thus it is possible to remove $e F_{ijkl}F^{ijkl} X^{(e)}$ and $e F_{aijk}F_b{}^{ijk} X^{(e) ab}$ in the $\mathcal{L}_1$ part. Also the field redefinition (\ref{eq:redefvl}) generates terms like $[eRX^{(e)}]$, but these are automatically taken into account if we consider generic ansatz of the effective action at $\mathcal{O}(R)$.

Second, let us redefine the 3-form field as
\begin{alignat}{3}
  A_{\mu\nu\rho} \;\longrightarrow\; A_{\mu\nu\rho} - 6 X^{(A)}_{\mu\nu\rho}, \label{eq:redefA}
\end{alignat}
where $X^{(A)}_{\mu\nu\rho}$ represents $\ell_p^6$ times mass dimension six terms. Then the Lagrangian density $\mathcal{L} = \mathcal{L}_0 + \ell_p^6 \mathcal{L}_1$ up to the next leading order changes as
\begin{alignat}{3}
  \mathcal{L} \;\longrightarrow\; 
  \mathcal{L} &= \mathcal{L}_0 + \ell_p^6 \mathcal{L}_1 
  - 6 e E(A)^{\mu\nu\rho} X^{(A)}_{\mu\nu\rho}, \notag
  \\
  &= \mathcal{L}_0 + \ell_p^6 \mathcal{L}_1 - e (D_d F^{dabc} - \tfrac{1}{8 \cdot 144} \epsilon_{11}^{abcijklmnop} F_{ijkl} F_{mnop})
  X^{(A)}_{abc}. \label{eq:redefLA}
\end{alignat}
Thus it is possible to remove $e D_d F^{dabc} X^{(A)}_{abc}$ in the $\mathcal{L}_1$ part. The field redefinition (\ref{eq:redefLA}) generates terms like $[\epsilon_{11} F^2X^{(A)}]$, but these are automatically taken into account if we consider generic ansatz of the effective action in the Fig.~\ref{fig:O(R0)}.

Third, let us redefine the Majorana gravitino as
\begin{alignat}{3}
  \psi_\mu \;\longrightarrow\; \psi_\mu - \tfrac{1}{9} \gamma_\mu X^{(\psi)} 
  - \big( \tfrac{4}{9} \delta^\nu_\mu - \tfrac{1}{18} \gamma_\mu{}^\nu \big) X^{(\psi)}_\nu, \label{eq:redefpsi}
\end{alignat}
where $X^{(\psi)}$ and $X^{(\psi)}_\nu$ represent $\ell_p^6$ times mass dimension 6.5 terms. Then the Lagrangian density up to the next leading order changes as
\begin{alignat}{3}
  \mathcal{L} \;\longrightarrow\; 
  \mathcal{L} &= \mathcal{L}_0 + \ell_p^6 \mathcal{L}_1 
  + \tfrac{1}{9} e \overline{X^{(\psi)}} \gamma_\mu E(\psi)^{\mu}
  - e \overline{X^{(\psi)}_\mu} \big( \tfrac{4}{9} E(\psi)^{\mu} 
  - \tfrac{1}{18} \gamma^{\mu\nu} E(\psi)_\nu \big), \notag
  \\
  &= \mathcal{L}_0 + \ell_p^6 \mathcal{L}_1
   - e \overline{X^{(\psi)}} \gamma^{ab} \psi_{ab}
   - e \overline{X^{(\psi)}_a} \gamma_b \psi^{ab}. \label{eq:redefLpsi}
\end{alignat}
Thus we can add linear combinations of $e \overline{X^{(\psi)}} \gamma^{ab} \psi_{ab}$ and 
$e \overline{X^{(\psi)}_a} \gamma_b \psi^{ab}$ to $\mathcal{L}_1$. In general, $X^{(\psi)}$ and $X^{(\psi)}_a$ contain gamma matrices, and it is necessary to use relations of
\begin{alignat}{3}
  &\gamma^{a_1\cdots a_{n-1}} \gamma^b \psi_{ab}
  = \gamma^{a_1\cdots a_{n-1} b} \psi_{ab}
  - \sum_{i=1}^{n-1} (-1)^{n-i} \gamma^{a_1 \cdots \check{a}_i \cdots a_{n-1}} \psi_a{}^{a_i}, \label{eq:gp1}
  \\
  &\gamma^{a_1\cdots a_{n-2}} \gamma^{ab} \psi_{ab}
  - 2 \sum_{i=1}^{n-2} (-1)^{n-i} \gamma^{a_1 \cdots \check{a}_i \cdots a_{n-2}} 
  \gamma^a \psi^{a_i}{}_a \notag
  \\[-0.2cm]
  &= \gamma^{a_1\cdots a_{n-2}ab} \psi_{ab} - 2 \!\!\sum_{i,j=1(i< j)}^{n-2} (-1)^{i+j} 
  \gamma^{a_1 \cdots \check{a}_i \cdots \check{a}_j \cdots a_{n-2}} \psi^{a_i a_j}, \label{eq:gp2}
\end{alignat}
where a symbol $\check{a}_i$ represents that $a_i$ is missing. Now we set 
\begin{alignat}{3}
  \overline{X^{(\psi)}} &= \overline{X^{(\psi)}_{a_1\cdots a_{n-2}}} 
  \gamma^{a_1\cdots a_{n-2}}, \notag
  \\
  \overline{X^{(\psi)}_a} &= -2 \overline{X^{(\psi)}_{a_1\cdots a_{n-2}}} 
  \sum_{i=1}^{n-2} (-1)^{n-i} \gamma^{a_1 \cdots \check{a}_i \cdots a_{n-2}} \delta^{a_i}_a
  + \overline{X^{(\psi)}{}_{a a_1\cdots a_{n-1}}} \gamma^{a_1\cdots a_{n-1}}. \label{eq:setX}
\end{alignat}
Then the Lagrangian density (\ref{eq:redefLpsi}) is evaluated as
\begin{alignat}{3}
  \mathcal{L} 
  &= \mathcal{L}_0 + \ell_p^6 \mathcal{L}_1 
   - e \overline{X^{(\psi) a}{}_{a_1\cdots a_{n-1}}} 
   \Big\{ \gamma^{a_1\cdots a_{n-1} b} \psi_{ab}
  - \sum_{i=1}^{n-1} (-1)^{n-i}  
  \gamma^{a_1 \cdots \check{a}_i \cdots a_{n-1}} \psi_a{}^{a_i} \Big\} \notag
  \\[-0.2cm]
  &\qquad\qquad\;\;\;
  - e \overline{X^{(\psi)}_{a_1\cdots a_{n-2}}} \Big\{ \gamma^{a_1\cdots a_{n-2}ab} \psi_{ab} 
  - 2 \!\!\sum_{i,j=1(i< j)}^{n-2} (-1)^{i+j}
  \gamma^{a_1 \cdots \check{a}_i \cdots \check{a}_j \cdots a_{n-2}} \psi^{a_i a_j} \Big\}. \label{eq:redefLpsi2}
\end{alignat}
This means that terms like $e \overline{X^{(\psi) a}{}_{a_1\cdots a_{n-1}}} \gamma^{a_1\cdots a_{n-1} b} \psi_{ab}$ and $e \overline{X^{(\psi)}_{a_1\cdots a_{n-2}}} \gamma^{a_1\cdots a_{n-2}ab} \psi_{ab}$ can be removed out of the $\mathcal{L}_1$ part.

Finally, let us consider the field redefinitions of the Majorana gravitino, which generate left hand sides of the eqs.~(\ref{eq:Dpsi1_1_1})-(\ref{eq:Dpsi7_7_7}) in the appendix \ref{app:appHDT}. By using these equations, it is possible to add right hand sides of the eqs.~(\ref{eq:Dpsi1_1_1})-(\ref{eq:Dpsi7_7_7}), and remove 
\begin{alignat}{3}
  &e\overline{X^{(D\psi) ab}_{a_1\cdots a_{n-1}}} 
  \gamma^{a_1\cdots a_{n-1}c} \mathcal{D}_a \psi_{bc},
  &\quad
  &e\overline{X^{(D\psi) ab}_{a_1\cdots a_{n-1}}} 
  \gamma^{a_1\cdots a_{n-1}c} \mathcal{D}_c \psi_{ab},
  &\quad
  &e\overline{X^{(D\psi) a}_{a_1\cdots a_n}} 
  \gamma^{a_1\cdots a_n} \mathcal{D}^b \psi_{ab}, \notag
  \\
  &e\overline{X^{(D\psi) a}_{a_1\cdots a_{n-2}}} 
  \gamma^{a_1\cdots a_{n-2}bc} \mathcal{D}_a \psi_{bc},
  &\quad
  &e\overline{X^{(D\psi) a}_{a_1\cdots a_{n-2}}} 
  \gamma^{a_1\cdots a_{n-2}bc} \mathcal{D}_b \psi_{ca},
  &\quad
  &e\overline{X^{(D\psi)}_{a_1\cdots a_{n-1}}} 
  \gamma^{a_1\cdots a_{n-1}a} \mathcal{D}^b \psi_{ab}, \notag
  \\
  &e\overline{X^{(D\psi)}_{a_1\cdots a_{n-3}}} 
  \gamma^{a_1\cdots a_{n-3}abc} \mathcal{D}_a \psi_{bc}, \label{eq:removeDpsi}
\end{alignat} 
out of the $\mathcal{L}_1$ part. Here $[\overline{X^{(D\psi)}}]$s represent  $\ell_p^6$ times mass dimension 6.5 terms.

In summary, as far as the Fig.~\ref{fig:O(R0)part} is concerned, it is possible to remove following terms out of the effective action.
\begin{alignat}{3}
  B_1 &: e D_d\hat{F}^{dabc} (D\hat{F})^3, \notag
  \\
  B_2 &: e \epsilon_{11}\hat{R}D_d\hat{F}^{dabc} (D\hat{F})^2, \notag
  \\
  F_1 &: e D_d F^{dabc} DF \overline{\psi_2}\gamma\mathcal{D}\psi_2, &\;
  &e(DF)^2\overline{\psi_{ab}}\gamma^{a_1\cdots a_{n-1}b}\mathcal{D}\psi_2, &\;
  &e(DF)^2\overline{\psi_{ab}}\gamma^{a_1\cdots a_{n-2}ab}\mathcal{D}\psi_2, \notag
  \notag
  \\
  &\;\;\;
  e(DF)^2\overline{\psi_2} \gamma^{a_1\cdots a_{n-1}c} \mathcal{D}_a \psi_{bc}, &\;
  &e(DF)^2\overline{\psi_2} \gamma^{a_1\cdots a_{n-1}c} \mathcal{D}_c \psi_{ab}, &\;
  &e(DF)^2\overline{\psi_2} \gamma^{a_1\cdots a_n} \mathcal{D}^b \psi_{ab}, \notag
  \\
  &\;\;\;
  e(DF)^2\overline{\psi_2} \gamma^{a_1\cdots a_{n-2}bc} \mathcal{D}_a \psi_{bc}, &\;
  &e(DF)^2\overline{\psi_2} \gamma^{a_1\cdots a_{n-2}bc} \mathcal{D}_b \psi_{ca}, &\;
  &e(DF)^2\overline{\psi_2} \gamma^{a_1\cdots a_{n-1}a} \mathcal{D}^b \psi_{ab}, \notag
  \\
  &\;\;\;
  e(DF)^2\overline{\psi_2} \gamma^{a_1\cdots a_{n-3}abc} \mathcal{D}_a \psi_{bc}, \label{eq:reducedact}
  \\
  F_2 &: e D_d F^{dabc} (DF)^2\bar{\psi}\gamma\psi_2, &\;
  &e (DF)^3\bar{\psi} \gamma^{a_1\cdots a_{n-1} b} \psi_{ab}, &\;
  &e (DF)^3\bar{\psi} \gamma^{a_1\cdots a_{n-2}ab} \psi_{ab}, \notag
  \\
  F_3 &: e D_d F^{dabc} F(DF)^2\bar{\psi}\gamma\psi. \notag
\end{alignat}
Examples of these terms are written in the eqs.~(\ref{eq:B1_2}), (\ref{eq:B2_2}) and (\ref{eq:F1[11]}). When we use the field redefinition ambiguities, we simply set coefficients of these terms to be zero.

\subsection{The corrections to the local supersymmetry transformations} \label{subsec:correctsusy}

Since we require the invariance under the local supersymmetry, we should take into account corrections to the transformations. Let us express the local supersymmetry transformations up to the next leading order as
\begin{alignat}{3}
  &\delta e^a{}_{\mu} = \bar{\epsilon} \gamma^a \psi_\mu 
  + \ell_p^6 \delta_1 e^a{}_\mu, \quad
  \delta \psi_\mu = 2 \mathcal{D}_\mu \epsilon 
  + \ell_p^6 \delta_1 \psi_\mu, \quad
  \delta_0 A_{\mu\nu\rho} = -3 \bar{\epsilon} \gamma_{[\mu\nu} \psi_{\rho]} 
  + \ell_p^6 \delta_1 A_{\mu\nu\rho}. \label{eq:susytr2}
\end{alignat}
Then, up to the total derivative terms, the transformation of the supergravity part of the Lagrangian density (\ref{eq:sugra}) at the next leading order is written as
\begin{alignat}{3}
  \delta_1 \mathcal{L}_0 &= e E(e)^a{}_\mu \delta_1 e^\mu{}_a 
  + e \overline{\delta_1\psi_\mu} E(\psi)^\mu
  + e E(A)^{\mu\nu\rho} \delta_1 A_{\mu\nu\rho}. \label{eq:susytrcorr}
\end{alignat}
As like the previous section, by appropriately choosing $\delta_1 e^\mu{}_a$, $\delta_1\psi_\mu$ and
$\delta_1 A_{\mu\nu\rho}$, it is possible to generate any combinations of the equations of motion. Thus, with the aid of the appendix \ref{app:appHDT}, the above transformation is written as
\begin{alignat}{3}
  \delta_1 \mathcal{L}_0 =
  &- e \big(F_{ijkl}F^{ijkl} \!-\! 144 R \big) \bar{\epsilon} Y^{(e)} 
  - e \big( F_{aijk} F_b{}^{ijk} \!-\! 12 R_{ab} \!-\! 12 \eta_{ab} R \big) 
  \bar{\epsilon} Y^{(e) ab} \qquad\qquad\qquad\quad \notag
  \\[0.1cm]
  &- e \big( D_d F^{dabc} - \tfrac{1}{8 \cdot 144} \epsilon_{11}^{abcijklmnop} 
  F_{ijkl} F_{mnop} \big) \bar{\epsilon} Y^{(A)}_{abc} \notag
\end{alignat}
\begin{alignat}{3}
  &- e Y^{(\psi) a}{}_{a_1\cdots a_{n-1}} 
  \bar{\epsilon} \Big\{ \gamma^{a_1\cdots a_{n-1}b} \psi_{ab} 
  - \sum_{i=1}^{n-1} (-1)^{n-i} \gamma^{a_1 \cdots \check{a}_i \cdots a_{n-1}} 
  \psi_a{}^{a_i} \Big\} \notag
  \\
  &- e Y^{(\psi)}_{a_1\cdots a_{n-2}} \bar{\epsilon} 
  \Big\{ \gamma^{a_1\cdots a_{n-2}ab} \psi_{ab}
  - 2 \!\!\sum_{i,j=1(i< j)}^{n-2} (-1)^{i+j}
  \gamma^{a_1 \cdots \check{a}_i \cdots \check{a}_j \cdots a_{n-2}} \psi^{a_i a_j} \Big\}
  \qquad\qquad\quad \notag
  \\
  &- e Y^{(D\psi) ab}_{a_1\cdots a_{n-1}} \bar{\epsilon} 
  \Big\{ \gamma^{a_1\cdots a_{n-1}c} \mathcal{D}_a \psi_{bc} 
  - \sum_{i=1}^{n-1} (-1)^{n-i} \gamma^{a_1 \cdots \check{a}_i \cdots a_{n-1}} 
  \mathcal{D}_a \psi_b{}^{a_i} \notag
  \\[-0.2cm]
  &\quad\;
  - \gamma^{a_1\cdots a_{n-1}} \big( - \tfrac{1}{6} F_{ija}{}^c \gamma^{ij} \psi_{bc}
  - \tfrac{1}{144} F_{ijkl} \gamma^{ijkl} \psi_{ab}
  + \tfrac{1}{36} F_{ijk}{}^c \gamma_a{}^{ijk} \psi_{bc} \big) \Big\} \qquad \notag
  \\
  &- e Y^{(D\psi) ab}_{a_1 \cdots a_{n-1}} \bar{\epsilon} \Big\{ \gamma^{a_1 \cdots a_{n-1} c} 
  \mathcal{D}_{c} \psi_{ab} - \sum_{i=1}^{n-1} (-1)^{n-i} 
  \gamma^{a_1 \cdots \check{a}_i \cdots a_{n-1}}\mathcal{D}^{a_i} \psi_{ab} \notag
  \\[-0.1cm]
  &\quad\;
  -  \gamma^{a_1 \cdots a_{n-1}} \big( \tfrac{1}{4} R_{ijab} \gamma^c \gamma^{ij} \psi_c 
  + \gamma^c DF_{ab} \psi_c + \gamma^c F^2_{ab} \psi_c \notag
  \\[-0.1cm]
  &\qquad\qquad\qquad\;\;
  + \tfrac{1}{3} F_{ij[a}{}^c \gamma^{ij} \psi_{b]c}
  + \tfrac{1}{72} F_{ijkl} \gamma^{ijkl} \psi_{ab}
  + \tfrac{1}{18} F_{ijk}{}^c \gamma^{ijk}{}_{[a} \psi_{b]c} \big) \Big\} \notag
  \\[0.1cm]
  &- e Y^{(D\psi) a}_{a_1\cdots a_n} \bar{\epsilon} \Big\{ \gamma^{a_1\cdots a_n} 
  \mathcal{D}^b \psi_{ab} \notag
  \\
  &\quad\;
  - \gamma^{a_1\cdots a_n} \big( \tfrac{1}{4} R_{ijab}\gamma^{ij} \psi^b 
  \!+\! DF_{ab} \psi^b 
  \!+\! F^2_{ab} \psi^b 
  \!+\! \tfrac{1}{3} F_{ijka} \gamma^i \psi^{jk}
  \!+\! \tfrac{1}{12} F^{ijkl} \gamma_{aij} \psi_{kl}
  \!-\! \tfrac{1}{18}  F^{ijkl} \gamma_{ijk} \psi_{al} \big) \Big\} \notag
  \\
  &- e Y^{(D\psi) a}_{a_1\cdots a_{n-2}} 
  \bar{\epsilon} \Big\{ \gamma^{a_1\cdots a_{n-2}bc} \mathcal{D}_a \psi_{bc}
  - 2 \!\!\!\sum_{i,j=1(i<j)}^{n-2} \!\!\! (-1)^{i+j} 
  \gamma^{a_1 \cdots \check{a}_i \cdots \check{a}_j \cdots a_{n-2}} 
  \mathcal{D}_a \psi^{a_i a_j}  \label{eq:susytrcorr2}
  \\[-0.1cm]
  &\quad\;
  - \gamma^{a_1\cdots a_{n-2}} \big( - \tfrac{2}{3} F_{ijka} \gamma^i \psi^{jk}
  + \tfrac{1}{9} F^{ijkl} \gamma_{ijk} \psi_{al} - \tfrac{1}{6} F^{ijkl} 
  \gamma_{aij} \psi_{kl} \big) \notag
  \\[-0.1cm]
  &\quad\;
  + 2 \sum_{i=1}^{n-2} (\!-\! 1)^{n-i} \gamma^{a_1 \cdots \check{a}_i \cdots a_{n-2}}
  \big( \!\!-\! \tfrac{1}{6} F_{ija}{}^b \gamma^{ij} \psi^{a_i}{}_b
  \!-\! \tfrac{1}{144} F_{ijkl} \gamma^{ijkl} \psi_{a}{}^{a_i} 
  \!+\! \tfrac{1}{36} F_{ijk}{}^b \gamma_a{}^{ijk} \psi^{a_i}{}_b \big) \Big\} \notag
  \\[-0.1cm]
  &- e Y^{(D\psi) a}_{a_1\cdots a_{n-2}} 
  \bar{\epsilon} \Big\{ \gamma^{a_1\cdots a_{n-2}bc} \mathcal{D}_b \psi_{ca} 
  - \!\!\!\!\! \sum_{i,j=1(i<j)}^{n-2} \!\!\!\!\! (-1)^{i+j} 
  \gamma^{a_1 \cdots \check{a}_i \cdots \check{a}_j \cdots a_{n-2}} 
  (\mathcal{D}^{a_i} \psi^{a_j}{}_{a} \!-\! \mathcal{D}^{a_j} \psi^{a_i}{}_{a}) \notag
  \\[-0.3cm]
  &\quad\;
  + \sum_{i=1}^{n-2} (-1)^{n-i} \gamma^{a_1 \cdots \check{a}_i \cdots a_{n-2}} \eta^{a_i b}
  \big( \tfrac{1}{4} R_{ijab} \gamma^c \gamma^{ij} \psi_c 
  + \gamma^c DF_{ab} \psi_c + \gamma^c F^2_{ab} \psi_c \notag
  \\[-0.3cm]
  &\qquad\qquad\qquad\qquad\qquad\qquad\qquad\;
  + \tfrac{1}{6} F_{ija}{}^c \gamma^{ij} \psi_{bc}
  + \tfrac{1}{144} F_{ijkl} \gamma^{ijkl} \psi_{ab}
  + \tfrac{1}{36} F_{ijk}{}^c \gamma^{ijk}{}_{a} \psi_{bc} \big) \notag
  \\[0.1cm]
  &\quad\;
  - \gamma^{a_1\cdots a_{n-2}} \big( \tfrac{1}{4} R_{ijab}\gamma^{ij} \psi^b 
  \!+\! DF_{ab} \psi^b \!+\! F^2_{ab} \psi^b 
  \!+\! \tfrac{1}{3} F_{ijka} \gamma^i \psi^{jk}
  \!-\! \tfrac{1}{18} F^{ijkl} \gamma_{ijk} \psi_{al}
  \!+\! \tfrac{1}{12} F^{ijkl} \gamma_{aij} \psi_{kl} \big) \Big\} \notag
  \\[0.2cm]
  &- e Y^{(D\psi)}_{a_1\cdots a_{n-1}} \bar{\epsilon} 
  \Big\{ \gamma^{a_1\cdots a_{n-1}a} \mathcal{D}^b \psi_{ab} \notag
  \\
  &\quad\;
  - \gamma^{a_1\cdots a_{n-1}} \big( \tfrac{1}{12} F_{ijkl} \gamma^{ij} \psi^{kl} 
  - \tfrac{1}{36} F_{klm}{}^i \gamma^{klm} \gamma^j \psi_{ij} \big) \notag
  \\[-0.1cm]
  &\quad\;
  - \sum_{i=1}^{n-1} (-1)^{n-i} \gamma^{a_1 \cdots \check{a}_i \cdots a_{n-1}} \eta^{a a_i} 
  \big( \tfrac{1}{4} R_{ijab}\gamma^{ij} \psi^b + DF_{ab} \psi^b + F^2_{ab} \psi^b \notag
  \\[-0.3cm]&\qquad\qquad\qquad\qquad\qquad\qquad\qquad\;\;
  + \tfrac{1}{3} F_{ijka} \gamma^i \psi^{jk}
  + \tfrac{1}{12} F^{ijkl} \gamma_{aij} \psi_{kl}
  - \tfrac{1}{18}  F^{ijkl} \gamma_{ijk} \psi_{al} \big) \Big\} \notag
  \\
  &- e Y^{(D\psi)}_{a_1\cdots a_{n-3}} \bar{\epsilon} 
  \Big\{ \gamma^{a_1\cdots a_{n-3}abc} \mathcal{D}_a \psi_{bc} 
  - 6 \!\! \sum_{i<j<k}^{n-3} \!\!\! (-1)^{n+i+j+k} 
  \gamma^{a_1 \cdots \check{a}_i \cdots \check{a}_j \cdots \check{a}_k \cdots a_{n-3}} 
  \mathcal{D}^{[a_i} \psi^{a_j a_k]} \notag
  \\[-0.3cm]
  &\quad\;
  - \sum_{i=1}^{n-3} (-1)^{n-i} \gamma^{a_1 \cdots \check{a}_i \cdots a_{n-3}} \eta^{a a_i} 
  \big( \tfrac{1}{2} R_{ijab}\gamma^{ij} \psi^b + 2 DF_{ab} \psi^b 
  + 2 F^2_{ab} \psi^b \big) \notag
  \\[-0.2cm]
  &\quad\;
  - 2 \sum_{i<j}^{n-3} (-1)^{i+j} 
  \gamma^{a_1 \cdots \check{a}_i \cdots \check{a}_j \cdots a_{n-3}} \eta^{a_i [a} \eta^{b] a_j}
  \big( \tfrac{1}{4} R_{ijab} \gamma^c \gamma^{ij} \psi_c 
  + \gamma^c DF_{ab} \psi_c + \gamma^c F^2_{ab} \psi_c \big) \Big\}. \notag
\end{alignat}
Here $\check{a}$ means that index $a$ is missing. $[Y^{(e)}]$s and $[Y^{(A)}]$ represent terms with mass dimension six, and $[Y^{(\psi)}]$s and $[Y^{(D\psi)}]$s do terms with mass dimension 6.5. In the above, it is useful to note the relation of
\begin{alignat}{3}
  \gamma^c DF_{ab} \psi_c &= - \tfrac{1}{144} D_{[a} F^{ijkl} \gamma_{b]ijklm} \psi^m
  + \tfrac{1}{18} D_{[a} F_{b]ijk} \gamma^{ijkl} \psi_l
  + \tfrac{1}{36} D_{[a} F^{ijkl} \gamma_{b]ijk} \psi_l \label{eq:gDFP}
  \\
  &\quad\,
  + \tfrac{1}{144} \gamma_{ijkl} D_{[a} F^{ijkl} \psi_{b]}
  - \tfrac{1}{6} D_{[a} F_{b]ijk} \gamma^{ij} \psi^k. \notag
\end{alignat}
The gamma matrix $\gamma^{(n-\text{odd})}$ is multiplied by $\gamma^c DF_{ab} \psi_c$, and $\gamma^{(n-\text{even})}$ is done by $DF_{ab} \psi^b$. The structure of the eq.~(\ref{eq:susytrcorr2}) will fix the corrections to the local supersymmetry transformations. 

Since we are considering the cancellation within the Fig.~\ref{fig:O(R0)part}, the classes of $V_1$, $V_2$, $V_4$ and $V_5$ are concerned in the eq.~(\ref{eq:susytrcorr2}). By appropriately setting $[Y^{(A)}]$, $[Y^{(\psi)}]$s and $[Y^{(D\psi)}]$s with arbitrary coefficients, we obtain nontrivial relations among terms in those classes. First, by choosing $[Y^{A}]=[DFDDF\gamma\psi_2]$, $[(DF)^2\gamma \mathcal{D}\psi_2]$, $[(DF)^3\gamma\psi]$ or $[FDFDDF\gamma\psi]$, it is possible to generate terms
which include $D_dF^{dabc}$ in $V_1$, $V_2$, $V_4$ or $V_5$. Second, by setting $[Y^{(\psi)}]=[(DF)^2DDF]$, we generate relations among terms in $V_1$. Third, by setting $[Y^{(D\psi)}]=[(DF)^3]$, we do relations among terms in $V_2$ and $V_4$. Note that we neglected other classes of variations which are not displayed in the Fig.~\ref{fig:O(R0)part}. Combining all these relations with arbitrary coefficients, it is possible to write $\delta_1 \mathcal{L}_0$ as
\begin{alignat}{3}
  \delta_1 \mathcal{L}_0 &=
  \begin{pmatrix}
    z_1 \!&\! z_2 \!&\! z_4 \!&\! z_5
  \end{pmatrix} 
  \begin{pmatrix}
    E_{V_1 V_1} & 0 & 0 & 0 \\
    0 & E_{V_2 V_2} & E_{V_2 V_4} & 0 \\
    0 & 0 & E_{V_4 V_4} & 0 \\
    0 & 0 & 0 & E_{V_5 V_5}
  \end{pmatrix} 
  \begin{pmatrix}
    V_1 \\ V_2 \\ V_4 \\ V_5 
  \end{pmatrix}. \label{eq:Ematrix}
\end{alignat}
Here $z_1[m,i], z_2[m,i], z_4[m,i]$ and $z_5[m,i]$ are arbitrary coefficients. If we can determine these coefficients, the structure of the corrections is understood.

\subsection{Cancellation of variations : Results} \label{sec:result}

At last, we sum the variations under the local supersymmetry (\ref{eq:Mmatrix}) and the corrections to the local supersymmetry (\ref{eq:Ematrix}), and impose the projection (\ref{eq:Pmatrix}) to the bases of the variations. Then the cancellation of variations is expressed as follows.
\begin{alignat}{3}
  0 &= \delta_0 \mathcal{L}_1 + \delta_1 \mathcal{L}_0 \notag
  \\
  &= \Bigg[ \!\!
  \begin{pmatrix}
    b_1\! & \!\!b_2\!\! & \!\!f_1\!\! & \!\!f_2\!\! & \!f_3
  \end{pmatrix} \!\!
  \begin{pmatrix}
    M_{B_1 V_1}\!\! & 0 & \!\!M_{B_1 V_4}\!\! & \!\!M_{B_1 V_5}\!\! \\
    M_{B_2 V_1}\!\! & 0 & 0 & \!\!M_{B_2 V_5} \\
    M_{F_1 V_1}\!\! & \!\!M_{F_1 V_2}\!\! & 0 & 0 \\
    M_{F_2 V_1}\!\! & \!\!M_{F_2 V_2}\!\! & \!\!M_{F_2 V_4}\!\! & 0 \\
    0 & 0 & \!\!M_{F_3 V_4}\!\! & \!\!M_{F_3 V_5}
  \end{pmatrix} \notag
  \\
  &\quad\;\;\; +
  \begin{pmatrix}
    z_1\! & \!\!z_2\!\! & \!\!z_4\!\! & \!z_5
  \end{pmatrix} \!\!
  \begin{pmatrix}
    E_{V_1 V_1}\!\! & 0 & 0 & 0 \\
    0 & \!\!E_{V_2 V_2}\!\! & \!E_{V_2 V_4}\!\! & 0 \\
    0 & 0 & \!\!E_{V_4 V_4}\!\! & 0 \\
    0 & 0 & 0 & \!\!E_{V_5 V_5}
  \end{pmatrix} \!\! \Bigg] \!
  \begin{pmatrix}
    P_{V_1 V_1}\!\! & 0 & 0 & 0 \\
    0 & \!\!P_{V_2 V_2}\!\! & \!\!P_{V_2 V_4}\!\! & 0 \\
    0 & 0 & \!\!P_{V_4 V_4}\!\! & 0 \\
    0 & 0 & 0 & \!\!P_{V_5 V_5}
  \end{pmatrix} \!\!\!
  \begin{pmatrix}
    V_1 \\ V_2 \\ V_4 \\ V_5 
  \end{pmatrix} \notag
  \\
  &= (b_1 M_{B_1 V_1} \!+\! b_2 M_{B_2 V_1} \!+\! f_1 M_{F_1 V_1} \!+\! f_2 M_{F_2 V_1}
  \!+\! z_1 E_{V_1 V_1}) P_{V_1 V_1} V_1 \notag
  \\
  &\quad
  + (f_1 M_{F_1 V_2} \!+\! f_2 M_{F_2 V_2} \!+\! z_2 E_{V_2 V_2}) P_{V_2 V_2} V_2 \notag
  \\
  &\quad
  + \big\{ ( b_1 M_{B_1 V_4} \!\!+\!\! f_2 M_{F_2 V_4} \!\!+\!\! f_3 M_{F_3 V_4}
  \!+\! z_2 E_{V_2 V_4} \!+\! z_4 E_{V_4 V_4}) P_{V_4 V_4} \label{eq:variations}
  \\
  &\quad\,
  + (f_1 M_{F_1 V_2} \!+\! f_2 M_{F_2 V_2} \!+\! z_2 E_{V_2 V_2}) P_{V_2 V_4 }\big\} 
  V_4 \notag
  \\
  &\quad
  + (b_1 M_{B_1 V_5} \!+\! b_2 M_{B_2 V_5} \!+\! f_3 M_{F_3 V_5} 
  \!+\! z_5 E_{V_5 V_5}) P_{V_5 V_5} V_5. \notag
\end{alignat}
Thus we obtain following linear equations for $b_1$, $b_2$, $f_1$, $f_2$, $f_3$, $z_1$, $z_2$, $z_4$ and $z_5$.
\begin{alignat}{3}
  0 &= (b_1 M_{B_1 V_1} \!+\! b_2 M_{B_2 V_1} \!+\! f_1 M_{F_1 V_1} \!+\! f_2 M_{F_2 V_1}
  \!+\! z_1 E_{V_1 V_1}) P_{V_1 V_1}, \notag
  \\
  0 &= (f_1 M_{F_1 V_2} \!+\! f_2 M_{F_2 V_2} \!+\! z_2 E_{V_2 V_2}) P_{V_2 V_2}, \notag
  \\
  0 &= (b_1 M_{B_1 V_4} \!\!+\!\! f_2 M_{F_2 V_4} \!\!+\!\! f_3 M_{F_3 V_4}
  \!+\! z_2 E_{V_2 V_4} \!+\! z_4 E_{V_4 V_4}) P_{V_4 V_4} \label{eq:cancel}
  \\
  &\quad\,
  \!+\! (f_1 M_{F_1 V_2} \!\!+\!\! f_2 M_{F_2 V_2} 
  \!+\! z_2 E_{V_2 V_2}) P_{V_2 V_4}, \notag
  \\
  0 &= (b_1 M_{B_1 V_5} \!+\! b_2 M_{B_2 V_5} \!+\! f_3 M_{F_3 V_5} 
  \!+\! z_5 E_{V_5 V_5}) P_{V_5 V_5}. \notag
\end{alignat}
It is possible to solve the above equations by using the Mathematica codes, and we show a solution for $b_1[1], \cdots, b_1[54]$ and $b_2[1], \cdots, b_2[22]$ explicitly.

By solving the eq.~(\ref{eq:cancel}) completely, 14 parameters out of $b_1[1], \cdots, b_1[24]$ can be determined as
\begin{alignat}{3}
  b_1[4] &= -\tfrac{4}{3} b_1[1] + \tfrac{2}{3} b_1[2] + b_1[3], \notag
  \\
  b_1[11] &= \tfrac{3}{8} b_1[2] - \tfrac{9}{4} b_1[5] - \tfrac{3}{4} b_1[6] 
  + \tfrac{1}{4 }b_1[8] - \tfrac{9}{4} b_1[10], \notag
  \\
  b_1[12]  &= - \tfrac{1}{36} b_1[2] + \tfrac{1}{2} b_1[5] + \tfrac{1}{18} b_1[6] 
  + \tfrac{1}{18} b_1[8] - \tfrac{1}{2} b_1[10], \notag
  \\
  b_1[13] &= - \tfrac{5}{8} b_1[2] + \tfrac{9}{4} b_1[5] + \tfrac{3}{4} b_1[6] 
  - \tfrac{1}{4} b_1[8] + \tfrac{9}{4} b_1[10], \notag
  \\
  b_1[15] &= - \tfrac{1}{2} b_1[2] + 9 b_1[5] + b_1[6] + 2 b_1[7] - b_1[8] + 9 b_1[10], \notag
  \\
  b_1[16] &= - b_1[1] - \tfrac{1}{8} b_1[2] - \tfrac{9}{4} b_1[5] + \tfrac{3}{4} b_1[6] 
  + \tfrac{1}{4} b_1[8] + \tfrac{9}{4} b_1[10], \notag
  \\
  b_1[17] &= - \tfrac{3}{16} b_1[2] + \tfrac{9}{8} b_1[5] + \tfrac{3}{8} b_1[6] 
  - \tfrac{1}{8} b_1[8] + \tfrac{9}{8} b_1[10] + \tfrac{1}{4} b_1[14], \label{eq:solb1}
  \\
  b_1[18] &= \tfrac{1}{8}b_1[2] - \tfrac{1}{4}b_1[6], \notag
  \\
  b_1[19] &= \tfrac{1}{8} b_1[2] - \tfrac{3}{4} b_1[5] - \tfrac{1}{4} b_1[6] - \tfrac{1}{4} b_1[8] 
  - \tfrac{3}{4} b_1[10] + \tfrac{1}{6} b_1[14], \notag
  \\
  b_1[20] &= \tfrac{1}{8} b_1[2] - \tfrac{3}{4} b_1[5] - \tfrac{1}{4} b_1[6] + \tfrac{1}{12} b_1[8] 
  + \tfrac{1}{2} b_1[9] - \tfrac{3}{4} b_1[10] - \tfrac{1}{6} b_1[14], \notag
  \\
  b_1[21] &= \tfrac{2}{3} b_1[1] - \tfrac{7}{24} b_1[2] + \tfrac{3}{4} b_1[5] + \tfrac{1}{4} b_1[6] 
  - \tfrac{5}{12} b_1[8] + \tfrac{3}{4} b_1[10], \notag
  \\
  b_1[22] &= \tfrac{1}{24} b_1[2] - \tfrac{1}{4} b_1[5] - \tfrac{1}{12} b_1[6] + \tfrac{1}{36} b_1[8] 
  - \tfrac{1}{4} b_1[10], \notag
  \\
  b_1[23] &= - \tfrac{1}{4} b_1[5] - \tfrac{1}{36} b_1[8], \notag
  \\
  b_1[24] &= - \tfrac{1}{2304} b_1[2] + \tfrac{3}{128} b_1[5] + \tfrac{1}{1152} b_1[6] 
  + \tfrac{1}{1152} b_1[8] - \tfrac{1}{128} b_1[10], \notag
\end{alignat}
and there remain 10 free parameters of $b_1[1], \cdots, b_1[3], b_1[5], \cdots, b_1[10], b_1[14]$. There are also 32 free parameters of $b_1[25], \cdots, b_1[56]$, but these are set to be zero by using the field redefinitions as explained in the subsection \ref{subsec:fieldredef}. 

In addition to the above $b_1[i]$, five parameters out of $b_2[1], \cdots, b_2[10]$ can be fixed as
\begin{alignat}{3}
  b_2[4] &= - \tfrac{1}{288} b_1[2] - \tfrac{1}{16} b_1[5] + \tfrac{1}{144} b_1[6] 
  + \tfrac{1}{144} b_1[8] - \tfrac{1}{16} b_1[10] - 2 b_2[1] - \tfrac{1}{2} b_2[3], \notag
  \\
  b_2[5] &= \tfrac{1}{2} b_2[2], \notag
  \\
  b_2[6] &= - \tfrac{1}{2592} b_1[2] - \tfrac{1}{144} b_1[5] + \tfrac{1}{1296} b_1[6] 
  + \tfrac{1}{1296} b_1[8] - \tfrac{1}{144} b_1[10] - \tfrac{1}{9} b_2[2], \label{eq:solb2}
  \\
  b_2[7] &= 2 b_2[1] - b_2[2] - \tfrac{1}{2} b_2[3], \notag
  \\
  b_2[8] &= \tfrac{1}{576} b_1[2] + \tfrac{1}{32} b_1[5] - \tfrac{1}{288} b_1[6] 
  - \tfrac{1}{288} b_1[8] + \tfrac{1}{32} b_1[10] + \tfrac{1}{2} b_2[2] + \tfrac{1}{2} b_2[3], \notag
\end{alignat}
and there remain 5 free parameters of $b_2[1], b_2[2], b_2[3], b_2[9], b_2[10]$. There are also 12 free parameters of $b_2[11],\cdots, b_2[22]$, but these become zero by using the field redefinitions. It should be remarked that parameters $f_1[m,i], f_2[m,i], f_3[m,i]$ and $z_1[m,i], z_2[m,i], z_4[m,i], z_5[m,i]$ can also be fixed simultaneously and there remain many free parameters. Furthermore, the field redefinition ambiguities impose additional constraints on these parameters, but the results of (\ref{eq:solb1}) and (\ref{eq:solb2}) remain the same.

Finally let us solve the eq.~(\ref{eq:cancel}) by setting $z_1=z_2=z_4=z_5=0$. This means that we do not assume the corrections to the local supersymmetry transformations. As a result, 15 parameters out of $b_1[1], \cdots, b_1[24]$ can be determined by the solution~(\ref{eq:solb1}) and
\begin{alignat}{3}
  b_1[10] = - \tfrac{1}{18} b_1[2] - b_1[5] + \tfrac{1}{9} b_1\![6] + \tfrac{1}{9} b_1[8].
  \label{eq:b1_10}
\end{alignat}
Explicit forms of $b_1[i]$ and $b_2[i]$ can be found in eqs.~(\ref{eq:solb1_2}) and (\ref{eq:solb2_2}) in the appendix \ref{app:anothersol}.

In summary, the variation of the generic Lagrangian density (\ref{eq:variations}) under the local supersymmetry is completely cancelled if we require nontrivial relations among $b_1, b_2, f_1, f_2, f_3$ and $z_1, z_2, z_4, z_5$. The solutions for $b_1$ and $b_2$ are given by the eqs.~(\ref{eq:solb1}) and (\ref{eq:solb2}), and these are also the solutions even if we use the field redefinitions in the effective action. We also solved the eq.~(\ref{eq:variations}) by setting $z_1=z_2=z_4=z_5=0$, and the results are given by the eqs.~(\ref{eq:solb1_2}) and (\ref{eq:solb2_2}) in the appendix \ref{app:anothersol}.

\section{Comparison with the Effective Action of Scattering Amplitudes} \label{sec:Scattering}

Let us compare the result (\ref{eq:solb1}) with the effective action obtained by evaluating scattering amplitudes of superparticles or supermembranes. The result of the scattering amplitudes is written by the eq.~(3.3) in ref. \cite{Peeters:2005tb}, and by using our notations it is expressed as
\begin{alignat}{3}
  &\mathcal{A}_{(DF)^4} = 2^{19}3^7 \times \notag
  \\[0.1cm]
  &\big\{\big( \tfrac{27}{2} \!+\! \tfrac{9}{2} \tilde{z}_1 \!+\! \tfrac{9}{2} \tilde{z}_2 
  \!-\! \tfrac{1}{2} \tilde{z}_5 \big) B_1[1] 
  + \big( 18 \!-\! 54 \tilde{z}_1 \!+\! 18 \tilde{z}_2 \!-\! 2 \tilde{z}_5 \big) B_1[2] \notag
  \\
  &\quad\,
  + \big( 12 \!-\! 192 \tilde{z}_3 \!+\! \tfrac{2}{3} \tilde{z}_6 \big) B_1[3]
  + \big( 6 \!-\! 42 \tilde{z}_1 \!+\! 6 \tilde{z}_2 \!-\! 192 \tilde{z}_3 \!-\! \tfrac{2}{3} \tilde{z}_5
  \!+\! \tfrac{2}{3} \tilde{z}_6 \big) B_1[4] \notag
  \\
  &\quad\,
  + \big( \!-\! 1 \!+\! 32 \tilde{z}_3 \!+\! 32 \tilde{z}_7 \big) B_1[5] 
  + \big( 18 \!-\! 18 \tilde{z}_1 \!-\! \tilde{z}_5 \big) B_1[6] \notag
  \\
  &\quad\,
  + \big( 9 \!+\! 9 \tilde{z}_1 \!-\! 9 \tilde{z}_2 \!+\! \tfrac{1}{2} \tilde{z}_4 \big) B_1[7] 
  + \big( 9 \!-\! 288 \tilde{z}_3 \!+\! 36 \tilde{z}_8 \big) B_1[8]
  + \big( \!-\! 6 \!-\! \tfrac{2}{3} \tilde{z}_6 \big) B_1[9]  \notag
  \\
  &\quad\,
  + \big( \!-\! 1 \!+\! \tilde{z}_1 \!-\! \tilde{z}_2 \!-\! 64 \tilde{z}_3 \!-\! 32 \tilde{z}_7 
  \!+\! 4 \tilde{z}_8 \big) B_1[10]
  + \big( \!-\! 9 \tilde{z}_1 \!+\! 9 \tilde{z}_2 \big) B_1[11]  \notag
  \\
  &\quad\,
  + \big( 1 \!+\! 32 \tilde{z}_3 \!+\! 32 \tilde{z}_7 \big) B_1[12]
  + \big( \!-\! \tfrac{9}{2} \!+\! \tfrac{45}{2} \tilde{z}_1 \!-\! \tfrac{27}{2} \tilde{z}_2 
  \!+\! \tfrac{1}{2} \tilde{z}_5 \big) B_1[13] \label{eq:scatter}
  \\
  &\quad\,
  + \big( 18 \!+\! 90 \tilde{z}_1 \!+\! 18 \tilde{z}_2 \!-\! 2 \tilde{z}_4 \!+\! 2 \tilde{z}_5 
  \!+\! 2 \tilde{z}_9 \big) B_1[14]
  + \big( 36 \tilde{z}_1 \!-\! 36 \tilde{z}_2 \!+\! \tilde{z}_4 \big) B_1[15] \notag
  \\
  &\quad\,
  + \big( \!-\! 9 \tilde{z}_1 \!-\! 9 \tilde{z}_2 \!-\! 288 \tilde{z}_3 \!-\! 144 \tilde{z}_7 
  \!+\!18 \tilde{z}_8 \big) B_1[16] 
  + \big(\tfrac{9}{2} \!+\! 27 \tilde{z}_1 \!-\! \tfrac{1}{2} \tilde{z}_4 \!+\! \tfrac{1}{2} \tilde{z}_5 
  \!+\! \tfrac{1}{2} \tilde{z}_9 \big) B_1[17] \notag
  \\
  &\quad\,
  + \big( \!-\! \tfrac{9}{4} \!-\! \tfrac{9}{4} \tilde{z}_1 \!+\! \tfrac{9}{4} \tilde{z}_2 \big) B_1[18]
  + \big(12 \tilde{z}_1 \!+\! 6 \tilde{z}_2 \!+\! 96 \tilde{z}_3 \!-\! \tfrac{1}{3} \tilde{z}_4 
  \!+\! \tfrac{1}{3} \tilde{z}_5 \!-\! 12 \tilde{z}_8 \!+\! \tfrac{1}{3} \tilde{z}_9 \big) B_1[19] \notag
  \\
  &\quad\,
  + \big( \!-\! 6 \!-\! 18 \tilde{z}_1 \!+\! \tfrac{1}{3} \tilde{z}_4 \!-\! \tfrac{1}{3} \tilde{z}_5
  \!-\! \tfrac{1}{3} \tilde{z}_6 \!-\! \tfrac{1}{3} \tilde{z}_9 \big) B_1[20]
  + \big( 3 \!+\! 15 \tilde{z}_1 \!-\! 3 \tilde{z}_2 \!+\! 96 \tilde{z}_3 \!-\! 12 \tilde{z}_8 \big) B_1[21] \notag
  \\
  &\quad\,
  + \big( \!-\! \tilde{z}_1 \!+\! \tilde{z}_2 \big) B_1[22] 
  + \big( \!-\! 8 \tilde{z}_7 \!-\! \tilde{z}_8 \big) B_1[23] 
  + \big( \tilde{z}_3 \!+\! \tilde{z}_7 \big) B_1[24] \big\}. \notag
\end{alignat}
Here $\tilde{z}_i \,(i=1,\cdots,9)$ are coefficients of $\tilde{Z}_i$, which are defined in the eq.~(A.7) in the ref.~\cite{Peeters:2005tb}. $\tilde{Z}_i \,(i=1,\cdots,9)$ vanish at the level of the four point amplitudes and $\tilde{z}_i$ are dealt as free parameters.

After some calculations, it is possible to show that the eq.~(\ref{eq:scatter}) can be reproduced from the eq.~(\ref{eq:solb1}) by choosing 10 free parameters as
\begin{alignat}{3}
  b_1[1] &= \tfrac{27}{2} + \tfrac{9}{2} \tilde{z}_1 + \tfrac{9}{2} \tilde{z}_2 
  - \tfrac{1}{2} \tilde{z}_5, 
  &b_1[2] &= 18 - 54 \tilde{z}_1 + 18 \tilde{z}_2 - 2 \tilde{z}_5, \notag
  \\
  b_1[3] &= 12 - 192 \tilde{z}_3 + \tfrac{2}{3} \tilde{z}_6,
  &b_1[5] &= - 1 + 32 \tilde{z}_3 + 32 \tilde{z}_7, \notag
  \\
  b_1[6] &= 18 - 18 \tilde{z}_1 - \tilde{z}_5, 
  &b_1[7] &= 9 + 9 \tilde{z}_1 - 9 \tilde{z}_2 + \tfrac{1}{2} \tilde{z}_4, \label{eq:b1sub}
  \\
  b_1[8] &= 9 - 288 \tilde{z}_3 + 36 \tilde{z}_8, 
  &b_1[9] &= - 6 - \tfrac{2}{3} \tilde{z}_6, \notag
  \\
  b_1[10] &= - 1 + \tilde{z}_1 - \tilde{z}_2 - 64 \tilde{z}_3 - 32 \tilde{z}_7 + 4 \tilde{z}_8, \quad
  &b_1[14] &= 18 + 90 \tilde{z}_1 + 18 \tilde{z}_2 - 2 \tilde{z}_4 + 2 \tilde{z}_5 
  + 2 \tilde{z}_9, \notag
\end{alignat}
up to the overall factor $2^{19}3^7$. Thus the effective action determined by the local supersymmetry is consistent with one obtained by the scattering amplitudes. If we substitute the eq.~(\ref{eq:b1sub}) into the eq.~(\ref{eq:solb2}), we obtain
\begin{alignat}{3}
  b_2[4] &= \tfrac{1}{4} - 2 b_2[1] - \tfrac{1}{2} b_2[3], 
  &b_2[5] &= \tfrac{1}{2} b_2[2], \notag
  \\
  b_2[6] &= \tfrac{1}{36} - \tfrac{1}{9} b_2[2], 
  &b_2[7] &= 2 b_2[1] - b_2[2] - \tfrac{1}{2} b_2[3], \label{eq:b2sub}
  \\
  b_2[8] &= - \tfrac{1}{8} + \tfrac{1}{2} b_2[2] + \tfrac{1}{2} b_2[3]. \qquad \notag
\end{alignat}
It is interesting to note that the coefficients $\tilde{z}_i$ disappear.

Finally, let us examine the solutions of the eq.~(\ref{eq:solb1_2}) and the eq.~(\ref{eq:solb2_2}), which are obtained by setting $z_1=z_2=z_4=z_5=0$. In order to reproduce the eq.~(\ref{eq:scatter}) from the eq.~(\ref{eq:solb1_2}), at least the eq.~(\ref{eq:b1_10}) should be satisfied. However, this contradicts with the eq.~(\ref{eq:b1sub}). Thus the assumptions $z_1=z_2=z_4=z_5=0$ are not consistent with the result of the scattering amplitudes. This means that we should take into account the corrections to the local supersymmetry transformations seriously.

\section{Conclusions and Discussions} \label{sec:Conclusion}

In this paper, we pursued the effective action of the M-theory with eight derivative terms and restricted its structure by imposing the local supersymmetry. Ideally, we will write down all possible eight derivative terms, evaluate the variations of these terms under the local supersymmetry, and examine the cancellation of these variations. There are a lot of classes in the effective action and the variations, however, we made the ansatz in the subsection \ref{subsec:Ansatz} so that the number of covariant derivatives in the terms becomes as small as possible.

As is well known, the effective action of the M-theory contains terms of $[eR^4]$ at the eight derivative order, and these terms played crucial roles to investigate corrections to black hole geometries and gauge/gravity duality. From the viewpoint of the local supersymmetry, however, it is reasonable to investigate $[e(D\hat{F})^4]$ terms, which do not contain the Riemann tensor at all. 
Following the ansatz, we explicitly figured out the terms in the effective action and their variations at $\mathcal{O}(R^0)$ in the subsection \ref{subsec:VarO(R0)}. The cancellation mechanism of these variations are shown in the Fig.~\ref{fig:O(R0)}. 

Since there are a lot of terms even at the level of $\mathcal{O}(R^0)$, we have restricted the ansatz for the effective action to  $B_1=[e(D\hat{F})^4]$, $B_2=[e\epsilon_{11}\hat{R}(D\hat{F})^3]$, $F_1=[e(DF)^2 \overline{\psi_2} \gamma \mathcal{D} \psi_2]$, $F_2=[e(DF)^3 \bar{\psi} \gamma \psi_2]$ and $F_3=[eF(DF)^3 \bar{\psi} \gamma \psi]$. In order to impose the local supersymmetry, we also classified their variations which are represented by $V_1=[e(DF)^2 DDF \bar{\epsilon} \gamma \psi_2]$, $V_2=[e(DF)^3 \bar{\epsilon} \gamma \mathcal{D} \psi_2]$, $V_4=[e(DF)^4 \bar{\epsilon} \gamma \psi]$ and $V_5=[eF(DF)^2 DDF \bar{\epsilon} \gamma \psi]$. The results are obtained by using the Mathematica codes, and are summarized in the Tables \ref{tb:B1}-\ref{tb:V5} in the section \ref{sec:ActionVariation}. The explicit forms of the variations with respect to the terms in $B_1$, $B_2$, $F_1$, $F_2$ and $F_3$ are displayed in the section \ref{sec:SusyVariation}, and actual calculations are done by employing the Mathematica codes.

Technical difficulties in the calculations of higher derivative terms arise form the Bianchi identities, the commutation relations and the dimension dependent identities. By using the Bianchi identity (\ref{eq:V2V4}) and the commutation relation (\ref{eq:V1V1}), the terms in $V_1$ and $V_2$ are related to the variations which include the Riemann tensor, and especially the terms in $V_2$ are related to those in $V_4$. The dimension dependent identities give relations among the terms which may include equations of motion for the eleven dimensional supergravity. Thus the independent terms of the variations should be carefully extracted by taking into account the projection matrix (\ref{eq:Pmatrix}).

Other technical difficulties come from the field redefinition ambiguities and the corrections to the local supersymmetry transformations. The former is used to reduce the terms in the effective action which include equations of motion for the eleven dimensional supergravity. The latter generates the terms in the variations which include equations of motion for the eleven dimensional supergravity, in a quite complicated way. Especially some of these relate the terms in $V_2$ with those in $V_4$, and we should carefully treat these corrections as like the eq.~(\ref{eq:Ematrix}). Of course, we heavily used the Mathematica codes to resolve these difficulties.

Finally we executed the cancellation of the variations in the end of the section \ref{sec:Cancel}. The bosonic parts of the results are shown by the eqs.~(\ref{eq:solb1}) and (\ref{eq:solb2}). The coefficients of $B_1$ are controlled by 10 parameters, and those of $B_2$ are done by 5 ones. The result obtained by imposing the local supersymmetry is consistent with that of the scattering amplitudes in the type IIA superstring theories and the M-theory. On the other hand, if we do not take into account the corrections to the local supersymmetry, the result contradicts with that of the scattering amplitudes. Thus the corrections to the local supersymmetry should be considered seriously.

In this paper we pursued the structure of $[e(D\hat{F})^4]$ and $[e\epsilon_{11}\hat{R}(D\hat{F})^3]$, which is a portion of the effective action at $\mathcal{O}(R^0)$ in the M-theory. As a near future work, it is natural to investigate the terms like $[e\epsilon_{11}\hat{F}^2(D\hat{F})^3]$ and $[e\hat{R}\hat{F}^2(D\hat{F})^2]$ in the Table \ref{fig:O(R0)}. After completing such terms, we try to complete the variations shown in the Table \ref{fig:O(R0)} and terms of higher powers with respect to the Riemann tensor. Since these procedures give more and more constraints on the form of the effective action, we expect that the local supersymmetry is enough to determine the structure of the effective action in the M-theory uniquely. Actually the superinvariant for $t_8 t_8 R^4$ in the M-theory were uniquely determined in ref.~\cite{Hyakutake:2007sm}, but the dimension dependent identities were neglected there. Thus the number of superinvariants in the M-theory is still in question, but it will be resolved in the future.

Recently it is pointed out that the Kaluza-Klein reduction of the $t_8 t_8 R^4$ in the eleven dimensions generates nontrivial terms, which include R-R massless fields and cannot be reproduced by the 1-loop 4 points scattering amplitudes in the type IIA superstring theory\cite{Aggarwal:2025lxf}. On the other hand, it is reported that the similar terms appear from the scattering amplitudes  in the type IIB superstring theory\cite{Liu:2025uqu}. Since the method of requiring the local supersymmetry is universal, it is also interesting to derive the effective actions in the type II superstring theories both at tree level and 1-loop order by imposing the local supersymmetry.

\section*{Acknowledgements}

I would like to thank S.~Ogushi for the collaboration in the early stage, and M.~Weissenbacher for useful discussions. I would also like to thank participants in ``2024 Kanto-NTU High Energy Physics Workshop'', especially H.~Kawai, S.~Kawamoto, N.~Ohta and K.~Yoshida. ``The NTHU and Ibaraki Univ. co-seminar'' was helpful to summarize my results. Discussions at the seminar at Toyota Technological Institute was also useful and I would like to thank J.~Sakamoto, R.~Suzuki, S.~Tomizawa and H.~Yoshino. This work was supported by Japan Society for the Promotion of Science (JSPS), Grant-in-Aid for Scientific Research (C) Grant Number 22K03613.

\newpage
\appendix
\section{Some Calculations for the Field Redefinition Ambiguities and the Corrections to the Local Supersymmetry}\label{app:appHDT}

In this appendix, we derive the relations of $[eX\gamma\mathcal{D}\psi_2]$ which are used in the subsections \ref{subsec:fieldredef} and \ref{subsec:correctsusy}. First, we use the equations in the (\ref{eq:RDFF2}) and obtain
\begin{alignat}{3}
  3 \gamma^c \mathcal{D}_{[a} \psi_{bc]} 
  &= \tfrac{1}{4} R_{ijab} \gamma^c \gamma^{ij} \psi_c 
  + \gamma^c DF_{ab} \psi_c + \gamma^c F^2_{ab} \psi_c \label{eq:aDpsi1}
  \\&\quad\,
  + \tfrac{1}{2} E(e)_{i[a} \gamma^i \psi_{b]} - \tfrac{1}{18} E(e)^i{}_i \gamma_{[a} \psi_{b]}
  + E(A)_{ij[a} \gamma^{ij} \psi_{b]} + \tfrac{1}{6} E(A)^{ijk} \gamma_{ijk[a} \psi_{b]}, \notag
  \\[0.1cm]
  3 \gamma^{bc} \mathcal{D}_{[a} \psi_{bc]} 
  &= \tfrac{1}{4} R_{ijab} \{\gamma^b, \gamma^c\} \gamma^{ij} \psi_c
  + \{\gamma^b, \gamma^c\} DF_{ab} \psi_c
  + \{\gamma^b, \gamma^c\} F^2_{ab} \psi_c \notag
  \\&\quad\,
  - \gamma^c \big( - \tfrac{1}{4} E(e)_{ai} \gamma^i + \tfrac{1}{36} E(e)^i{}_i \gamma_a
  - \tfrac{1}{2} E(A)_{aij} \gamma^{ij} 
  + \tfrac{1}{12} E(A)^{ijk} \gamma_{aijk} \big) \psi_c \notag
  \\&\quad\,
  + \gamma^b \big( \tfrac{1}{2} E(e)_{i[a} \gamma^i \psi_{b]} 
  - \tfrac{1}{18} E(e)^i{}_i \gamma_{[a} \psi_{b]}
  + E(A)_{ij[a} \gamma^{ij} \psi_{b]} 
  + \tfrac{1}{6} E(A)^{ijk} \gamma_{ijk[a} \psi_{b]} \big) \notag
  \\
  &= \tfrac{1}{2} R_{ijab}\gamma^{ij} \psi^b + 2 DF_{ab} \psi^b + 2 F^2_{ab} \psi^b
  \label{eq:aDpsi2}
  \\&\quad\,
  - \tfrac{1}{2} E(e)_{ai} \gamma^{ij} \psi_j
  + \tfrac{1}{2} E(e)_{ai} \psi^i
  + \tfrac{1}{18} E(e)^i{}_i \gamma_a{}^j \psi_j \notag
  \\&\quad\,
  - 2 E(A)_{aij} \gamma^{i} \psi^j
  + E(A)_{aij} \gamma^{ijk} \psi_k 
  + \tfrac{1}{2} E(A)^{ijk} \gamma_{aij} \psi_k
  - \tfrac{1}{6} E(A)^{ijk} \gamma_{aijkl} \psi^l, \notag
  \\[0.1cm]
  3 \gamma^{abc} \mathcal{D}_{[a} \psi_{bc]} 
  &= \tfrac{3}{2} E(e)_{ai} \gamma^i \psi^a 
  + \tfrac{9}{2} E(A)_{aij} \gamma^{ij} \psi^a. \label{eq:aDpsi3}
\end{alignat}
Then by using the eq.~(\ref{eq:aDpsi1}) and 
\begin{alignat}{3}
  \{ \gamma^c, F_a\} 
  &= - \tfrac{1}{36} F_{aijk} \{\gamma^c, \gamma^{ijk}\}
  + \tfrac{1}{288} F_{ijkl} \{\gamma^c,\gamma_a{}^{ijkl}\} \notag
  \\
  &= - \tfrac{1}{6} F_{ija}{}^c \gamma^{ij}
  + \tfrac{1}{144} \delta^c_a F_{ijkl} \gamma^{ijkl}
  + \tfrac{1}{36} F_{ijk}{}^c \gamma_a{}^{ijk}, \label{eq:gammaF}
\end{alignat}
it is possible to represent the terms of $[\gamma \mathcal{D}\psi_2]$ with one contraction as
\begin{alignat}{3}
  \gamma^c \mathcal{D}_a \psi_{bc}
  &= \{\gamma^c, F_a\} \psi_{bc} + (D_a - F_a) \gamma^c \psi_{bc} \notag
  \\
  &= - \tfrac{1}{6} F_{ija}{}^c \gamma^{ij} \psi_{bc}
  - \tfrac{1}{144} F_{ijkl} \gamma^{ijkl} \psi_{ab}
  + \tfrac{1}{36} F_{ijk}{}^c \gamma_a{}^{ijk} \psi_{bc}
  + (D_a - F_a) \gamma^c \psi_{bc}, \label{eq:Dpsi1}
  \\[0.2cm]
  \gamma^c \mathcal{D}_{c} \psi_{ab} 
  &= 3 \gamma^c \mathcal{D}_{[c} \psi_{ab]} - 2 \gamma^c \mathcal{D}_{[a} \psi_{b]c} \notag
  \\
  &= 3 \gamma^c \mathcal{D}_{[c} \psi_{ab]} - 2 \{\gamma^c, F_{[a}\} \psi_{b]c} 
  - 2 (D_{[a} - F_{[a}) \gamma^c \psi_{b]c} \notag
  \\
  &= \tfrac{1}{4} R_{ijab} \gamma^c \gamma^{ij} \psi_c 
  + \gamma^c DF_{ab} \psi_c + \gamma^c F^2_{ab} \psi_c \notag
  \\&\quad\,
  + \tfrac{1}{3} F_{ij[a}{}^c \gamma^{ij} \psi_{b]c}
  + \tfrac{1}{72} F_{ijkl} \gamma^{ijkl} \psi_{ab}
  + \tfrac{1}{18} F_{ijk}{}^c \gamma^{ijk}{}_{[a} \psi_{b]c} \label{eq:Dpsi2}
  \\&\quad\,
  + \tfrac{1}{2} E(e)_{i[a} \gamma^i \psi_{b]} - \tfrac{1}{18} E(e)^i{}_i \gamma_{[a} \psi_{b]}
  + E(A)_{ij[a} \gamma^{ij} \psi_{b]} + \tfrac{1}{6} E(A)^{ijk} \gamma_{ijk[a} \psi_{b]} \notag
  \\&\quad\,
  - 2 (D_{[a} - F_{[a}) \gamma^c \psi_{b]c}, \notag
  \\[0.2cm]
  \mathcal{D}^b \psi_{ab} 
  &= \tfrac{1}{2} \gamma^b \gamma^c \mathcal{D}_c \psi_{ab}
  + \tfrac{1}{2} \gamma^b \gamma^c \mathcal{D}_b \psi_{ac} \notag
  \\
  &= \tfrac{1}{4} R_{ijab}\gamma^{ij} \psi^b + DF_{ab} \psi^b + F^2_{ab} \psi^b \notag
  \\&\quad\,
  + \tfrac{1}{3} F_{ijka} \gamma^i \psi^{jk}
  + \tfrac{1}{12} F^{ijkl} \gamma_{aij} \psi_{kl}
  - \tfrac{1}{18}  F^{ijkl} \gamma_{ijk} \psi_{al} \label{eq:Dpsi3}
  \\&\quad\,
  - \tfrac{1}{4} E(e)_{ai} \gamma^{ij} \psi_j
  + \tfrac{1}{4} E(e)_{ai} \psi^i
  + \tfrac{1}{36} E(e)^i{}_i \gamma_{aj} \psi^j \notag
  \\&\quad\,
  -  E(A)_{aij} \gamma^{i} \psi^j
  + \tfrac{1}{2} E(A)_{aij} \gamma^{ijk} \psi_k 
  + \tfrac{1}{4} E(A)^{ijk} \gamma_{aij} \psi_k
  - \tfrac{1}{12} E(A)^{ijk} \gamma_{aijkl} \psi^l \notag
  \\&\quad\,
  + \tfrac{1}{12} F_{ija}{}^c \gamma^{ij} \gamma^b \psi_{bc}
  + \tfrac{1}{96} F_{ijkl} \gamma^{ijkl} \gamma^b \psi_{ab}
  - \tfrac{1}{72} F^{ijkc} \gamma_{aijk} \gamma^b \psi_{bc} \notag
  \\&\quad\,
  - \tfrac{1}{2} \gamma^b (D_{a} - F_{a}) \gamma^c \psi_{bc}
  + \gamma^c (D_c - F_c) \gamma^b \psi_{ab}. \notag
\end{alignat}
Similarly, by using the eq.~(\ref{eq:aDpsi2}) and the eq.~(\ref{eq:gammaF}), it is possible to represent the terms of $[\gamma \mathcal{D}\psi_2]$ with two contractions as
\begin{alignat}{3}
  \gamma^{bc} \mathcal{D}_a \psi_{bc}
  &= - \tfrac{2}{3} F_{ijka} \gamma^i \psi^{jk}
  + \tfrac{1}{9} F^{ijkl} \gamma_{ijk} \psi_{al}
  - \tfrac{1}{6} F^{ijkl} \gamma_{aij} \psi_{kl} \label{eq:Dpsi4}
  \\&\quad\,
  - \tfrac{1}{6} F_{ija}{}^c \gamma^{ij} \gamma^b \psi_{bc}
  - \tfrac{1}{144} F_{ijkl} \gamma^{ijkl} \gamma^b \psi_{ab}
  + \tfrac{1}{36} F_{ijk}{}^c \gamma_a{}^{ijk} \gamma^b \psi_{bc}
  + \gamma^b (D_a - F_a) \gamma^c \psi_{bc}, \notag
  \\[0.2cm]
  \gamma^{bc} \mathcal{D}_b \psi_{ca} 
  &= \tfrac{1}{2} \gamma^{bc} (3\mathcal{D}_{[b} \psi_{ca]} - \mathcal{D}_a \psi_{bc} ) \notag
  \\
  &= \tfrac{1}{4} R_{ijab}\gamma^{ij} \psi^b + DF_{ab} \psi^b + F^2_{ab} \psi^b \notag
  \\&\quad\,
  + \tfrac{1}{3} F_{ijka} \gamma^i \psi^{jk}
  - \tfrac{1}{18} F^{ijkl} \gamma_{ijk} \psi_{al}
  + \tfrac{1}{12} F^{ijkl} \gamma_{aij} \psi_{kl}  \label{eq:Dpsi5}
  \\&\quad\,
  - \tfrac{1}{4} E(e)_{ai} \gamma^{ij} \psi_j
  + \tfrac{1}{4} E(e)_{ai} \psi^i
  + \tfrac{1}{36} E(e)^i{}_i \gamma_a{}^j \psi_j \notag
  \\&\quad\,
  - E(A)_{aij} \gamma^{i} \psi^j
  + \tfrac{1}{2} E(A)_{aij} \gamma^{ijk} \psi_k 
  + \tfrac{1}{4} E(A)^{ijk} \gamma_{aij} \psi_k
  - \tfrac{1}{12} E(A)^{ijk} \gamma_{aijkl} \psi^l \notag
  \\&\quad\,
  + \tfrac{1}{12} F_{ija}{}^c \gamma^{ij} \gamma^b \psi_{bc}
  + \tfrac{1}{288} F_{ijkl} \gamma^{ijkl} \gamma^b \psi_{ab}
  - \tfrac{1}{72} F_{ijk}{}^c \gamma_a{}^{ijk} \gamma^b \psi_{bc}
  - \tfrac{1}{2} \gamma^b (D_a - F_a) \gamma^c \psi_{bc}, \notag
  \\[0.2cm]
  \gamma^a \mathcal{D}^b \psi_{ab} 
  &= \{\gamma^a, F^b\} \psi_{ab} + (D^b - F^b) \gamma^a \psi_{ab} \notag
  \\
  &= \tfrac{1}{12} F_{ijkl} \gamma^{ij} \psi^{kl} 
  - \tfrac{1}{36} F_{klm}{}^i \gamma^{klm} \gamma^j \psi_{ij}
  + (D^b - F^b) \gamma^a \psi_{ab}. \label{eq:Dpsi6}
\end{alignat}
Note that, as shown in the eq.~(\ref{eq:EOM2}), $\gamma^a \psi_{ab}$ is expressed as a linear combination of $E(\psi)^a$.

Next, the terms of $[\gamma \mathcal{D}\psi_2]$ which have one contraction are decomposed as
\begin{alignat}{3}
  \gamma^{a_1\cdots a_{n-1}c} \mathcal{D}_a \psi_{bc} 
  &= \gamma^{a_1\cdots a_{n-1}}\gamma^c \mathcal{D}_a \psi_{bc} 
  + \sum_{i=1}^{n-1} (-1)^{n-i} 
  \gamma^{a_1 \cdots \check{a}_i \cdots a_{n-1}} \mathcal{D}_a \psi_b{}^{a_i}, 
  \label{eq:Dpsi1_1}
  \\[-0.1cm]
  \gamma^{a_1 \cdots a_{n-1} c} \mathcal{D}_{c} \psi_{ab} 
  &= \gamma^{a_1 \cdots a_{n-1}} \gamma^c \mathcal{D}_{c} \psi_{ab} 
  + \sum_{i=1}^{n-1} (-1)^{n-i}
  \gamma^{a_1 \cdots \check{a}_i \cdots a_{n-1}}\mathcal{D}^{a_i} \psi_{ab}. 
  \label{eq:Dpsi2_2}
\end{alignat}
Similarly, the terms of $[\gamma \mathcal{D}\psi_2]$ which have two contractions are decomposed as
\begin{alignat}{3}
  \gamma^{a_1\cdots a_{n-2}bc} \mathcal{D}_a \psi_{bc} 
  &= \gamma^{a_1\cdots a_{n-2}} \gamma^{bc} 
  \mathcal{D}_a \psi_{bc}
  - 2 \sum_{i=1}^{n-2} (-1)^{n-i} \gamma^{a_1 \cdots \check{a}_i \cdots a_{n-2}} \gamma^b
  \mathcal{D}_a \psi^{a_i}{}_b \notag
  \\[-0.3cm]&\quad\,
  + 2 \!\! \sum_{i,j=1(i<j)}^{n-2} \!\!\! (-1)^{i+j} 
  \gamma^{a_1 \cdots \check{a}_i \cdots \check{a}_j \cdots a_{n-2}} 
  \mathcal{D}_a \psi^{a_i a_j},  \label{eq:Dpsi4_4}
  \\[0.1cm]
  \gamma^{a_1\cdots a_{n-2}bc} \mathcal{D}_b \psi_{ca} 
  &= \gamma^{a_1\cdots a_{n-2}} \gamma^{bc} \mathcal{D}_b \psi_{ca} \notag
  \\[-0.1cm]&\quad\,
  - \sum_{i=1}^{n-2} (-1)^{n-i} \gamma^{a_1 \cdots \check{a}_i \cdots a_{n-2}} \gamma^b
  (\mathcal{D}^{a_i} \psi_{ba} - \mathcal{D}_b \psi^{a_i}{}_{a})  \label{eq:Dpsi5_5}
  \\[-0.2cm]&\quad\,
  + \!\! \sum_{i,j=1(i<j)}^{n-2} \!\!\! (-1)^{i+j} 
  \gamma^{a_1 \cdots \check{a}_i \cdots \check{a}_j \cdots a_{n-2}} 
  (\mathcal{D}^{a_i} \psi^{a_j}{}_{a} - \mathcal{D}^{a_j} \psi^{a_i}{}_{a}), \notag
  \\[0cm]
  \gamma^{a_1\cdots a_{n-1}a} \mathcal{D}^b \psi_{ab} 
  &= \gamma^{a_1\cdots a_{n-1}} \gamma^a \mathcal{D}^b \psi_{ab}
  + \sum_{i=1}^{n-1} (-1)^{n-i} 
  \gamma^{a_1 \cdots \check{a}_i \cdots a_{n-1}} \mathcal{D}^b \psi^{a_i}{}_{b},
  \label{eq:Dpsi6_6}
\end{alignat}
and the term with three contractions is written as
\begin{alignat}{3}
  \gamma^{a_1\cdots a_{n-3}abc} \mathcal{D}_a \psi_{bc} 
  &= \gamma^{a_1\cdots a_{n-3}} \gamma^{abc} \mathcal{D}_a \psi_{bc}
  + 3 \sum_{i=1}^{n-3} (-1)^{n-i}\gamma^{a_1 \cdots \check{a}_i 
  \cdots a_{n-3}} \eta^{a_i a} \gamma^{bc} \mathcal{D}_{[a} \psi_{bc]} \notag
  \\[-0.1cm]&\quad\,
  + 6 \sum_{i<j}^{n-3} (-1)^{i+j} \gamma^{a_1 \cdots \check{a}_i 
  \cdots \check{a}_j \cdots a_{n-3}} \eta^{a_i a} \eta^{a_j b} \gamma^{c} 
  \mathcal{D}_{[a} \psi_{bc]} \label{eq:Dpsi7_7}
  \\[-0.1cm]&\quad\,
  + 6 \!\! \sum_{i<j<k}^{n-3} \!\!\! (-1)^{n+i+j+k} \gamma^{a_1 \cdots 
  \check{a}_i \cdots \check{a}_j \cdots \check{a}_k \cdots a_{n-3}} 
  \mathcal{D}^{[a_i} \psi^{a_j a_k]}. \notag
\end{alignat}

Finally, by using the eqs.~(\ref{eq:Dpsi1}) and  (\ref{eq:Dpsi1_1}), it is possible to express the terms with the derivative of field equation for the eleven dimensional supergravity as
\begin{alignat}{3}
  &\gamma^{a_1\cdots a_{n-1}} (D_a - F_a) \gamma^c \psi_{bc} \notag
  \\
  &= \gamma^{a_1\cdots a_{n-1}c} \mathcal{D}_a \psi_{bc} 
  - \sum_{i=1}^{n-1} (-1)^{n-i} \gamma^{a_1 \cdots \check{a}_i \cdots a_{n-1}} \mathcal{D}_a \psi_b{}^{a_i}
  \label{eq:Dpsi1_1_1}
  \\[-0.1cm]
  &\quad\,
  - \gamma^{a_1\cdots a_{n-1}} \big( - \tfrac{1}{6} F_{ija}{}^c \gamma^{ij} \psi_{bc}
  - \tfrac{1}{144} F_{ijkl} \gamma^{ijkl} \psi_{ab}
  + \tfrac{1}{36} F_{ijk}{}^c \gamma_a{}^{ijk} \psi_{bc} \big). \notag
\end{alignat}
By using the eqs.~(\ref{eq:Dpsi2}) and  (\ref{eq:Dpsi2_2}), we obtain
\begin{alignat}{3}
  &\gamma^{a_1 \cdots a_{n-1}} \big( \tfrac{1}{2} E(e)_{i[a} \gamma^i \psi_{b]} 
  - \tfrac{1}{18} E(e)^i{}_i \gamma_{[a} \psi_{b]} + E(A)_{ij[a} \gamma^{ij} \psi_{b]} \notag
  \\&\qquad\qquad\qquad\qquad
  + \tfrac{1}{6} E(A)^{ijk} \gamma_{ijk[a} \psi_{b]} - 2 (D_{[a} - F_{[a}) \gamma^c \psi_{b]c} \big) \notag
  \\
  &= \gamma^{a_1 \cdots a_{n-1} c} \mathcal{D}_{c} \psi_{ab} 
  - \sum_{i=1}^{n-1} (-1)^{n-i} \gamma^{a_1 \cdots \check{a}_i \cdots a_{n-1}}\mathcal{D}^{a_i} \psi_{ab} 
  \label{eq:Dpsi2_2_2}
  \\
  &\quad\,
  - \gamma^{a_1 \cdots a_{n-1}} \big( \tfrac{1}{4} R_{ijab} \gamma^c \gamma^{ij} \psi_c 
  + \gamma^c DF_{ab} \psi_c + \gamma^c F^2_{ab} \psi_c \notag
  \\&\qquad\qquad\qquad\qquad
  + \tfrac{1}{3} F_{ij[a}{}^c \gamma^{ij} \psi_{b]c}
  + \tfrac{1}{72} F_{ijkl} \gamma^{ijkl} \psi_{ab}
  + \tfrac{1}{18} F_{ijk}{}^c \gamma^{ijk}{}_{[a} \psi_{b]c} \big) . \notag
\end{alignat}
By using the eq.~(\ref{eq:Dpsi3}), we obtain
\begin{alignat}{3}
  &\gamma^{a_1\cdots a_n} 
  \big( - \tfrac{1}{4} E(e)_{ai} \gamma^{ij} \psi_j
  + \tfrac{1}{4} E(e)_{ai} \psi^i
  + \tfrac{1}{36} E(e)^i{}_i \gamma_{aj} \psi^j \notag
  \\&\qquad\quad\;\,
  -  E(A)_{aij} \gamma^{i} \psi^j
  + \tfrac{1}{2} E(A)_{aij} \gamma^{ijk} \psi_k 
  + \tfrac{1}{4} E(A)^{ijk} \gamma_{aij} \psi_k
  - \tfrac{1}{12} E(A)^{ijk} \gamma_{aijkl} \psi^l \notag
  \\&\qquad\quad\;\,
  + \tfrac{1}{12} F_{ija}{}^c \gamma^{ij} \gamma^b \psi_{bc}
  + \tfrac{1}{96} F_{ijkl} \gamma^{ijkl} \gamma^b \psi_{ab}
  - \tfrac{1}{72} F^{ijkc} \gamma_{aijk} \gamma^b \psi_{bc} \notag
  \\&\qquad\quad\;\,
  - \tfrac{1}{2} \gamma^b (D_{a} - F_{a}) \gamma^c \psi_{bc}
  + \gamma^c (D_c - F_c) \gamma^b \psi_{ab} \big) \notag
  \\
  &= \gamma^{a_1\cdots a_n} \mathcal{D}^b \psi_{ab}
  - \gamma^{a_1\cdots a_n} \big( \tfrac{1}{4} R_{ijab}\gamma^{ij} \psi^b 
  + DF_{ab} \psi^b + F^2_{ab} \psi^b 
  \label{eq:Dpsi3_3_3}
  \\&\qquad\qquad\qquad\qquad\qquad\quad
  + \tfrac{1}{3} F_{ijka} \gamma^i \psi^{jk}
  + \tfrac{1}{12} F^{ijkl} \gamma_{aij} \psi_{kl}
  - \tfrac{1}{18}  F^{ijkl} \gamma_{ijk} \psi_{al} \big). \notag
\end{alignat}
By using the eqs.~(\ref{eq:Dpsi1}), (\ref{eq:Dpsi4}) and  (\ref{eq:Dpsi4_4}), we obtain
\begin{alignat}{3}
  &\gamma^{a_1\cdots a_{n-2}} 
  \big( \!\!-\! \tfrac{1}{6} F_{ija}{}^c \gamma^{ij} \gamma^b \psi_{bc}
  \!-\! \tfrac{1}{144} F_{ijkl} \gamma^{ijkl} \gamma^b \psi_{ab} 
  \!+\! \tfrac{1}{36} F_{ijk}{}^c \gamma_a{}^{ijk} \gamma^b \psi_{bc}
  \!+\! \gamma^b (D_a \!\!-\!\! F_a) \gamma^c \psi_{bc} \big) \notag
  \\
  &- 2 \sum_{i=1}^{n-2} (-1)^{n-i} 
  \gamma^{a_1 \cdots \check{a}_i \cdots a_{n-2}} (D_a - F_a) \gamma^b \psi^{a_i}{}_b \notag
  \\[-0.2cm]
  &= \gamma^{a_1\cdots a_{n-2}bc} \mathcal{D}_a \psi_{bc}
  - 2 \!\!\!\sum_{i,j=1(i<j)}^{n-2} \!\!\! (-1)^{i+j} 
  \gamma^{a_1 \cdots \check{a}_i \cdots \check{a}_j \cdots a_{n-2}} 
  \mathcal{D}_a \psi^{a_i a_j} \notag
  \\
  &\quad\,
  - \gamma^{a_1\cdots a_{n-2}} 
  \big( - \tfrac{2}{3} F_{ijka} \gamma^i \psi^{jk}
  + \tfrac{1}{9} F^{ijkl} \gamma_{ijk} \psi_{al}
  - \tfrac{1}{6} F^{ijkl} \gamma_{aij} \psi_{kl} \big) \label{eq:Dpsi4_4_4}
  \\
  &\quad\,
  + 2 \sum_{i=1}^{n-2} (\!-\! 1)^{n-i} 
  \gamma^{a_1 \cdots \check{a}_i \cdots a_{n-2}}
  \big( \!\!-\! \tfrac{1}{6} F_{ija}{}^b \gamma^{ij} \psi^{a_i}{}_b
  \!-\! \tfrac{1}{144} F_{ijkl} \gamma^{ijkl} \psi_{a}{}^{a_i} 
  \!+\! \tfrac{1}{36} F_{ijk}{}^b \gamma_a{}^{ijk} \psi^{a_i}{}_b \big) . \notag
\end{alignat}
By using the eqs.~(\ref{eq:Dpsi1}), (\ref{eq:Dpsi2}), (\ref{eq:Dpsi5}) and  (\ref{eq:Dpsi5_5}), we obtain
\begin{alignat}{3}
  &\gamma^{a_1\cdots a_{n-2}} 
  \big( - \tfrac{1}{4} E(e)_{ai} \gamma^{ij} \psi_j
  + \tfrac{1}{4} E(e)_{ai} \psi^i
  + \tfrac{1}{36} E(e)^i{}_i \gamma_a{}^j \psi_j \notag
  \\&\qquad
  - E(A)_{aij} \gamma^{i} \psi^j
  + \tfrac{1}{2} E(A)_{aij} \gamma^{ijk} \psi_k 
  + \tfrac{1}{4} E(A)^{ijk} \gamma_{aij} \psi_k
  - \tfrac{1}{12} E(A)^{ijk} \gamma_{aijkl} \psi^l \notag
  \\&\qquad
  + \tfrac{1}{12} F_{ija}{}^c \gamma^{ij} \gamma^b \psi_{bc}
  + \tfrac{1}{288} F_{ijkl} \gamma^{ijkl} \gamma^b \psi_{ab}
  - \tfrac{1}{72} F_{ijk}{}^c \gamma_a{}^{ijk} \gamma^b \psi_{bc}
  - \tfrac{1}{2} \gamma^b (D_a - F_a) \gamma^c \psi_{bc} \big) \notag
  \\[-0.1cm]
  &- \sum_{i=1}^{n-2} (-1)^{n-i} 
  \gamma^{a_1 \cdots \check{a}_i \cdots a_{n-2}} \eta^{a_i b}
  \big( \tfrac{1}{2} E(e)_{i[a} \gamma^i \psi_{b]} 
  - \tfrac{1}{18} E(e)^i{}_i \gamma_{[a} \psi_{b]}
  + E(A)_{ij[a} \gamma^{ij} \psi_{b]} \notag
  \\[-0,3cm]&\qquad\qquad\qquad\qquad\qquad\qquad\qquad\qquad\quad
  + \tfrac{1}{6} E(A)^{ijk} \gamma_{ijk[a} \psi_{b]} 
  - (D_a - F_a) \gamma^c \psi_{bc} \big) \notag
  \\[-0.1cm]
  &= \gamma^{a_1\cdots a_{n-2}bc} \mathcal{D}_b \psi_{ca} 
  - \!\!\!\!\! \sum_{i,j=1(i<j)}^{n-2} \!\!\!\!\! (-1)^{i+j} 
  \gamma^{a_1 \cdots \check{a}_i \cdots \check{a}_j \cdots a_{n-2}} 
  (\mathcal{D}^{a_i} \psi^{a_j}{}_{a} \!-\! \mathcal{D}^{a_j} \psi^{a_i}{}_{a}) \notag
  \\[-0.3cm]
  &\quad\,
  + \sum_{i=1}^{n-2} (-1)^{n-i} 
  \gamma^{a_1 \cdots \check{a}_i \cdots a_{n-2}} \eta^{a_i b}
  \big( \tfrac{1}{4} R_{ijab} \gamma^c \gamma^{ij} \psi_c 
  + \gamma^c DF_{ab} \psi_c + \gamma^c F^2_{ab} \psi_c \label{eq:Dpsi5_5_5}
  \\[-0.3cm]
  &\qquad\qquad\qquad\qquad\qquad\qquad\qquad
  + \tfrac{1}{6} F_{ija}{}^c \gamma^{ij} \psi_{bc}
  + \tfrac{1}{144} F_{ijkl} \gamma^{ijkl} \psi_{ab}
  + \tfrac{1}{36} F_{ijk}{}^c \gamma^{ijk}{}_{a} \psi_{bc} \big) \notag
  \\[0.1cm]
  &\quad\,
  - \gamma^{a_1\cdots a_{n-2}} \big( \tfrac{1}{4} R_{ijab}\gamma^{ij} \psi^b 
  \!+\! DF_{ab} \psi^b \!+\! F^2_{ab} \psi^b 
  \!+\! \tfrac{1}{3} F_{ijka} \gamma^i \psi^{jk}
  \!-\! \tfrac{1}{18} F^{ijkl} \gamma_{ijk} \psi_{al}
  \!+\! \tfrac{1}{12} F^{ijkl} \gamma_{aij} \psi_{kl} \big). \notag
\end{alignat}
By using the eqs.~(\ref{eq:Dpsi3}), (\ref{eq:Dpsi6}) and  (\ref{eq:Dpsi6_6}), we obtain
\begin{alignat}{3}
  &\gamma^{a_1\cdots a_{n-1}} (D^b - F^b) \gamma^a \psi_{ab} \notag
  \\[-0.1cm]
  &+ \sum_{i=1}^{n-1} (-1)^{n-i} \gamma^{a_1 \cdots \check{a}_i \cdots a_{n-1}} \eta^{a a_i} 
  \big( - \tfrac{1}{4} E(e)_{ai} \gamma^{ij} \psi_j
  + \tfrac{1}{4} E(e)_{ai} \psi^i
  + \tfrac{1}{36} E(e)^i{}_i \gamma_{aj} \psi^j \notag
  \\[-0.3cm]&\qquad\qquad
  -  E(A)_{aij} \gamma^{i} \psi^j
  + \tfrac{1}{2} E(A)_{aij} \gamma^{ijk} \psi_k 
  + \tfrac{1}{4} E(A)^{ijk} \gamma_{aij} \psi_k
  - \tfrac{1}{12} E(A)^{ijk} \gamma_{aijkl} \psi^l \notag
  \\&\qquad\qquad
  + \tfrac{1}{12} F_{ija}{}^c \gamma^{ij} \gamma^b \psi_{bc}
  + \tfrac{1}{96} F_{ijkl} \gamma^{ijkl} \gamma^b \psi_{ab}
  - \tfrac{1}{72} F^{ijkc} \gamma_{aijk} \gamma^b \psi_{bc} \notag
  \\&\qquad\qquad
  - \tfrac{1}{2} \gamma^b (D_{a} - F_{a}) \gamma^c \psi_{bc}
  + \gamma^c (D_c - F_c) \gamma^b \psi_{ab} \big) \notag
  \\[0.2cm]
  &= \gamma^{a_1\cdots a_{n-1}a} \mathcal{D}^b \psi_{ab} 
  - \gamma^{a_1\cdots a_{n-1}} 
  \big( \tfrac{1}{12} F_{ijkl} \gamma^{ij} \psi^{kl} 
  - \tfrac{1}{36} F_{klm}{}^i \gamma^{klm} \gamma^j \psi_{ij} \big) \notag
  \\[-0.1cm]
  &\quad\,
  - \sum_{i=1}^{n-1} (-1)^{n-i} 
  \gamma^{a_1 \cdots \check{a}_i \cdots a_{n-1}} \eta^{a a_i} 
  \big( \tfrac{1}{4} R_{ijab}\gamma^{ij} \psi^b + DF_{ab} \psi^b + F^2_{ab} \psi^b
  \label{eq:Dpsi6_6_6}
  \\[-0.3cm]&\qquad\qquad\qquad\qquad\qquad\qquad
  + \tfrac{1}{3} F_{ijka} \gamma^i \psi^{jk}
  + \tfrac{1}{12} F^{ijkl} \gamma_{aij} \psi_{kl}
  - \tfrac{1}{18}  F^{ijkl} \gamma_{ijk} \psi_{al} \big) . \notag
\end{alignat}
By using the eqs.~(\ref{eq:aDpsi1}), (\ref{eq:aDpsi2}), (\ref{eq:aDpsi3}) and (\ref{eq:Dpsi7_7}), we obtain
\begin{alignat}{3}
  &\gamma^{a_1\cdots a_{n-3}} 
  \big( \tfrac{1}{2} E(e)_{ai} \gamma^i \psi^a 
  + \tfrac{3}{2} E(A)_{aij} \gamma^{ij} \psi^a \big) \notag
  \\[-0.1cm]
  &+ \sum_{i=1}^{n-3} (-1)^{n-i} \gamma^{a_1 \cdots \check{a}_i \cdots a_{n-3}} \eta^{a a_i} 
  \big( - \tfrac{1}{2} E(e)_{ai} \gamma^{ij} \psi_j
  + \tfrac{1}{2} E(e)_{ai} \psi^i
  + \tfrac{1}{18} E(e)^i{}_i \gamma_a{}^j \psi_j \notag
  \\[-0.3cm]&\qquad\qquad\qquad
  - 2 E(A)_{aij} \gamma^{i} \psi^j
  + E(A)_{aij} \gamma^{ijk} \psi_k 
  + \tfrac{1}{2} E(A)^{ijk} \gamma_{aij} \psi_k
  - \tfrac{1}{6} E(A)^{ijk} \gamma_{aijkl} \psi^l \big) \notag
  \\[-0.2cm]
  &+ 2 \sum_{i<j}^{n-3} (-1)^{i+j} 
  \gamma^{a_1 \cdots \check{a}_i \cdots \check{a}_j \cdots a_{n-3}} \eta^{a_i [a} \eta^{b] a_j}
  \big( \tfrac{1}{2} E(e)_{i[a} \gamma^i \psi_{b]} 
  - \tfrac{1}{18} E(e)^i{}_i \gamma_{[a} \psi_{b]} \notag
  \\[-0.3cm]
  &\qquad\qquad\qquad\qquad\qquad\qquad\qquad\qquad
  + E(A)_{ij[a} \gamma^{ij} \psi_{b]} 
  + \tfrac{1}{6} E(A)^{ijk} \gamma_{ijk[a} \psi_{b]} \big) \notag
  \\
  &= \gamma^{a_1\cdots a_{n-3}abc} \mathcal{D}_a \psi_{bc} 
  - 6 \!\! \sum_{i<j<k}^{n-3} \!\!\! (-1)^{n+i+j+k} 
  \gamma^{a_1 \cdots \check{a}_i \cdots \check{a}_j \cdots \check{a}_k \cdots a_{n-3}} 
  \mathcal{D}^{[a_i} \psi^{a_j a_k]} \notag
  \\[-0.3cm]
  &\quad
  - \sum_{i=1}^{n-3} (-1)^{n-i}
  \gamma^{a_1 \cdots \check{a}_i \cdots a_{n-3}} \eta^{a a_i} 
  \big( \tfrac{1}{2} R_{ijab}\gamma^{ij} \psi^b + 2 DF_{ab} \psi^b + 2 F^2_{ab} \psi^b \big) 
  \label{eq:Dpsi7_7_7}
  \\[-0.2cm]
  &\quad\,
  - 2 \sum_{i<j}^{n-3} (-1)^{i+j} 
  \gamma^{a_1 \cdots \check{a}_i \cdots \check{a}_j \cdots a_{n-3}} \eta^{a_i [a} \eta^{b] a_j}
  \big( \tfrac{1}{4} R_{ijab} \gamma^c \gamma^{ij} \psi_c 
  + \gamma^c DF_{ab} \psi_c + \gamma^c F^2_{ab} \psi_c \big). \notag
\end{alignat}


\section{Solution of the Eq.~(\ref{eq:cancel}) with $z_i=0$}\label{app:anothersol}


Let us solve the eq.~(\ref{eq:cancel}) by setting $z_1=z_2=z_4=z_5=0$. This means that we do not assume corrections to the local supersymmetry transformations. The parameters of $b_1[1], \cdots, b_1[56]$ can be determined as 
\begin{alignat}{3}
  b_1\![4] &\!=\! - \tfrac{4}{3} b_1\![1] \!\!+\!\! \tfrac{2}{3} b_1\![2] 
  \!\!+\!\! b_1\![3], \notag
  &b_1\![10] &\!=\! -\tfrac{1}{18} b_1\![2] \!\!-\!\! b_1\![5] 
  \!\!+\!\! \tfrac{1}{9} b_1\![6] \!\!+\!\! \tfrac{1}{9} b_1\![8], \notag
  \\
  b_1\![11] &\!=\! \tfrac{1}{2} b_1\![2] \!\!-\!\! b_1\![6], 
  &b_1\![12]  &\!=\! b_1\![5], \notag
  \\
  b_1\![13] &\!=\! - \tfrac{3}{4} b_1\![2] \!\!+\!\! b_1\![6], 
  &b_1\![15] &\!=\! - b_1\![2] \!\!+\!\! 2 b_1\![6] \!\!+\!\! 2 b_1\![7], \notag
  \\
  b_1\![16] &\!=\! - b_1\![1] \!\!-\!\! \tfrac{1}{4} b_1\![2] 
  \!\!-\!\! \tfrac{9}{2} b_1\![5] \!\!+\!\! b_1\![6] \!\!+\!\! \tfrac{1}{2} b_1\![8], 
  &b_1\![17] &\!=\! - \tfrac{1}{4} b_1\![2] \!\!+\!\! \tfrac{1}{2} b_1\![6] 
  \!\!+\!\! \tfrac{1}{4} b_1\![14], \notag
  \\
  b_1\![18] &\!=\! \tfrac{1}{8}b_1\![2] \!\!-\!\! \tfrac{1}{4}b_1\![6],
  &b_1\![19] &\!=\! \tfrac{1}{6} b_1\![2] \!\!-\!\! \tfrac{1}{3} b_1\![6] 
  \!\!-\!\! \tfrac{1}{3} b_1\![8] \!\!+\!\! \tfrac{1}{6} b_1\![14], \notag
  \\
  b_1\![20] &\!=\! \tfrac{1}{6} b_1\![2] \!\!-\!\! \tfrac{1}{3} b_1\![6] 
  \!\!+\!\! \tfrac{1}{2} b_1\![9] \!\!-\!\! \tfrac{1}{6} b_1\![14], 
  &b_1\![21] &\!=\! \tfrac{2}{3} b_1\![1] \!\!-\!\! \tfrac{1}{3} b_1\![2] 
  \!\!+\!\! \tfrac{1}{3} b_1\![6] \!\!-\!\! \tfrac{1}{3} b_1\![8], \notag
  \\
  b_1\![22] &\!=\! \tfrac{1}{18} b_1\![2] \!\!-\!\! \tfrac{1}{9} b_1\![6], 
  &b_1\![23] &\!=\! - \tfrac{1}{4} b_1\![5] \!\!-\!\! \tfrac{1}{36} b_1\![8], \notag
  \\
  b_1\![24] &\!=\! \tfrac{1}{32} b_1\![5],
  &b_1\![25] &\!=\! \tfrac{1}{3} b_1\![2] \!\!+\!\! \tfrac{4}{3} b_1\![6] 
  \!\!-\!\! \tfrac{1}{3} b_1\![14], \notag
  \\
  b_1\![26] &\!=\! - \tfrac{4}{3} b_1\![2] \!\!+\!\! \tfrac{8}{3} b_1\![6] 
  \!\!+\!\! \tfrac{4}{3} b_1\![14], 
  &b_1\![27] &\!=\! - b_1\![2] \!\!+\!\! 2 b_1\![6] \!\!+\!\! b_1\![14], \notag
  \\
  b_1\![28] &\!=\! - \tfrac{1}{3} b_1\![2] \!\!+\!\! \tfrac{2}{3} b_1\![6] 
  \!\!+\!\! \tfrac{1}{3} b_1\![14], 
  &b_1\![29] &\!=\! \tfrac{1}{2} b_1\![2] \!\!-\!\! b_1\![6], \notag
  \\
  b_1\![30] &\!=\! \tfrac{1}{3} b_1\![2] \!\!-\!\! 3 b_1\![5] 
  \!\!-\!\! \tfrac{2}{3} b_1\![6] \!\!-\!\! \tfrac{1}{3} b_1\![8], 
  &b_1\![31] &\!=\! - \tfrac{1}{4} b_1\![2] \!\!+\!\! b_1\![6] 
  \!\!+\!\! \tfrac{1}{4} b_1\![14], \notag
  \\
  b_1\![32] &\!=\! \tfrac{5}{6} b_1\![2] \!\!+\!\! 3 b_1\![5] \!\!-\!\! \tfrac{4}{3} b_1\![6] 
  \!\!+\!\! \tfrac{1}{3} b_1\![8] \!\!+\!\! \tfrac{1}{6} b_1\![14], 
  &b_1\![33] &\!=\! - b_1\![2] \!\!+\!\! 2 b_1\![6], \label{eq:solb1_2}
  \\
  b_1\![34] &\!=\! - \tfrac{1}{3} b_1\![2] \!\!+\!\! \tfrac{2}{3} b_1\![6], 
  &b_1\![35] &\!=\! \tfrac{1}{2} b_1\![2] \!\!-\!\! b_1\![6], \notag
  \\
  b_1\![36] &\!=\! \tfrac{1}{9} b_1\![2] \!\!-\!\! b_1\![5] \!\!-\!\! \tfrac{2}{9} b_1\![6] 
  \!\!-\!\! \tfrac{1}{9} b_1\![8],
  &b_1\![37] &\!=\! - \tfrac{1}{36} b_1\![2] \!\!-\!\! \tfrac{3}{4} b_1\![5] 
  \!\!+\!\! \tfrac{1}{18} b_1\![6] \!\!+\!\! \tfrac{1}{36} b_1\![8], \notag
  \\
  b_1\![38] &\!=\! - 2 b_1\![5], 
  &b_1\![39] &\!=\! \tfrac{1}{4} b_1\![5], \notag
  \\
  b_1\![40] &\!=\! \tfrac{1}{2} b_1\![5], 
  &b_1\![41] &\!=\! - b_1\![2] \!\!-\!\! 3 b_1\![3] \!\!+\!\! 2 b_1\![6] 
  \!\!+\!\! 2 b_1\![8] \!\!-\!\! b_1\![14], \notag
  \\
  b_1\![42] &\!=\! - 3 b_1\![5] \!\!-\!\! \tfrac{1}{3} b_1\![8], 
  &b_1\![43] &\!=\! - \tfrac{1}{3} b_1\![2] \!\!+\!\! 3 b_1\![5] 
  \!\!+\!\! \tfrac{2}{3} b_1\![6] \!\!+\!\! \tfrac{1}{3} b_1\![8], \notag
  \\
  b_1\![44] &\!=\! \tfrac{1}{3} b_1\![2] \!\!-\!\! 3 b_1\![5] 
  \!\!-\!\! \tfrac{2}{3} b_1\![6] \!\!-\!\! \tfrac{1}{3} b_1\![8], 
  &b_1\![45] &\!=\! - 2 b_1\![1] \!\!+\!\! \tfrac{5}{4} b_1\![2] 
  \!\!-\!\! b_1\![6] \!\!+\!\! \tfrac{1}{4} b_1\![14], \notag
  \\
  b_1\![46] &\!=\! -2 b_1\![1] \!\!+\!\! \tfrac{3}{2} b_1\![2] \!\!-\!\! 2 b_1\![6], 
  &b_1\![47] &\!=\! - \tfrac{1}{3} b_1\![2] \!\!+\!\! \tfrac{2}{3} b_1\![6] 
  \!\!-\!\! 2 b_1\![7] \!\!-\!\! \tfrac{2}{3} b_1\![14], \notag
  \\
  b_1\![48] &\!=\! \tfrac{2}{3} b_1\![2] \!\!+\!\! b_1\![3] \!\!-\!\! \tfrac{4}{3} b_1\![6] 
  \!\!-\!\! \tfrac{2}{3} b_1\![8] \!\!+\!\! b_1\![9], 
  &b_1\![49] &\!=\! - \tfrac{3}{2} b_1\![3] \!\!-\!\! b_1\![6] \!\!+\!\! b_1\![8], \notag
  \\
  b_1\![50] &\!=\! - b_1\![2] \!\!+\!\! 4 b_1\![6] \!\!+\!\! b_1\![14], 
  &b_1\![51] &\!=\! - b_1\![2] \!\!+\!\! 4 b_1\![6] \!\!+\!\! b_1\![14], \notag
  \\
  b_1\![52] &\!=\! \tfrac{5}{3} b_1\![2] \!\!+\!\! 2 b_1\![3] \!\!+\!\! 6 b_1\![5] 
  \!\!-\!\! \tfrac{4}{3} b_1\![6] \!\!-\!\! \tfrac{2}{3} b_1\![8] \!\!+\!\! \tfrac{1}{3} b_1\![14], \;\;
  &b_1\![53] &\!=\! - b_1\![2] \!\!-\!\! b_1\![14], \notag
  \\
  b_1\![54] &\!=\! 6 b_1\![5] \!\!-\!\! \tfrac{4}{3} b_1\![6] \!\!+\!\! \tfrac{4}{3} b_1\![7] 
  \!\!+\!\! \tfrac{2}{3} b_1\![8] \!\!+\!\! 2 b_1\![9],
  &b_1\![55] &\!=\! - 2 b_1\![1] \!\!+\!\! b_1\![2] \!\!+\!\! 2 b_1\![7] 
  \!\!+\!\! \tfrac{1}{2} b_1\![14], \notag
  \\
  b_1\![56] &\!=\! \tfrac{2}{3} b_1\![1] \!\!+\!\! \tfrac{1}{6} b_1\![2] 
  \!\!+\!\! 3 b_1\![5] \!\!-\!\! \tfrac{2}{3} b_1\![6] \!\!-\!\! \tfrac{1}{3} b_1\![8]. \notag
\end{alignat}
Thus we have 9 free parameters, $b_1[1], \cdots, b_1[3], b_1[5], \cdots, b_1[9], b_1[14]$. At the same time, the parameters $b_2[1], \cdots, b_2[22]$ are fixed as
\begin{alignat}{3}
  b_2[4] &= - 2 b_2[1] - \tfrac{1}{2} b_2[3],
  &b_2[5] &= \tfrac{1}{2} b_2[2],
  &b_2[6] &= - \tfrac{1}{9} b_2[2], \notag
  \\
  b_2[7] &= 2 b_2[1] - b_2[2] - \tfrac{1}{2} b_2[3], 
  &b_2[8] &= \tfrac{1}{2} b_2[2] + \tfrac{1}{2} b_2[3], 
  &b_2[9] &= \tfrac{1}{2} b_2[2] + \tfrac{1}{2} b_2[3], \notag
  \\ 
  b_2[10] &= \tfrac{1}{6} b_2[2] + \tfrac{1}{6} b_2[3], 
  &b_2[11] &= 2 b_2[1], 
  &b_2[12] &= - \tfrac{1}{2} b_2[1], \notag
  \\
  b_2[13] &= - \tfrac{1}{2} b_2[3], 
  &b_2[14] &= - \tfrac{4}{3} b_2[2], 
  &b_2[15] &= \tfrac{2}{3} b_2[2], \label{eq:solb2_2}
  \\ 
  b_2[16] &= - \tfrac{2}{3} b_2[2], 
  &b_2[17] &= 4 b_2[1] - 2 b_2[2] - 2 b_2[3], 
  &\;\;b_2[18] &= \tfrac{2}{3} b_2[2], \notag
  \\
  b_2[19] &= - \tfrac{1}{2} b_2[2] - \tfrac{1}{2} b_2[3], 
  &b_2[20] &= 0, 
  &b_2[21] &= - \tfrac{1}{12} b_2[2], \notag
  \\ 
  b_2[22] &= - \tfrac{1}{12} b_2[2] - \tfrac{1}{12} b_2[3]. \qquad \notag
\end{alignat}
And we have 3 additional parameters, $b_2[1], b_2[2], b_2[3]$.


\section{Tips for the Mathematica Codes}\label{app:Code}


In this appendix, we explain the Mathematica codes used in this paper. The codes in the appendices \ref{app:basis}, \ref{app:bi}, \ref{app:ddi} and \ref{app:eomid} are mainly defined in a notebook of ``ActVarBase.nb''. Some codes in the appendices \ref{app:bi} and \ref{app:eomid} are written in a notebook of ``ActVarBiaEom.nb''. The code in the appendix \ref{app:gamma} is defined in ``GammaProduct.nb''. The codes in appendix \ref{app:varsusy} are defined in notebooks of ``V1\_SusyTr\_(DF)2DDFeP2.nb'', ``V2\_SusyTr\_(DF)3eDP2.nb'', ``V4\_SusyTr\_(DF)4e.nb'' and ``V5\_SusyTr\_F(DF)2DDFeP.nb''. There are more notebooks to complete the results in this paper, but we just put ``ActVarBase.nb'', ``ActVarBiaEom.nb'' and ``GammaProduct.nb'' in \cite{Mathcodes}.

In order to explain details of the codes, let us pick up the variation $V_1=[e(DF)^2DDF\bar{\epsilon}\gamma\psi_2]$. As an example of the TNS in $V_1$, let us choose
\begin{alignat}{3}
  \text{TNS} = eD_{f_1}F_{f_2 f_3 f_4 f_5}D_{f_6}F_{f_7 f_8 f_9 f_{10}}
  D_{f_{11}}D_{f_{12}}F_{f_2 f_7 f_{11} f_{12}}\bar{\epsilon}\gamma_{f_1 f_5 f_6 f_8 f_9 f_{10}}\psi_{f_3 f_4}. \label{eq:TNS}
\end{alignat}
From this, it is possible to extract the IND like
\begin{alignat}{3}
  &\text{IND} = \big\{\{f_1\},\{f_2,f_3,f_4,f_5\},\{f_6\},\{f_7,f_8,f_9,f_{10}\},\{f_{11}\},\{f_{12}\}, \notag
  \\
  &\qquad\quad\;\,\{f_2,f_7,f_{11},f_{12}\},\{f_3,f_4\},\{f_1,f_5,f_6,f_8,f_9,f_{10}\}\big\}, \label{eq:IND}
\end{alignat}
and the RNK is written by
\begin{alignat}{3}
  \text{RNK}=\{1,4,1,4,1,1,4,2,\text{LEG}\}, \qquad \text{LEG}=6. \label{eq:RNK}
\end{alignat}
Here we put the indices of the gamma matrix at the end of the list, because it affects the running time of the Mathematica codes when LEG is close to 11. The adjacency matrix is defined so that $(i,j)$ component is equal to the number of intersections between $i$th set and $j$th set. The representative matrix of the adjacency matrices of the IND, which we call the ADJ, is given by
\begin{alignat}{3}
  \text{ADJ} &= 
  \begin{pmatrix}
    0 & 0 & 0 & 0 & 0 & 0 & 0 & 0 & 1 \\
    0 & 0 & 0 & 0 & 0 & 0 & 1 & 2 & 1 \\
    0 & 0 & 0 & 0 & 0 & 0 & 0 & 0 & 1 \\
    0 & 0 & 0 & 0 & 0 & 0 & 1 & 0 & 3 \\
    0 & 0 & 0 & 0 & 0 & 0 & 1 & 0 & 0 \\
    0 & 0 & 0 & 0 & 0 & 0 & 1 & 0 & 0 \\
    0 & 1 & 0 & 1 & 1 & 1 & 0 & 0 & 0 \\
    0 & 2 & 0 & 0 & 0 & 0 & 0 & 0 & 0 \\
    1 & 1 & 1 & 3 & 0 & 0 & 0 & 0 & 0 
  \end{pmatrix}. \label{eq:ADJ}
\end{alignat}
The ADJs are symmetric matrices and their diagonal components are zero, since $[F]$, $[\gamma]$ and $[\psi_2]$ are antisymmetric tensors. By definition, the sum of each row or each column of the ADJ is equal to the corresponding rank of the TNS. Since the order of two $[DF]$s can be exchangeable in the (\ref{eq:TNS}), several adjacency matrices correspond to the same index structure in general. Thus we should consider the conjugacy class of the adjacency matrices to obtain the ADJ, which is obtained by dividing the adjacency matrices by the permutation group of the identical index structure. In this case, the permutation group of the indices is given by
\begin{alignat}{3}
  \text{PRM} = \{\{1,2,3,4\},\{3,4,1,2\}\}. \label{eq:PRM}
\end{alignat}
Each number represents the position of the row (column). The first set does not change the adjacency matrices, but the second set exchange the first row (column) with third one and the second row (column) with the fourth one.

In order to classify all possible terms in $V_1$, we trace the above procedure in the reverse order. That is, first we classify all possible representative matrices, the ADJs, of the adjacency matrices. Second, we figure out the basis of the INDs. Third, we write down the Bianchi identities, the commutation relations and the dimension dependent identities in terms of the basis of the INDs. Finally we solve these identities and relations to obtain the basis for the TNSs.

\subsection{Generate the basis for tensors} \label{app:basis}

First, let us list up all possible ADJs. This is done by using the command \textsf{Adjmat}.\\

\noindent
\textsf{Adjmat[rank0\_, pm0\_]}
\begin{quote}
  Generates a list of representatives of the adjacency matrices. \textsf{rank0}$=$RNK and \textsf{pm0}$=$PRM. 
  This command uses \textsf{Genadjmat[...]} to generate adjacency matrices by
  filling each row step by step. \textsf{Noradjmat[...]} is used to select representatives of the conjugacy classes. 
\end{quote}

\hspace{-0.2cm}
\textsf{Genadjmat[matl0\_, rank0\_, p0\_]}
\begin{quote}
  Generates adjacency matrices of \textsf{p0}-th row. 
  \textsf{matl0} is a list of adjacency matrices which are filled up to $(\textsf{p0}-1)$-th row. 
  \textsf{rank0}$=$RNK.
\end{quote}

\hspace{-0.2cm}
\textsf{Noradjmat[matl0\_, pm0\_]}
\begin{quote}
  Select representative matrices of the conjugacy classes.
  \textsf{matl0} is the adjacency matrices and \textsf{pm0}$=$PRM.
\end{quote}

\vspace{0.2cm}
The index list IND is easily produced by the representative matrix ADJ. However, we have not imposed tensor symmetries yet, so we use builtin commands \textsf{TensorContract} and \textsf{TensorReduce} to assign an unique label to the IND. This can be executed by the command \textsf{Canlist}.\\

\noindent
\textsf{Canlist[matl0\_, tns0\_, rotl0\_]}
\begin{quote}
  Returns \{\textsf{indl, canl, hlist}\}. 
  The \textsf{indl} is a list of INDs. The \textsf{canl} is a list of H[$h$]s for the \textsf{indl}, 
  and $h$ is a compressed number which is uniquely mapped from an information on the contractions of the IND. 
  The \textsf{hlist} is a set of $h$s.
  The \textsf{matl0} is a list of ADJs and the \textsf{tns0} characterizes the tensor symmetries.
  The \textsf{rotl0} will be used if we represent $[F]$ in terms of gauge field $[A]$.
  This command uses \textsf{MatInd[...]} to generate INDs from ADJs.
  \textsf{CanForm[...]} is used to generate H[$h$]s from INDs. 
\end{quote}

\hspace{-0.2cm}
\textsf{MatInd[adjmat0\_, fn0\_, ind0\_]}
\begin{quote}
  Generates the IND from the \textsf{adjmat0}$=$ADJ. 
  The \textsf{fn0} is a list of characters $f_i$ used for the IND. 
  The \textsf{ind0} is an initial value of IND wih $f_i=0$. 
\end{quote}

\hspace{-0.2cm}
\textsf{CanForm[indf0\_, fn0\_, tns0\_, rotl0\_]}
\begin{quote}
  Generates a canonical form H[$h$] for the IND. 
  The \textsf{indf0} is a flattened version of the IND.
  The \textsf{fn0} is a list of characters $f_i$ used for the IND. 
  The \textsf{tns0} characterizes tensor symmetries which are written by the Mathematica notation.
  The \textsf{rotl0} will be nontrivial if we represent $[F]$ in terms of gauge field $[A]$.
\end{quote}

The list of INDs generated by the \textsf{Canlist} basically gives the basis for the INDs. If we represent $F_{abcd}$ in terms of gauge field $A_{abc}$, however, we should carefully reduce the number of the list. In order to obtain the basis for the INDs and a map from H[$h$] to the term in the basis, we execute \textsf{Baselist}. \\

\noindent
\textsf{Baselist[indl0\_, canl0\_, hlist0\_, V0\_]}
\begin{quote}
  Returns \{\textsf{indl, dig, rep}\}.
  The \textsf{indl} is the basis for the INDs. The \textsf{dig} is a number of digits for maximum of $h$s.
  The \textsf{rep} is the map from H[$h$] to the terms in the \textsf{indl}.
  \{\textsf{indl0}, \textsf{canl0}, \textsf{hlist0}\} is a list generated by the \textsf{Canlist}. 
  \textsf{V0} is a label for the IND, and we assign \textsf{V0} $=$ V1 for $V_1$, for example.
\end{quote}

\vspace{0.2cm}
Now we know the map from H[$h$] into the terms in the basis, so it is possible to classify the IND in terms of the basis.
To execute this, we use \textsf{ClassifyInd}.\\

\noindent
\textsf{ClassifyInd[basel0\_, indf0\_, fn0\_, tens0\_, rotl0\_]}
\begin{quote}
  Classifies \textsf{indf0}, flattened version of the IND, in terms of the basis of \textsf{basel0} 
  which is generated by the \textsf{Baselist}.
  The \textsf{fn0} is a list of characters $f_i$ used for the IND. 
  The \textsf{tns0} characterizes the tensor symmetries in the Mathematica language.
  The \textsf{rotl0} is used if we represent $[F]$ in terms of gauge field $[A]$.
\end{quote}

\subsection{Generate the Bianchi identities and the commutation relations} \label{app:bi}

By using the command \textsf{Baselist}, it is possible to generate the basis for the INDs. However, the Bianchi identities and the commutation relations have not been imposed yet. In the case of $V_1$, we should take into account the Bianchi identities for two $[DF]$s and $[DDF]$, and the commutation relations for $[DDF]$. In the case of $V_2$, we should do the Bianchi identities for three $[DF]$s and $[\mathcal{D}\psi_2]$, and so on. A list of the Bianchi identities is constructed by using a command \textsf{Bianchilist}, which invokes a sub-routine \textsf{Bianchi}.\\

\noindent
\textsf{Bianchilist[basel0\_, rank0\_, psl0\_, fn0\_, tns0\_, rotl0\_, V0\_]}
\begin{quote}
  Generate a list of the Bianchi identities for the basis \textsf{basel0} of the INDs.
  The arguments are almost the same as the command \textsf{Bianchi}.
  \textsf{V0} is the label for the IND.
\end{quote}

\hspace{-0.2cm}
\noindent
\textsf{Bianchi[basel0\_, indf0\_, rank0\_, psl0\_, fn0\_, tns0\_, rotl0\_]}
\begin{quote}
  Generate the Bianchi identities and the commutation relations for the \textsf{indf0}=IND.
  The \textsf{basel0} is the basis for the INDs and \textsf{rank0}$=$RNK.
  The \textsf{psl0} is a list of positions of indices with respect to $[R]$, $[DF]$, $[DD]$ and $[\mathcal{D}\psi_2]$.
  The Bianchi idendities for the $[DD]$ contain \textsf{VBD[$i$]}, 
  and those for the $[\mathcal{D}\psi_2]$ do \textsf{VBP[$i$]}.
  The \textsf{fn0} is a list of characters $f_i$ used for IND,
  The \textsf{tns0} characterizes the tensor symmetries, and
  the \textsf{rotl0} is used if we represent $[F]$ in terms of the gauge field $[A]$.
  This command uses the \textsf{ClassifyInd[...]}.
\end{quote}

\vspace{0.2cm}
In the above codes, the terms in the second line of the eq.~(\ref{eq:V2V4}) are  symbolically abbreviated as $3\, \textsf{VBP[$i$]}$, and the terms in the second and third lines of eq.~(\ref{eq:V1V1}) are done as $2\, \textsf{VBD[$i$]}$. Here $i$ runs from 1 to \#TNS for the fixed LEG. In order to express the \textsf{VBP[$i$]} in terms of the other classes of the variations, we use a command \textsf{VBPDP2toDFP}, which is defined in the notebook of ``ActVarBiaEom.nb''. Since the \textsf{VBD[$i$]} includes the Riemann tensor, it is irrelevant as far as the Fig.~\ref{fig:O(R0)} is concerned. \\

\noindent
\textsf{VBPDP2toDFP[rank0\_, basel0\_, bia0\_, var0\_, V0\_, Vvar0\_]}
\begin{quote}
  Generate a replacement list of \{\textsf{VBP[$n,i$]} $\to$ \textsf{Vvar0}[$m,j$]s, $\cdots$\}. 
  Here $n \,(m)$ represents the value of LEG and $i \,(j)$ takes from 1 to \#TNS for each $n\,(m)$.
  In the case of $V_2$, \textsf{V0}$=$V2 and \textsf{Vvar0}$=$V4.
  The \textsf{rank0} is the RNK of \textsf{V0}[LEG]. The \textsf{basel0} is the basis for the \textsf{V0}[LEG],
  and the \textsf{var0} is the basis for the \textsf{Vvar0}.
  The \textsf{bia0} is the list of the Bianchi identities generated by the command \textsf{Bianchilist} for \textsf{V0}.
\end{quote}

\subsection{Generate the dimension dependent identities} \label{app:ddi}

Let us consider the dimension dependent identities. Note that the TNS (\ref{eq:TNS}) has twelve contracted indices, so if these twelve indices are completely antisymmetrized in eleven dimensions, we obtain a relation like
\begin{alignat}{3}
  eD_{[f_1}F_{f_2 f_3 f_4 f_5}D_{f_6}F_{f_7 f_8 f_9 f_{10}}
  D_{f_{11}}D_{f_{12}]}F_{f_2 f_7 f_{11} f_{12}}\bar{\epsilon}\gamma_{f_1 f_5 f_6 f_8 f_9 f_{10}}\psi_{f_3 f_4} = 0. \label{eq:TNSanti}
\end{alignat}
Of course, the above is trivially zero because of the Bianchi identity $D_{[e}F_{abcd]}=0$, but in this way it is possible to generate the dimension dependent identities.

In principle, if there are more than eleven contracted indices in the term, it is possible to generate dimension dependent identities by making any set of twelve indices antisymmetric. This procedure, however, is not realistic because a number of ways of antisymmetrization becomes very large when the number of contracted indices in the term is much larger than twelve. Thus we will execute another method to generate the dimension dependent identities. That is, we figure out possible ways to insert eleven numbers $\{0,1,\cdots,10\}$ into the contracted indices of the IND. If we find a combination of INDs which vanish identically after the insertions, this gives a dimension dependent identity. The dimensionality is specified since we insert eleven numbers definitely.

We explain our method by taking the IND (\ref{eq:IND}) as the example. Since there are twelve contracted indices in the IND, the multiplicity of $\{0,1,\cdots,10\}$ is expressed like
\begin{alignat}{3}
  \text{MP} = \{ \{0\} \to 4, \{1, 2, 3, 4, 5, 6, 7, 8, 9, 10\} \to 2 \}. \label{eq:MP}
\end{alignat}
This means that we put 0 four times and $1,2,\cdots,10$ two times into the contracted indices. MP is constructed so that smaller number has larger multiplicity, and the multiplicity is even and positive. In the case of twelve contracted indices, the MP is uniquely given by the above, but in general we have a list of MPs.

We insert $\{0,1,\cdots,10\}$ into the contracted indices by following the multiplicity list of the MP. The way of insertion is expressed by NUM, and initially it is given by a vacant IND like
\begin{alignat}{3}
  \text{NUM} = \{ \{\}, \{\}, \{\}, \{\}, \{\}, \{\}, \{\}, \{\}, \{\} \}, \label{eq:NUM0}
\end{alignat}
since no number is inserted at this moment. In the followings, we always consider a list of \{MP, NUM\}s. The initial list of \{MP, NUM\}s is obtained by a command \textsf{MpNumlistSet}.\\

\noindent
\textsf{MpNumlistSet[lfn0\_, rank0\_]}
\begin{quote}
  Generate a list of \{MP, NUM\}s, but the NUMs are empty initially.
  The \textsf{lfn0} is a number of pairs of contracted indices, and \textsf{rank0}$=$RNK.
\end{quote}

\vspace{0.2cm}
The insertion of numbers are executed by extracting them from the MP and adding them to the NUM step by step. After the insertion is completed, the MP becomes empty and the NUM is fully occupied. As an example of the first step, let us consider the ways to insert numbers from the MP to the 9-th position in the NUM (the gamma matrix with 6 indices).
There are two possibilities.
\begin{alignat}{3}
  \{\text{MP}, \text{NUM}\} &= 
  \big\{ \{\{0\} \to 3, \{1, 2, 3, 4, 5\} \to 1, \{6, 7, 8, 9, 10\} \to 2\}\}, \notag
  \\
  &\qquad
  \{ \{\}, \{\}, \{\}, \{\}, \{\}, \{\}, \{\}, \{\}, \{0, 1, 2, 3, 4, 5\} \} \big\},
  \\
  \{\text{MP}, \text{NUM}\} &= 
  \big\{ \{\{0\} \to 4, \{1, 2, 3, 4, 5, 6\} \to 1, \{7, 8, 9, 10\} \to 2\}\}, \notag
  \\
  &\qquad
  \{ \{\}, \{\}, \{\}, \{\}, \{\}, \{\}, \{\}, \{\}, \{1, 2, 3, 4, 5, 6\} \} \big\}. \notag
\end{alignat}
Here we inserted numbers $\{1,\cdots,10\}$ in ascending order because these are in the same multiplicity category and indistinguishable. Notice that the insertion of numbers generates a lot of \{MP, NUM\}s in general. The procedure of the insertion can be done by using a command \textsf{InsertNum}.\\

\noindent
\textsf{InsertNum[rank0\_, p0\_, \{mp0\_, num0\_\}]}
\begin{quote}
  Generate a list of \{MP, NUM\}s by inserting numbers in the MP=\textsf{mp0} 
  into the \textsf{p0}-th position in the NUM=\textsf{num0}. \textsf{rank0}$=$RNK.
\end{quote}

\vspace{0.2cm}
Basically it is possible to apply the command \textsf{InsertNum} to all positions in the NUM. It should be noticed, however, that some set of tensors in the TNS are symmetric under the exchange. In the case of $V_1$, two $[DF]$s are symmetric under the exchange. Then we use a command \textsf{MpNumlistWoS} to fill the 5,6,7,8,9-th positions in the NUM, which we call positions without symmetry. On the other hand, we use a command \textsf{MpNumlistSym} to generate 1,2,3,4-th positions in the NUM, which we call positions with symmetry. An example of the NUM is written as
\begin{alignat}{3}
  \text{NUM} &= 
  \{ \{6\}, \{6,\!7,\!8,\!9\}, \{10\}, \{7,\!8,\!9,\!10\}, \{0\}, \{0\}, \{0,\!1,\!2,\!3\}, 
  \{4,\!5\}, \{0,\!1,\!2,\!3,\!4,\!5\} \}. \label{eq:NUM}
\end{alignat}
There will be a lot of NUMs in general.

\vspace{0.2cm}
\noindent
\textsf{MpNumlistWoS[rank0\_, mpnuml0\_, pl0\_]}
\begin{quote}
  Generate a list of \{MP, NUM\}s by inserting numbers into the positions \textsf{pl0} without symmetry.
  \textsf{rank0}$=$RNK. The \textsf{mpnuml0} is a list of \{MP, NUM\}s before the insertion.
\end{quote}

\noindent
\textsf{MpNumlistSym[rank0\_, mpnuml0\_, pl0\_, sym0\_]}
\begin{quote}
  Generate a list of \{MP, NUM\}s by inserting numbers into the positions \textsf{pl0} with symmetry.
  \textsf{rank0}$=$RNK. The \textsf{mpnuml0} is the list of \{MP, NUM\}s
  obtained by running the command \textsf{MpNumlistWoS}. 
  The \textsf{sym0} represents a partition of $[DF]$s, $\{5,5\}$, in the case of $V_1$. 
  This command uses a command \textsf{InsertNumSym} in a sub-routine.
\end{quote}

\hspace{-0.2cm}
\textsf{InsertNumSym[rank0\_, mp0\_, pl0\_, sym0\_]}
\begin{quote}
  Generate a list of \{MP, NUM\}s by inserting numbers in the MP=\textsf{mp0} into the positions \textsf{pl0} with symmetry.
  The NUMs are vacant initially. \textsf{rank0}$=$RNK. The \textsf{sym0} represents a partition of the symmetric parts. 
\end{quote}

\vspace{0.2cm}
Now we obtained a list of NUM[$a$], where $a=1,\cdots,\#\text{NUM}$. Let us pick up one NUM[$a$]  and try to fit it into the basis of the INDs. In practice, we insert the NUM[$a$] into a linear combination of INDs like
\begin{alignat}{3}
  \sum_{i=1}^{\#\text{IND}} \text{Ci}[i] \, \text{IND}[i] \bigg|_{\text{NUM}[a]} = 
  \bigg( \sum_{i=1}^{\#\text{IND}} c_{ai} \text{Ci}[i] \bigg) \text{NUM}[a]. \label{eq:Ci}
\end{alignat}
Here Ci[$i$] is a coefficient of the IND[$i$] and $c_{ai}$ is the number obtained by fitting the NUM[$a$] into the IND[$i$]. Almost all of $c_{ai}$ are zero. We need to evaluate the above equation for all NUMs, and this can be done by using a command \textsf{CfNum}. \\

\noindent
\textsf{CfNum[fn0\_, indl0\_, indpsl0\_, indkey0\_, num0\_, numps0\_, symrep0\_, 
asym0\_, dvl0\_]}
\begin{quote}
  Generates $\sum_i c_{ai} \text{Ci}[i]$ by fitting \textsf{num0} = NUM[$a$] into 
  a linear combination of the \textsf{indl0}, which is the basis of the INDs. 
  The \textsf{fn0} is a list of characters $f_i$ used for the IND.
  The \textsf{indpsl0} is a set of lists generated by the command \textsf{Makeindps[...]}.
  The \textsf{indkey0} assigns a number to each list in the \textsf{indpsl0}.
  The \textsf{numps0} is a set of lists generated by the command \textsf{Makenumps[...]}.
  The \textsf{symrep0} with \textsf{asym0} gives information on symmetric and antisymmetric
  properties of the TNS. The \textsf{asym0} is nontrivial only for $F_3$.
  The \textsf{dvl0} represents all possible ways to divide given numbers into pairs.
\end{quote}

\hspace{-0.2cm}
\textsf{Makeindps[fn0\_, ind0\_]}
\begin{quote}
  Generate a list of positions of contracted indices for \textsf{ind0} = IND. 
  The \textsf{fn0} is a list of characters $f_i$ used for IND.
  The positions of $f_i$s are labelled by $1,2,3,\cdots$ from the top set of the \textsf{ind0}. 
  The list is composed of pairs of positions for $f_i$s. In the case of the IND (\ref{eq:IND}), 
  this command returns $\{\{1,9\},\{2,7\},\{2,8\},\{2,8\},\{2,9\},\{3,9\},$\\
  $\{4,7\},\{4,9\},\{4,9\},\{4,9\},\{5,7\},\{6,7\}\}$.
\end{quote}

\hspace{-0.2cm}
\textsf{Makenumps[num0\_]}
\begin{quote}
  Generate a list of positions of numbers for \textsf{num0} = NUM. 
  The positions of numbers $\{0,1,\cdots,10\}$ are labelled by $1,2,3,\cdots$ from the top set
  of the \textsf{num0}. The list is composed of sets of positions of numbers. 
  In the case of the NUM (\ref{eq:NUM}), this command returns $\{\{5,6,7,9\},\{7,9\},\{7,9\},\{7,9\},\{8,9\},\{8,9\},\{1,2\},\{2,4\},$\\$\{2,4\},\{2,4\},\{3,4\}\}$.
\end{quote}

\vspace{0.2cm}
So far we represented any linear combination of the INDs by inserting NUM[$a$]s. If such linear combination corresponds to some dimension dependent identity, all coefficients of NUM[$a$]s should vanish identically. Thus, in order to obtain the dimension dependent identities, we should solve
\begin{alignat}{3} 
  \sum_{i=1}^{\#\text{IND}} c_{ai} \text{Ci}[i] =0, \qquad 
  (a=1,\cdots,\#\text{NUM}) \label{eq:eqsCi}
\end{alignat}
for the variables $\text{Ci}[i]$. The number of independent solutions is equal to that of the dimension dependent identities. Note that the number of equations, \#NUM, to be solved is quite huge in general, so we divide \#NUM equations into several sets of equations like $\{\text{Eq}^{(1)}=0, \cdots, \text{Eq}^{(p)}=0, \text{Eq}^{(p+1)}=0, \cdots \}$. We will solve these sets recursively by using a command \textsf{CfSolve}. A solution of $\{\text{Eq}^{(1)}=0, \cdots, \text{Eq}^{(p)}=0\}$ is called $\text{Ci}^{(p)}[i]$, and by using a projection matrix $M^{(p)}$, it is expressed as $\text{Ci}^{(p)} = M^{(p)} \text{Ci}$ in the matrix notation. Then we substitute this solution into Eq$^{(p+1)}=0$ and solve it with respect to Ci. The solution is expressed as $\text{Ci} = M'{}^{(p)} \text{Ci}$ and we obtain $M^{(p+1)} = M^{(p)} M'{}^{(p)}$.\\

\noindent
\textsf{CfSolve[eq0\_, matcf0\_, cfl0\_]}
\begin{quote}
  \textsf{eq0} $=$ Eq$^{(p+1)}$, \textsf{matcf0} $= M^{(p)}$ and \textsf{cfl0} $=$ Ci.
  In this command, \textsf{eq0}$=0$ is solved by substituting 
  $\text{Ci}^{(p)} = M^{(p)} \text{Ci}$, and $M'{}^{(p)}$ is obtained.
  This command returns the projection matrix $M^{(p+1)} = M{}^{(p)} M'{}^{(p)}$.
\end{quote}

\subsection{Generate the corrections to the local supersymmetry} \label{app:eomid}

As explained in the section \ref{subsec:correctsusy}, when we take into account the corrections to the local supersymmetry transformations, these contribute to the variation of the Lagrangian density. Note that these terms partially contain equations of motion for the eleven dimensional supergravity, so it is possible to rewrite those terms as like the eq.~(\ref{eq:susytrcorr2}). The corrections with respect to the variations \textsf{V0} $=V_1, V_2, V_4, V_5$ are generated by a command \textsf{EOMidentity}. \\

\noindent
\textsf{EOMidentity[basel0\_, rank0\_, psl0\_, var0\_, V0\_, VEOM0\_]}
\begin{quote}
  This command consists of \textsf{P2toP2EOM}, \textsf{DP2toDP2EOM} and \textsf{DFtoDFEOM},
  which generate the corrections $[eY^{(\psi)}\bar{\epsilon}\gamma\psi_2]$, 
  $[eY^{(D\psi)}\bar{\epsilon}\gamma\mathcal{D}\psi_2]$ and 
  $[e(DF)\bar{\epsilon} Y^{(A)}]$ in the eq.~(\ref{eq:susytrcorr2}), respectively. 
  The \textsf{psl0} is a list of positions for $\{\psi_2, \mathcal{D}\psi_2, DF\text{s}\}$.
  The \textsf{basel0} is a basis of \textsf{V0}[$n$], and the \textsf{var0} is a set of 
  the bases of all \textsf{V0}[$n$]s. Here $n$ takes values in the LEG.
  The \textsf{rank0} is the RNK of \textsf{V0}[$n$].
  The \textsf{VEOM0} is a symbol for the corrections except for the \textsf{var0}.
\end{quote}

\hspace{-0.2cm}
\textsf{P2toP2EOM[basel0\_, rank0\_, psDF0\_, var0\_, V0\_, VEOM0\_]}
\begin{quote}
  This program generates the corrections of $[eY^{(\psi)}\bar{\epsilon}\gamma\psi_2]$.
  The \textsf{psDF0} represents a list of positions for $[DF]$s, and this is used to extract 
  the corrections of $[e(DF)\bar{\epsilon} Y^{(A)}]$ at the same time.
  In this paper, we use this command for \textsf{V0}$=$V1.
\end{quote}

\hspace{-0.2cm}
\textsf{DP2toDP2EOM[basel0\_, rank0\_, psDF0\_, var0\_, V0\_, VEOM0\_]}
\begin{quote}
  This program generates the corrections of 
  $[eY^{(D\psi)}\bar{\epsilon}\gamma\mathcal{D}\psi_2]$.
  The \textsf{psDF0} represents a list of positions for $[DF]$s, and this is used to extract 
  the corrections of $[e(DF)\bar{\epsilon} Y^{(A)}]$ at the same time.
  In this paper, we use this for \textsf{V0}$=$V2.
\end{quote}

\hspace{-0.2cm}
\textsf{DFtoDFEOM[basel0\_, rank0\_, psDF0\_, var0\_, V0\_, VEOM0\_]}
\begin{quote}
  This program generates the corrections of $[e(DF)\bar{\epsilon} Y^{(A)}]$. 
  In this paper, we use this for \textsf{V0}$=$V4, V5.
\end{quote}

\vspace{0.2cm}
In the above codes, some of the terms in the eq.~(\ref{eq:susytrcorr2}), which are not the same type as \textsf{V0}, are  symbolically represented by \textsf{VEOM0[$n$,$j$]}. In order to express \textsf{VEOM0[$n$,$j$]} in terms of the other variations explicitly, we use a command \textsf{EOMDP2toDFP}, which is defined in the notebook of ``ActVarBiaEom.nb''.\\

\noindent
\textsf{EOMDP2toDFP[rank0\_, psDF0\_, basel0\_, eom0\_, var0\_, V0\_, Vvar0\_, VEOM0\_]}
\begin{quote}
  Generate a list of \{\textsf{VEOM0[$n$,$j$]} $\to$ \textsf{Vvar0}[$n'$,$j'$]s, $\cdots$\}. 
  \textsf{rank0}$=$RNK. The \textsf{psDF0} represents a list of positions for $[DF]$s. 
  The \textsf{basel0} is a basis of \textsf{V0}[$n$], and
  the \textsf{var0} is a set of the bases of all \textsf{Vvar0}[$n'$]s.
  The \textsf{eom0} is the list generated by the \textsf{EOMidentity}, 
  and the \textsf{VEOM0} is the symbol for the other corrections.
\end{quote}

\subsection{The gamma matrix} \label{app:gamma}

When we execute the variations of the basis under the local supersymmetry, we often calculate a product of two gamma matrices. The formula of this product is given by
\begin{alignat}{3}
  \gamma^{a_1\cdots a_m}\gamma_{b_1\cdots b_n} &= 
  \sum_{k=m+n-11}^{\text{Min}(m,n)} (-1)^{\tfrac{1}{2}k(2m-k-1)}
  {}_mC_k \,{}_nC_k \, k!\, \delta^{[a_1\cdots a_k}_{[b_1\cdots b_k} 
  \gamma^{a_{k+1}\cdots a_m]}{}_{b_{k+1}\cdots b_n]}, \label{eq:gammaprod}
\end{alignat}
where $\delta^{a_1\cdots a_k}_{b_1\cdots b_k} 
= \delta^{a_1}_{[b_1}\delta^{a_2}_{b_2}\cdots\delta^{a_k}_{b_k]} 
= \frac{1}{k!}(\delta^{a_1}_{b_1}\delta^{a_2}_{b_2}\cdots\delta^{a_k}_{b_k}
-\delta^{a_1}_{b_2}\delta^{a_2}_{b_1}\cdots\delta^{a_k}_{b_k}+\cdots)$.
Note that 11 in the minimum of $k$ is the dimension of the spacetime. The product of two gamma matrices is executed by the following command.\\

\noindent
\textsf{GammaGamma[gind1\_, gind2\_]}
\begin{quote}
  Expand a product of two gamma matrices. The \textsf{gind1} or the \textsf{gind2} represents an index list 
  of each gamma matrix.
\end{quote}

\subsection{Generate variations of the terms in the effective action under the local supersymmetry} \label{app:varsusy}

The variations of the basis for the effective action under the local supersymmetry are done by using the Mathematica codes.
The eqs.~(\ref{eq:deltaB1}), (\ref{eq:deltaB2}), (\ref{eq:deltaF1}), (\ref{eq:deltaF2}) and (\ref{eq:deltaF3}) are generated by following codes. \\

\noindent
\textsf{B1toV1[indl\_]}, \textsf{B1toV4[indl\_]}, \textsf{B1toV5[indl\_]}
\begin{quote}
  The \textsf{indl} is the basis of $B_1$. These codes generate variations of $B_1$
  under the local supersymmetry, which belong to $V_1$, $V_4$ and $V_5$, respectively.
\end{quote}

\vspace{0.2cm}
\noindent
\textsf{B2toV1[indl\_]}, \textsf{B2toV5[indl\_]}
\begin{quote}
  The \textsf{indl} is the basis of $B_2$. These codes generate variations of $B_2$
  under the local supersymmetry, which belong to $V_1$ and $V_5$, respectively.
\end{quote}

\vspace{0.2cm}
\noindent
\textsf{F1toV1[indl\_]}, \textsf{F1toV2[indl\_]}
\begin{quote}
  The \textsf{indl} is the basis of $F_1$. These codes generate variations of $F_1$ 
  under local supersymmetry, which belong to $V_1$ and $V_2$, respectively.
\end{quote}

\vspace{0.2cm}
\noindent
\textsf{F2toV1[indl\_]}, \textsf{F2toV2[indl\_]}, \textsf{F2toV4[indl\_]}
\begin{quote}
  The \textsf{indl} is the basis of $F_2$. These codes generate variations of $F_2$ 
  under the local supersymmetry, which belong to $V_1$, $V_2$ and $V_4$, respectively.
\end{quote}

\vspace{0.2cm}
\noindent
\textsf{F3toV4[indl\_]}, \textsf{F3toV5[indl\_]}
\begin{quote}
  The \textsf{indl} is the basis of $F_3$. These codes generate variations of $F_3$ 
  under the local supersymmetry, which belong to $V_4$ and $V_5$, respectively.
\end{quote}

The codes which generate variations of $V_1$, $V_2$, $V_4$ and $V_5$ are written in notebooks of ``V1\_SusyTr\_(DF)2DDFeP2.nb'', ``V2\_SusyTr\_(DF)3eDP2.nb'', ``V4\_SusyTr\_(DF)4e.nb'' and ``V5\_SusyTr\_F(DF)2DDFeP.nb'', respectively.

\end{document}